\documentclass[fleqn,10pt]{wlscirep}
\usepackage[utf8]{inputenc} 
\usepackage[T1]{fontenc}    
\usepackage{hyperref}       
\usepackage{url}            
\usepackage{booktabs}       
\usepackage{amsfonts}       
\usepackage{nicefrac}       
\usepackage{microtype}      
\usepackage{xcolor}
\usepackage{colortbl}
\usepackage{array}
\usepackage{algorithm}
\usepackage{algpseudocode}        
\usepackage{graphicx}
\usepackage{caption}
\usepackage{subcaption}
\usepackage{multirow}
\usepackage{pifont}
\usepackage{gensymb}
\usepackage{ulem}
\usepackage{siunitx}
\usepackage{appendix}

\DeclareSIUnit{\molar}{M}
\newcommand{\cmark}{\ding{51}}
\newcommand{\xmark}{\ding{55}}
\DeclareCaptionLabelFormat{Cont}{#1~#2~(continued)}
\title{Accelerating Inhibitor Discovery With A Deep Generative Foundation Model: Validation for SARS-CoV-2 Drug Targets}

\author[1,+]{Vijil Chenthamarakshan}
\author[1,+]{Samuel C.~Hoffman}
\author[2,3]{C. David Owen}
\author[2,3]{Petra Lukacik}
\author[2,3]{Claire Strain-Damerell}
\author[2,3]{Daren Fearon}
\author[4]{Tika R. Malla}
\author[4]{Anthony Tumber}
\author[4]{Christopher J. Schofield}
\author[5]{Helen M.E. Duyvesteyn}
\author[5]{Wanwisa Dejnirattisai}
\author[5]{Loic Carrique}
\author[5]{Thomas S. Walter}
\author[6]{ Gavin R. Screaton}
\author[7]{Tetiana Matviiuk}
\author[1]{Aleksandra Mojsilovic}
\author[8,9]{Jason Crain}
\author[2,3,*]{Martin A. Walsh}
\author[5,*]{David I. Stuart}
\author[1,*]{Payel Das}

\affil[1]{IBM Research, Thomas J.~Watson Research Center, Yorktown Heights, New York, United States of America}
\affil[2]{Diamond Light Source Ltd., Harwell Science and Innovation Campus, OX11 0DE, Didcot, United Kingdom}
\affil[3]{Research Complex at Harwell, Harwell Science and Innovation Campus, OX11 0FA, Didcot, United Kingdom}
\affil[4]{Chemistry Research Laboratory, Department of Chemistry and the Ineos Oxford Institute for Antimicrobial Research, University of Oxford, 12 Mansfield Road, OX1 3TA, Oxford, United Kingdom}
\affil[5]{Division of Structural Biology, University of Oxford, The Wellcome Centre for Human Genetics, Headington, Oxford, United Kingdom}
\affil[6]{Wellcome Centre for Human Genetics, Nuffield Department of Medicine, University of Oxford, Oxford OX3 7BN, United Kingdom}
\affil[7]{Enamine Ltd, Chervonotkatska St, 67, Kyiv, 02094 Ukraine}
\affil[8]{IBM Research Europe, Hartree Centre, Daresbury, WA4 4AD, United Kingdom}
\affil[9]{Department of Biochemistry, University of Oxford, Oxford, OX1 3QU, United Kingdom}
\affil[*]{martin.walsh@diamond.ac.uk,david.stuart@strubi.ox.ac.uk, daspa@us.ibm.com}

\affil[+]{these authors contributed equally to this work}

\usepackage{amsmath,amsfonts,bm}

\def\eqref#1{equation~\ref{#1}}

\def\1{\bm{1}}

\def\rva{{\mathbf{a}}}

\def\rvp{{\mathbf{p}}}
\def\rvq{{\mathbf{q}}}

\def\rvx{{\mathbf{x}}}

\def\rvz{{\mathbf{z}}}

\DeclareMathAlphabet{\mathsfit}{\encodingdefault}{\sfdefault}{m}{sl}
\SetMathAlphabet{\mathsfit}{bold}{\encodingdefault}{\sfdefault}{bx}{n}

\newcommand{\E}{\mathbb{E}}

\begin{abstract}

The discovery of novel inhibitor molecules for emerging drug-target proteins is widely acknowledged as a challenging inverse design problem: Exhaustive exploration of the vast chemical search space is impractical, especially when the target structure or active molecules are unknown. Here we validate experimentally the broad utility of a deep generative framework trained at-scale on protein sequences, small molecules, and their mutual interactions --- that is unbiased toward any specific target. As demonstrators, we consider two dissimilar and relevant SARS-CoV-2 targets: the main protease  and the spike protein (receptor binding domain, RBD). To perform target-aware design of novel inhibitor molecules,  a protein sequence-conditioned  sampling on the generative foundation model is performed. Despite using only the target sequence information, and without performing any target-specific adaptation of the generative model, micromolar-level  inhibition was observed in in vitro experiments for two candidates out of only four synthesized for each target. The most potent spike RBD inhibitor also exhibited activity against several variants in live virus neutralization assays. These results therefore establish that a single, broadly deployable generative foundation model for accelerated hit discovery is effective and efficient, even in the most general case where neither target structure nor binder information is available.
\end{abstract}
\begin{document}

\flushbottom
\maketitle

\thispagestyle{empty}

\section*{Introduction}

\begin{figure}[t]
    \centering
    \includegraphics[width=\textwidth]{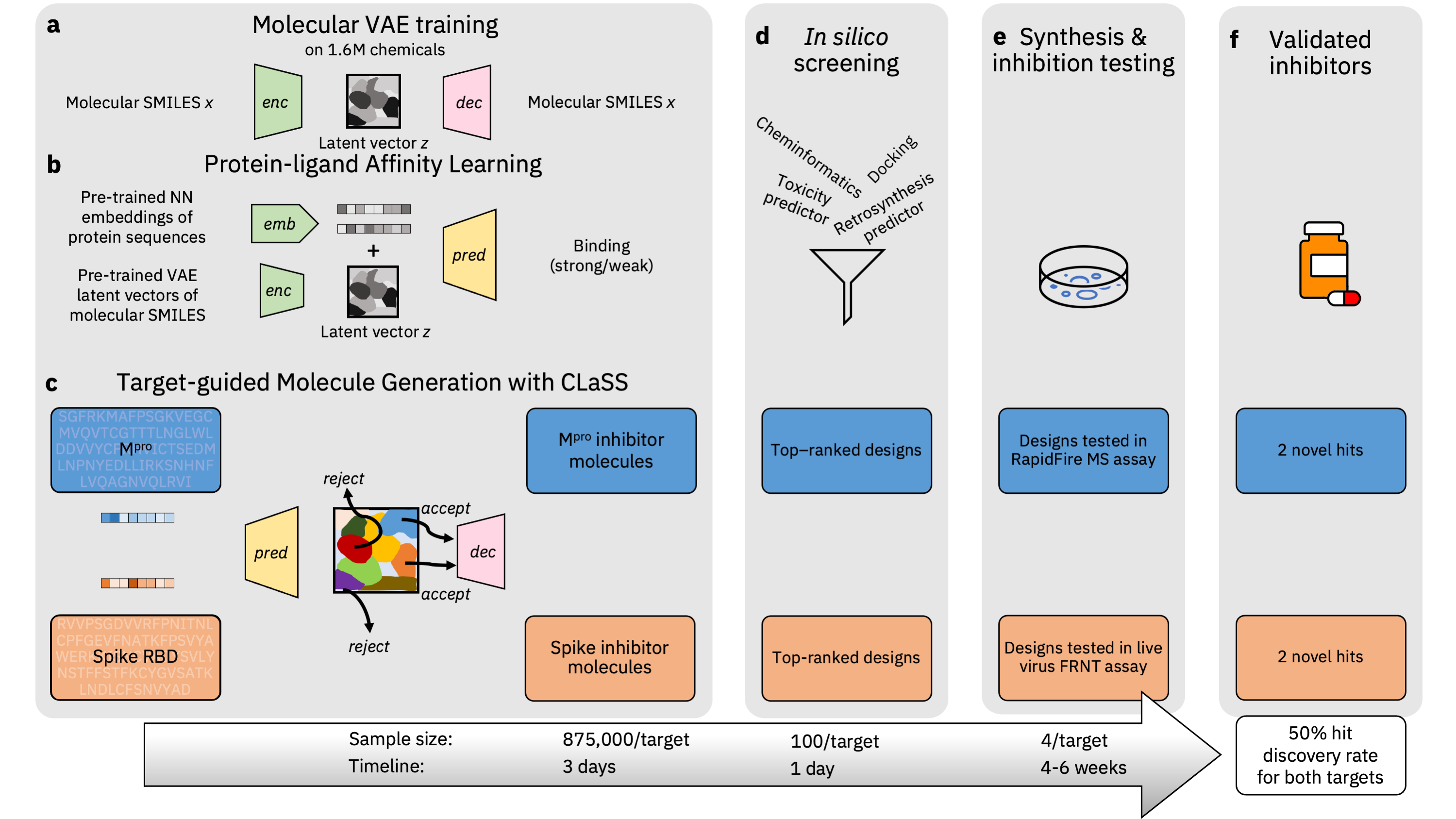}
    \caption{\textbf{Overview of our inhibitor discovery workflow driven by CogMol, a  sequence-guided deep generative foundation model.}
    (\textbf{a--b}) illustrate molecular Variational AutoEncoder (VAE) training on large-scale chemical SMILES ($\rvx$) data  and mapping of  existing protein-ligand  affinity relations on the VAE latent space ($\rvz$) by training a binding predictor, respectively.  For the latter, we leverage pre-trained neural network (NN) embeddings of a large volume of protein sequences. (\textbf{c}) shows a schematic representation of  Controllable Latent Space Sampling or CLaSS, which samples from the model of VAE latent vectors by using the guidance from  a set of molecular property predictors (e.g., protein binding), such that for a given target protein sequence, sampled $\rvz$ vectors corresponding to strong target binding affinity are accepted while vectors corresponding to weak target binding affinity are rejected. The accepted $\rvz$ vectors are then decoded into molecular SMILES.
    (\textbf{d}) Candidates are then ranked and filtered according to chemical properties, docking score to target structure, and predicted retrosynthetic feasibility and toxicity. (\textbf{e}) A small set of prioritized molecules are synthesised, followed by wet lab testing in  specific \textit{in vitro}  assays to confirm target inhibition. (\textbf{f}) In the present case, for each target, of the four molecules tested, two showed promising levels of inhibition. We also report approximate sample sizes and timeline for each stage of our discovery workflow. Note the timeline does not include the training and testing of the generative and predictive machine learning models.}
    \label{fig:pipeline_overview}
\end{figure}

 \textit{De novo} molecular design, the proposing of novel compounds with desired properties, is a
challenging problem with applications in drug discovery and materials engineering. For instance,  a key objective in the drug discovery workflow is to identify candidate molecules, known as hits, that can interact with and inhibit  a known drug-target protein with measurable activity. Searching for hit compounds that serve as the  chemical starting points for further design of drug candidates typically involves  high-throughput screening of libraries containing standard chemical compounds or smaller chemical fragments.  Success rates for this method of hit discovery are between 0.5 and 1 percent~\cite{lloyd2020high}, depending upon the size of the  library screened (typically on the order of $10^4$ entries) and target characteristics. This low success rate is in part due to the immense search space,  now estimated to span between $10^{33}$--$10^{80}$ feasible molecules~\cite{polishchuk2013estimation}, from which only a minute fraction typically possesses the traits sought. Exhaustive enumeration of this vast chemical space is infeasible and selection of compounds to be screened in priority is highly complicated.

In addition to the need for thousands of screening experiments, the initial selection of the library frequently requires detailed structural information on the target protein of interest, which is often not readily available. Further, discovery is often performed using hand-crafted rules and heuristics to link existing fragments and/or to avoid impractical synthetic pathways. Many hit discovery approaches tend to focus on compounds that have similar molecular structures to known hits, whereas more promising compounds could be found in other, previously less explored, molecular structures.
Finally,  hit discovery can be expensive, due to the cost of infrastructure, compounds, and reagents. Consequently, the cost of developing a single new drug is high, reaching up to \$2.8 billion, while the duration from concept to market typically exceeds a decade~\cite{dimasi2016innovation}.
Therefore, a more efficient  approach is urgently needed, to enable distillation of novel and promising molecules from the vast chemical space.  
This  approach will enable experimental validation of a  small selection of  candidates, resulting in a higher hit discovery  rate at reduced time and cost.

Deep learning-based generative models have the potential to enable discovery of novel molecules with desired functionality in a  ``rule-free'' manner, as they aim to first learn a  dense, continuous representation (hereafter referred to as a latent vector)  of known chemicals and then modify the latent vectors to decode into new molecules.  
Such models thus  offer access to previously unexplored chemical space unrestricted by conscious human bias.
However, for the task of target-specific drug-like inhibitor design, an ``inverse molecular design''\cite{zunger2018inverse} approach must be utilized, where the navigation through the learned  chemical representation is guided by  molecular property attributes, such as target inhibition activity and drug-likeness. In the case of designing inhibitors against a new target,  a sufficient amount of  exemplar molecules is required, which is likely unavailable and requires  costly and time-consuming screening experiments to obtain.  As the majority of existing deep generative frameworks (see Sousa, et al.~\cite{review1_sousa2021generative} for a review of generative deep learning for targeted molecule design)  still rely on learning from  target-specific  libraries of binder compounds, they  limit exploration beyond a fixed  library of known and monolithic  molecules, while  preventing  generalization of the machine learning framework toward more novel targets.
As a result, while some studies~\cite{Zhavoronkov2019natbio, merk2018novo, grisoni2021combining}  that use  deep generative models for target-specific inhibitor design  have been experimentally validated, to our knowledge,  demonstrations of of those models to tackle  validated inhibitor discovery  across  dissimilar protein targets, without having access to detailed target-specific prior data (e.g., target structure or  binder library), have not been reported.

Our work demonstrates the real-world applicability of a single, unified inhibitor design framework, based on a deep generative foundation model, across different target proteins simultaneously. The framework  only requires  more readily available  target sequence information to guide the design. Further, the work considers (i) off-target binding of the designed hits  to account for potential downstream adverse effects; (ii) identifying hits even in the case of unknown binders and/or  target structure; (iii)
prioritizing compounds that are readily synthesizable. We employ CogMol~\cite{chenthamarakshan2020cogmol}, a deep generative model,  to propose novel and chemically viable inhibitor designs for two important and distinct SARS-CoV-2 targets --- the main protease (M\textsuperscript{pro}) and the receptor binding domain (RBD) of the spike (S) protein.  The deep  generative framework, built upon large-scale data of chemical molecules, protein sequences, and protein-ligand binding data,  serves as a generative foundation model for target-aware inhibitor molecule design without any further finetuning on target-specific data and can extrapolate to new target sequences not present in the original training data.  The CogMol framework is conceptually similar to recent emergence of ``Foundation Models'' \cite{bommasani2021opportunities, IBMblog}, which are pre-trained on a broad set of unlabeled data and can be used for different downstream tasks with minimal fine-tuning.
A set of novel molecules targeting SARS-CoV-2 proteins,  which was designed  by CogMol, was shared under the Creative Commons  license in April 2020 in the IBM COVID-19 Molecule Explorer platform\cite{cogmolexplorer}. 
Here, we provide the first experimental validation of the broad utility and readiness of the CogMol deep generative framework,   by synthesizing and testing the inhibitory activity of a number of prioritized designs against SARS-CoV-2 M\textsuperscript{pro} and RBD of the S protein.
We further demonstrate the applicability of the binding affinity predictor model used in the CogMol framework by subjecting it to virtual screening of a library of lead-like chemicals and successfully identifying three compounds that were ultimately confirmed to be bound at the active site of the M\textsuperscript{pro} by crystallographic analysis, one of which showed micromolar inhibition.


To our knowledge, the present study provides the first  validated demonstration of a single generative machine intelligence framework that can propose novel and promising inhibitors for different  drug-target proteins with a high success rate, while only using protein sequence information during design. The demonstrated broad-spectrum antiviral activity of the designed spike inhibitor against the SARS-CoV-2 variants of concern further establishes the potential of such a deep generative framework to accelerate and automate the hit discovery  cycle, a process known to suffer from low yield and high attrition rates.

\definecolor{Gray}{RGB}{231, 230, 230}

\begin{figure}[]
\centering
\begin{tabular}{cc}
    \includegraphics[scale=0.3]{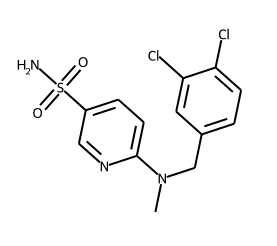} & \includegraphics[scale=0.3]{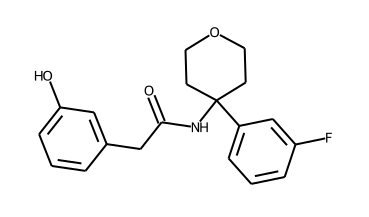} \\
    Z68337194 & Z1633315555 \\
    \scriptsize\texttt{CN(Cc1ccc(Cl)c(Cl)c1)c1ccc(S(N)(=O)=O)cn1} & \scriptsize\texttt{O=C(Cc1cccc(O)c1)NC1(c2cccc(F)c2)CCOCC1} \\
    \addlinespace[-0.5em]
    \multicolumn{2}{c}{\includegraphics[scale=0.3]{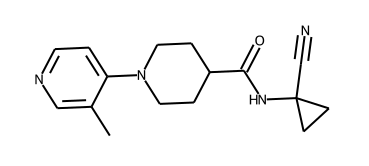}} \\
    \multicolumn{2}{c}{Z1365651030} \\
    \multicolumn{2}{c}{\scriptsize\texttt{Cc1cnccc1N1CCC(C(=O)NC2(C\#N)CC2)CC1}}\\
    \addlinespace[1.5em]

    \includegraphics[scale=0.3]{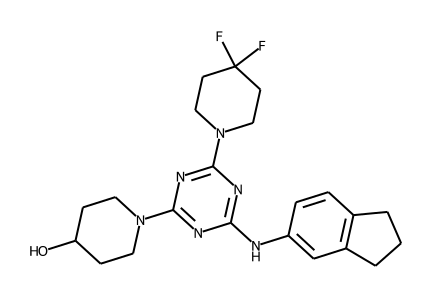} & \includegraphics[scale=0.3]{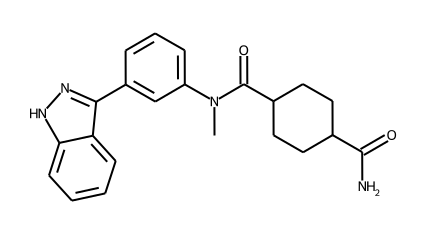} \\
    GXA70 & GXA104 \\
    \scriptsize\texttt{OC1CCN(c2nc(Nc3ccc4c(c3)CCC4)nc(N3CCC(F)(F)CC3)n2)CC1} & \scriptsize\texttt{CN(C(=O)C1CCC(C(N)=O)CC1)c1cccc(-c2n[nH]c3ccccc23)c1} \\
    \addlinespace[-0.5em]
    \includegraphics[scale=0.3]{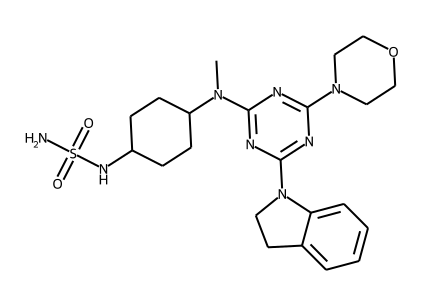} & \includegraphics[scale=0.3]{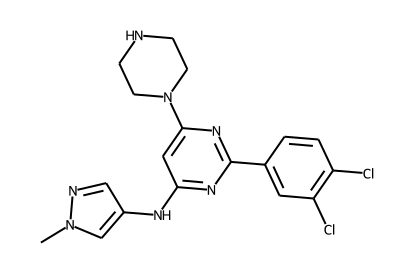} \\
    GXA112 & GXA56 \\
    \scriptsize\texttt{CN(c1nc(N2CCOCC2)nc(N2CCc3ccccc32)n1)C1CCC(NS(N)(=O)=O)CC1} & \scriptsize\texttt{Cn1cc(Nc2cc(N3CCNCC3)nc(-c3ccc(Cl)c(Cl)c3)n2)cn1} \\
    
    \addlinespace[2.5em]
    \includegraphics[scale=0.3]{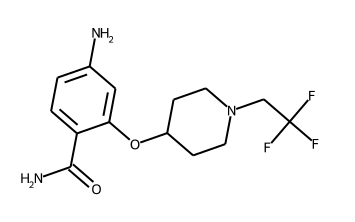} & \includegraphics[scale=0.3]{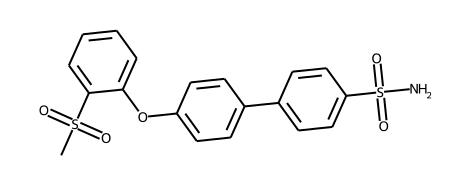} \\
    GEN626 & GEN725\\
    \scriptsize\texttt{NC(=O)c1ccc(N)cc1OC1CCN(CC(F)(F)F)CC1} & \newline \scriptsize\texttt{CS(=O)(=O)c1ccccc1Oc1ccc(-c2ccc(S(N)(=O)=O)cc2)cc1} \\
    \addlinespace[-0.5em]
    \includegraphics[scale=0.3]{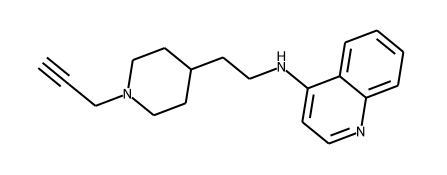} & \includegraphics[scale=0.3]{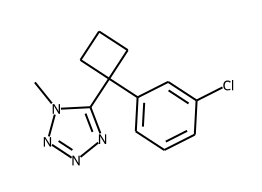} \\
    GEN727 & GEN777 \\
    \scriptsize\texttt{C\#CCN1CCC(CCNc2ccnc3ccccc23)CC1} & \scriptsize\texttt{Cn1nnnc1C1(c2cccc(Cl)c2)CCC1} \\
\end{tabular}
\caption{\textbf{\textit{De novo}-designed and commercially sourced molecules.} Molecules with the prefix ``Z'' are molecules from the Enamine Advanced Collection catalog targeting M\textsuperscript{pro} (top). Molecules with the prefix ``GXA'' are generated candidates targeting M\textsuperscript{pro} (middle) while those with the prefix ``GEN'' are generated candidates targeting the spike RBD (bottom).}
\label{fig:all_mols}
\end{figure}

\section*{Results}

\subsection*{Attribute-conditioned molecule generation with a deep generative model }
The overall  inhibitor discovery  pipeline is described in Figure \ref{fig:pipeline_overview} and consists of three main steps:  (a--c) candidate design in a target-conditioned manner using the deep generative  framework, (d) \textit{in silico} screening for candidate prioritization, and (e) wet lab validation of prioritized molecules. For \textit{de novo} molecule design,   we  used the deep generative framework CogMol as a foundation, which  enables the design of inhibitor molecules for different targets, without requiring training or fine-tuning the model on target-specific data. 
Hereafter, we refer  to machine-designed novel compounds as  \textit{de novo} compounds throughout rest of the paper.

CogMol works as follows: first, it uses   a variational autoencoder (VAE)~\cite{kingma2013auto}, a popular class of deep learning-based generative models,   as the generative foundation (Figure \ref{fig:pipeline_overview}a).
A VAE is comprised of a pair of neural nets -- the  encoder-decoder  pair. The encoder neural network maps the simplified molecular-input line-entry system (SMILES)\cite{weininger1988smiles} string of a molecule into a low-dimensional representation.
We will denote the encoder as $q_\phi(\rvz|\rvx)$, where $\rvz$ is a latent encoding of input SMILES $\rvx$ and $\phi$ represents the encoder parameters. The decoder $p_\theta(\rvx|\rvz))$, which is also a neural network,  then converts the  latent vector $\rvz$ back into the reconstructed SMILES $\rvx$. The  encoder in a VAE is probabilistic in nature as it outputs latent encodings  that are consistent with a Gaussian distribution. The decoder is therefore stochastic --- it samples from the latent distribution to produce an output $\rvx$. The encoder-decoder pair is trained end-to-end to optimize two objectives simultaneously. The first objective includes minimizing a loss term to ensure  accurate reconstruction of an input SMILES  from the corresponding  latent embedding. The second objective consists of a regularization term  to constrain the latent encodings  to a standard normal distribution. The resulting  latent space is continuous,   enabling smooth interpolation as well as random sampling of new molecules  from the latent space.   To  learn meaningful latent molecular representations that have  general knowledge about diverse chemicals,
in CogMol the  VAE is  trained on more than  one and half million small molecules from
public  databases (see the Material and Methods section for details).

Once the chemical latent representation is learned,  CogMol performs attribute-conditioned sampling on that representation to generate entirely new molecular entities with properties biased toward the design specifications. Specifically, the goal is to design novel drug-like molecules with a high binding affinity to the target protein of interest.  Two $\rvz$-based property predictors are used: a drug-likeness (QED) predictor and a  target-molecule binding (strong/weak) predictor. Both predictors used the  $\rvz$ encodings of molecules as input. For the binding predictor,   the protein sequence embeddings from a pre-existing  deep neural net~\cite{Alley_2019} was concatenated with the molecular latent encodings and trained on the general  protein-molecule binding affinity data available in the BindingDB database (Figure \ref{fig:pipeline_overview}b). Performance of the attribute predictors is reported in the Material and Methods section.

Given a target protein sequence of interest, those two predictors are used together to sample molecules with desired properties from the latent space, by using the CLaSS sampling method proposed by Das, et al.~\cite{das2021accelerated}. CLaSS relies on a rejection sampling schema to accept/reject molecules, while sampling from a density model of the $\rvz$ embeddings. Acceptance/rejection criteria are determined by the output probabilities of property predictors.
See the Materials and Methods and the Supplementary Details sections for further details on CLaSS.

Note,  the CogMol generative framework relies on a chemical VAE, a  protein sequence encoder, and a set of molecular property predictors,  all of which are pre-trained on large amount of broad data --- i.e., chemical SMILES,  protein sequences,  and available protein-ligand binding affinities. The  generative framework thus has important information already encoded about  protein sequence homologies,  chemical similarities, and protein-drug binding relations. This allows the framework to serve as a foundation,  as it is instantly adaptable to     different targets, without any further model retraining or fine-tuning on  target-specific data. The approach further  saves  time and cost associated with generating target-specific  binder libraries or resolving the target structure, which are typically considered as privileged information, i.e., not broadly available. The model can  also extrapolate to a target that does not share high similarity with the training data. This is indeed the case for the SARS-CoV-2 targets considered (see SI Table \ref{tab:fasta}) where the lowest Expect value
, a measure of sequence homology (lower values indicate high homology),  with respect to the BindingDB protein sequences is  0.51 (query coverage = 40\%) and 1.9 (query coverage = 26\%) for M\textsuperscript{pro} and spike RBD, respectively.  This analysis implies  that both targets are not significantly similar to the protein sequences in the BindingDB database that was used for training the binding predictor, spike RBD being more distinct than M\textsuperscript{pro}; nor do  they share any significant sequence, structure, or functional similarity to each other.

\subsection*{Candidate prioritization from the machine-designed ligand library}

The next stage includes \textit{in silico} screening of generated candidates (Figure~\ref{fig:pipeline_overview}d) to prioritize them for synthesis and wet lab evaluation. For practical considerations, we sought to keep the number of prioritized machine-designed \textit{de novo} compounds to be synthesized and tested very small --- around 10 for each target, as opposed to screening thousands of existing chemicals  in  a more traditional set-up,  as synthesis of novel chemicals is costly and time-consuming, particularly during a global pandemic. Careful analysis, including machine learning based retrosynthesis predictions, was conducted to define this set. We used a combination of  physicochemical properties (estimated using cheminformatics),   target-molecule binding free energy predicted by docking simulations, and retrosynthesis and toxicity predictions by using machine learning.  For retrosynthesis prediction, we used the IBM RXN platform \cite{schwaller2020predicting} that is based on a transformer neural network trained on chemical reaction data. For toxicity prediction, an in-house neural network-based model trained on publicly available \textit{in vitro} and clinical toxicity data was used. See Material and Methods for details of candidate filtering and prioritization criteria. At the end of the \textit{in silico} screening, the number of candidates per target is around 100, which was further narrowed down to around 10 per target by using the  discretion of Enamine Ltd., the chemical manufacturer. Feasibility of the predicted reaction schema, as evaluated by organic synthetic chemist experts, as well as commercial availability and cost of the predicted reactants, was used to finalize the candidate synthesis list. The final four candidates for each target were chosen based on the synthesis cost and delivery time, as provided by Enamine.

\begin{figure}[t]
    \centering
    \begin{subfigure}[b]{0.4\textwidth}
        \centering
        \begin{tabular}{@{}lc@{}}
            \toprule
            Candidate   & IC$_{50}$ (\si{\micro\molar}) \\ \midrule
            GXA70       & 43           \\
            GXA104      & ---          \\
            GXA112      & 34.2         \\
            GXA56       & ---          \\ \addlinespace[1em]
            Z68337194$^\dagger$   & 35.5         \\
            Z1633315555$^\dagger$ & ---          \\
            Z1365651030$^\dagger$ & ---          \\
            \bottomrule
        \end{tabular}
        \caption{Experimental inhibition}
        \label{tab:IC50}
    \end{subfigure}
    \begin{subfigure}[b]{0.4\textwidth}
        \centering
        \includegraphics[height=12em]{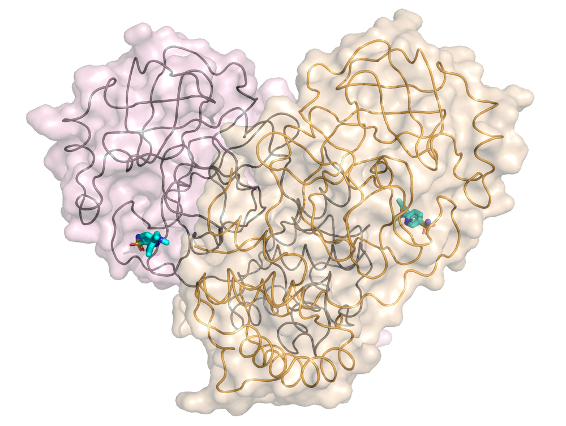}
        \caption{Macro view}
        \label{fig:struct_mpro194_macro}
    \end{subfigure}
    \begin{subfigure}[b]{0.4\textwidth}
        \centering
        \includegraphics[height=12em]{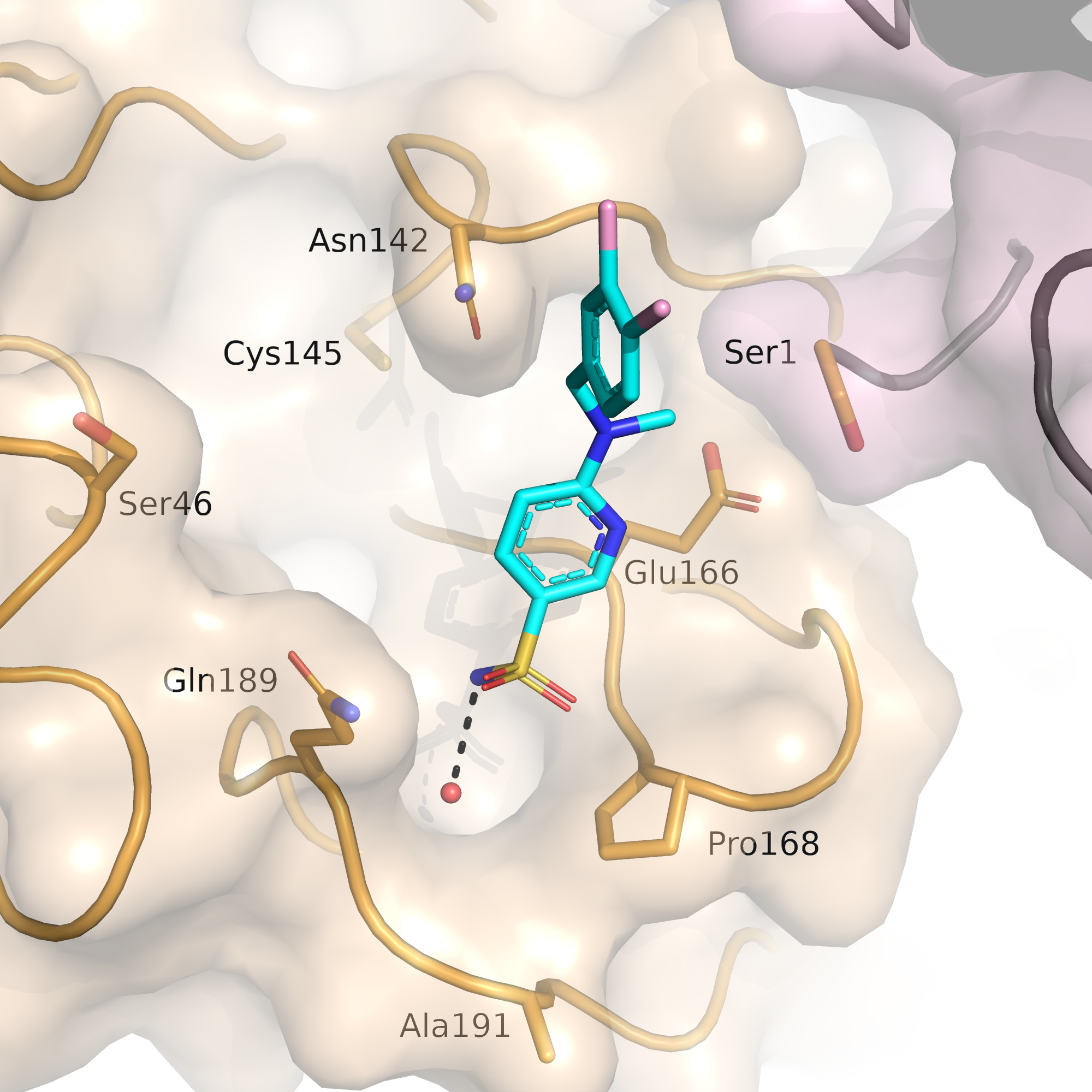}
        \caption{Pocket view}
        \label{fig:struct_mpro194_pocket}
    \end{subfigure}
    \begin{subfigure}[b]{0.4\textwidth}
        \centering
        \includegraphics[height=12em]{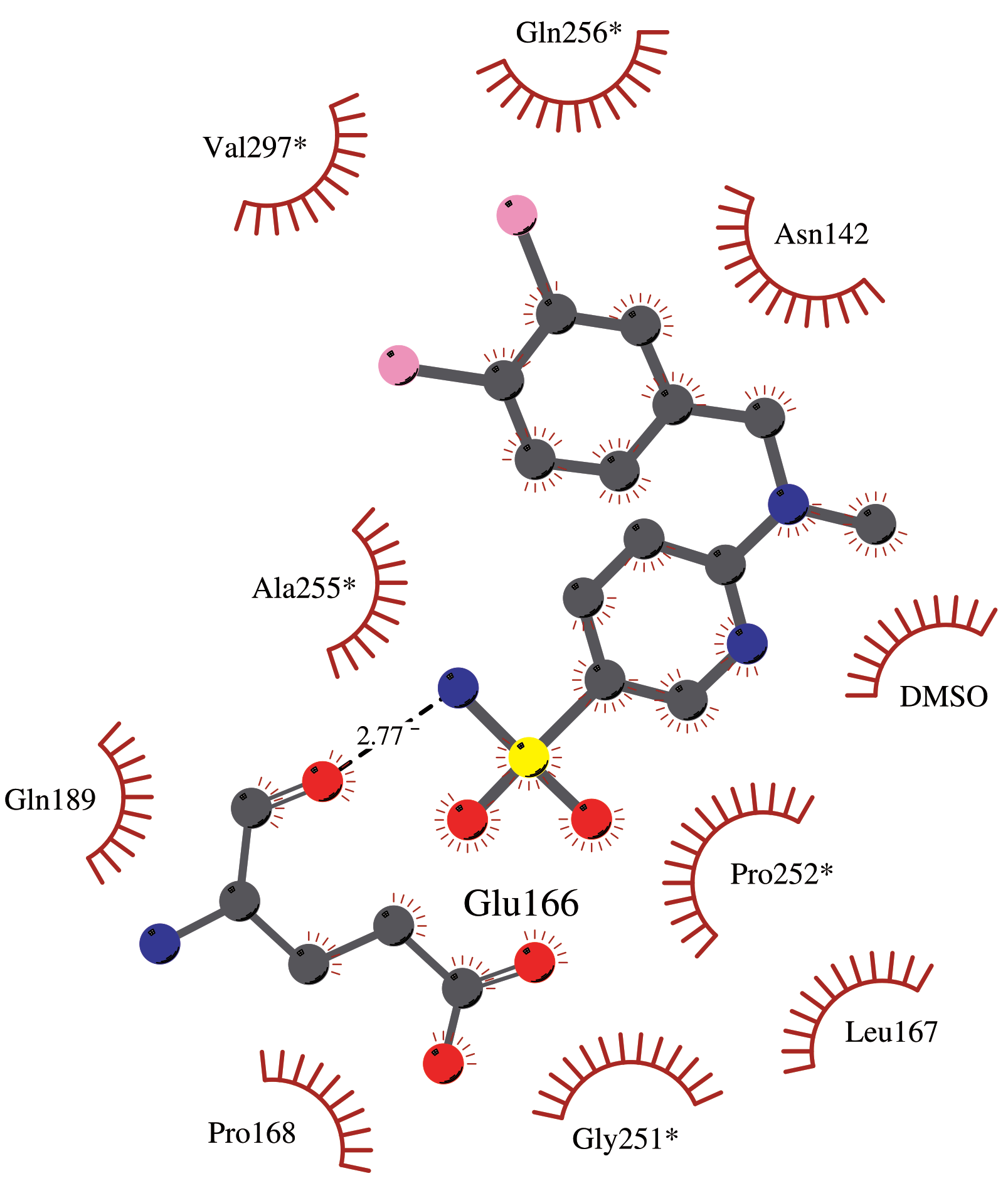}
        \caption{Residue interaction map}
        \label{fig:struct_mpro194_ligplot}
    \end{subfigure}
    \caption{\textbf{Inhibition of SARS-CoV-2 M\textsuperscript{pro} by machine-designed \textit{de novo} and commercially sourced compounds.}   (\textbf{a}) Half maximal inhibitory concentration (IC$_{50}$) from RapidFire MS experiments for \textit{de novo} and commercial M\textsuperscript{pro} inhibitor candidates. A ``---'' indicates no inhibition was detected. Candidates marked with $^\dagger$ had successful crystal structures determined. (\textbf{b-d}) Crystal structure of the SARS-CoV-2 M\textsuperscript{pro} in complex with Z68337194. (\textbf{b}) Ribbon representation with transparent surface of the M\textsuperscript{pro} dimer colored in wheat and light pink to delineate each protomer. The active site of each protomer is shown with Z68337194 in stick representation. (\textbf{c}) Surface representation showing the overall binding mode of Z68337194 at the active site of M\textsuperscript{pro}. (\textbf{d}) Schematic representation of the interactions of Z68337194 with M\textsuperscript{pro}. Residues indicated with * are from a symmetry related M\textsuperscript{pro} protomer. }
    \label{fig:struct_mpro194}
\end{figure}

\subsection*{Synthesis of \textit{de novo} compounds}   
Figure  \ref{fig:all_mols} lists the eight \textit{de novo} compounds designed by the generative machine learning framework that were synthesized (See SI Tables \ref{tab:mpro-denovo}--\ref{tab:rbd-denovo}  for the  predicted molecular properties).
Details of the experimental synthesis protocols is provided in Methods and SI Section \ref{sec:synthesis_steps}.
We also provide a comparison between the predicted and the actual retrosynthetic pathways for those eight machine-designed compounds in SI Table \ref{tab:synth-comp}.  Five were synthesized using the top predicted pathway of IBM RXN. For two compounds, GEN626 and GEN777, predictions were found to be unsuccessful, so alternative pathways as designed by Enamine were employed (see Methods for details). For GXA104, reactants included in the RXN prediction were not available, so an alternative route was employed. Overall, these results show the usefulness of machine learning-based retrosynthesis predictions for reliably identifying plausible candidates and recommending viable synthesis routes.

\subsection*{Experimental validation of M\textsuperscript{pro} inhibition of \textit{de novo}  and commercially sourced compounds}
Enzymatic inhibition by the \textit{de novo} M\textsuperscript{pro}-specific molecules was measured by solid phase extraction purification linked to  mass spectrometry (RapidFire MS)~\cite{malla2021mass}. The results are presented in Figure \ref{tab:IC50}. 
Out of the four \textit{de novo}  compounds tested for this target, GXA70 and GXA112 both showed M\textsuperscript{pro} inhibition in the micromolar range,  with IC$_{50}$ values of \SI{43}{\micro\molar} and \SI{34.2}{\micro\molar}, respectively. These low micromolar inhibition is  considered  to be a  good baseline for initial hit discovery,   similar to those  used in the prior  studies \cite{morris2021discovery,
glaab2021pharmacophore, merk2018novo, zhang2021potent}. This implies a 50\% success rate of hit discovery for M\textsuperscript{pro}. To be noted, prior studies do leverage knowledge of existing active molecules, which is not the case in the present work, as the goal here is to simulate the scenario of targeting less explored proteins.

We further tested the generalizability of the pIC$_{50}$ predictor (trained directly on the molecular SMILES and protein sequences) by validating  predictions on selected commercially available lead-like compounds from the Enamine Advanced Collection~\cite{enamineadvanced}. For this purpose, we selected the top three Enamine compounds based on their predicted pIC$_{50}$.  One of these Enamine compounds showed inhibition (IC$_{50}$ = \SI{35.5}{\micro\molar}).
Based on these results, we co-crystallised M\textsuperscript{pro} in the presence of this compound (ID Z68337194) and successfully obtained crystals (see SI Table \ref{tab:crystal_data}). The crystal structure determined revealed Z68337194 bound in the active site pocket. Structures of the other two commercially available compounds selected based on the pIC$_{50}$ predictions were also found bound to the active site of M\textsuperscript{pro}, although these compounds showed no detectable inhibition of M\textsuperscript{pro} using the RapidFire mass spectrometry-based assay.

Detailed analysis of the structure obtained for the complex of M\textsuperscript{pro} with Z68337194  (see Figure \ref{fig:struct_mpro194_macro}--\subref{fig:struct_mpro194_ligplot}) reveals that the sulphonamide group sits in the P4 subsite~\cite{Jin2020}  and the amine forms an electrostatic interaction with the backbone carbonyl of Glu166. This interaction mimics that made by the P4 site amide of nirmatrelvir (PF-07321332)~\cite{Abdelnabi_pfizer} (see SI Figure \ref{fig:nirmatrelvir}). Z68337194  occupancy refines to approximately 50\%. In the active site, shifts are observed in the positions of Pro168, Leu167, Glu166, and Met165 to accommodate ligand binding. The compound does not sit deeply in the active site and does not interact with the catalytic machinery, providing opportunities to elaborate upon the compound in order to take advantage of further subsites. In the captured crystal form, the active site sits at the interface between symmetry related protein monomers and as a result a symmetry related molecule provides additional interactions --- primarily a stacking interaction between the ligand phenylamine ring and Pro252. Additionally, a hydrophobic pocket in the symmetry mate formed primarily by Gln256 and Val297 accommodates the chlorinated ring.

\subsection*{Experimental validation of spike-based pseudovirus and live virus inhibition of \textit{de novo} compounds}
\begin{figure}[h]
    \centering
    \begin{subfigure}[b]{0.45\textwidth}
        \centering
        \includegraphics[width=\textwidth]{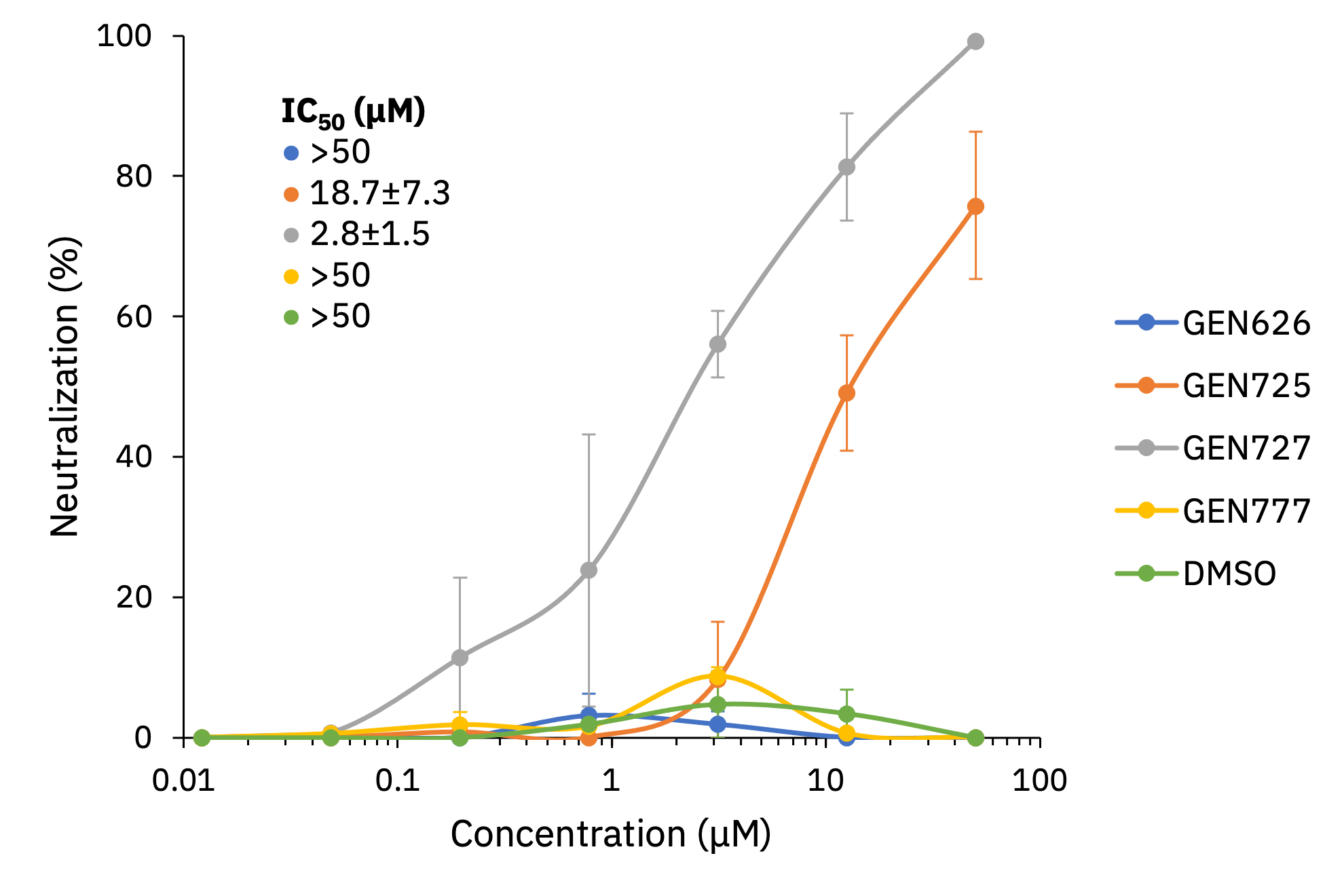}
        \caption{Pseudoviral neutralization --- Victoria}
        \label{fig:pseudo_compounds}
    \end{subfigure}
    \begin{subfigure}[b]{0.45\textwidth}
        \centering
        \includegraphics[width=\textwidth]{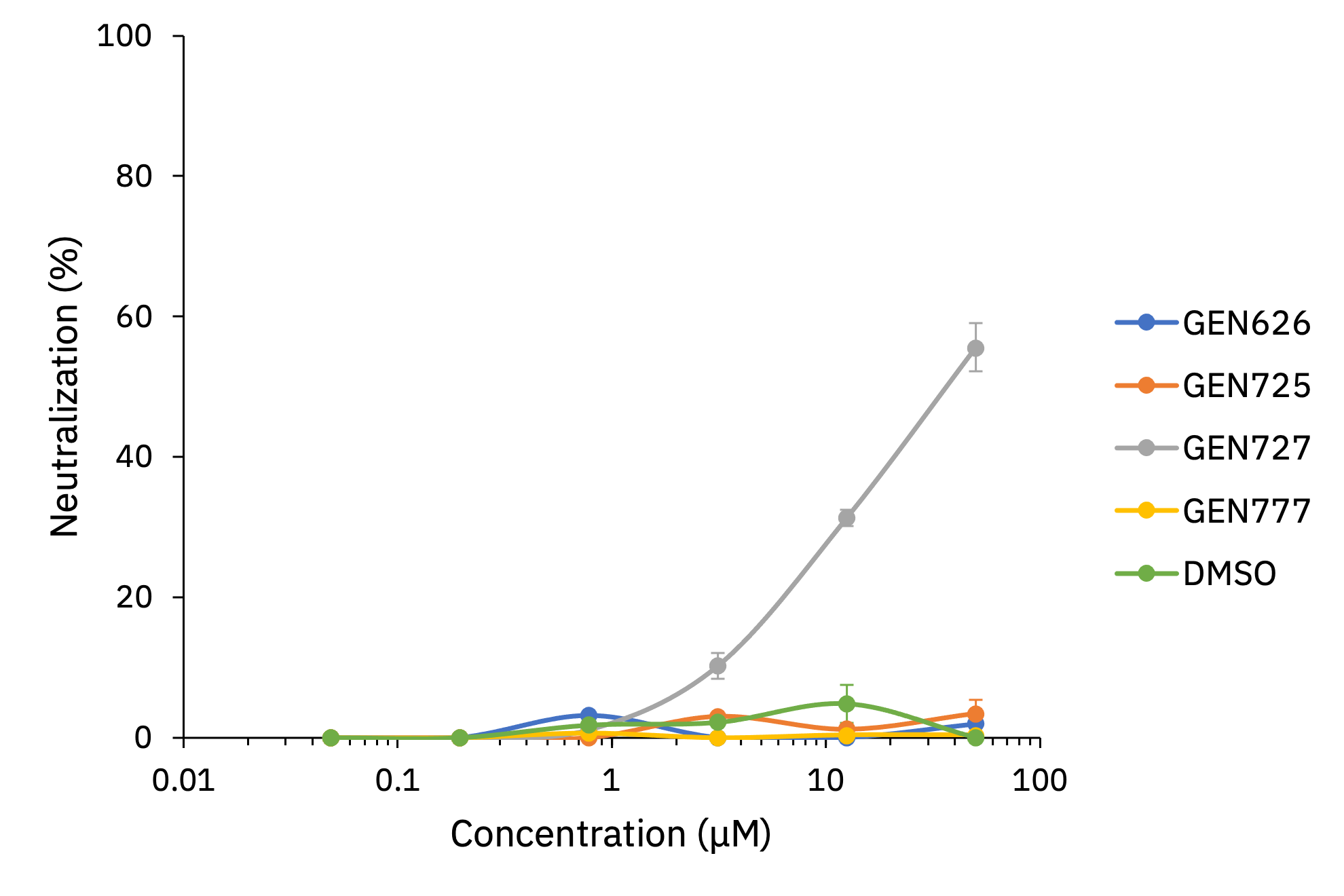}
        \caption{Live virus neutralization --- Victoria}
        \label{fig:live_compounds}
    \end{subfigure}
    \begin{subfigure}[b]{0.45\textwidth}
        \centering
        \includegraphics[width=\textwidth]{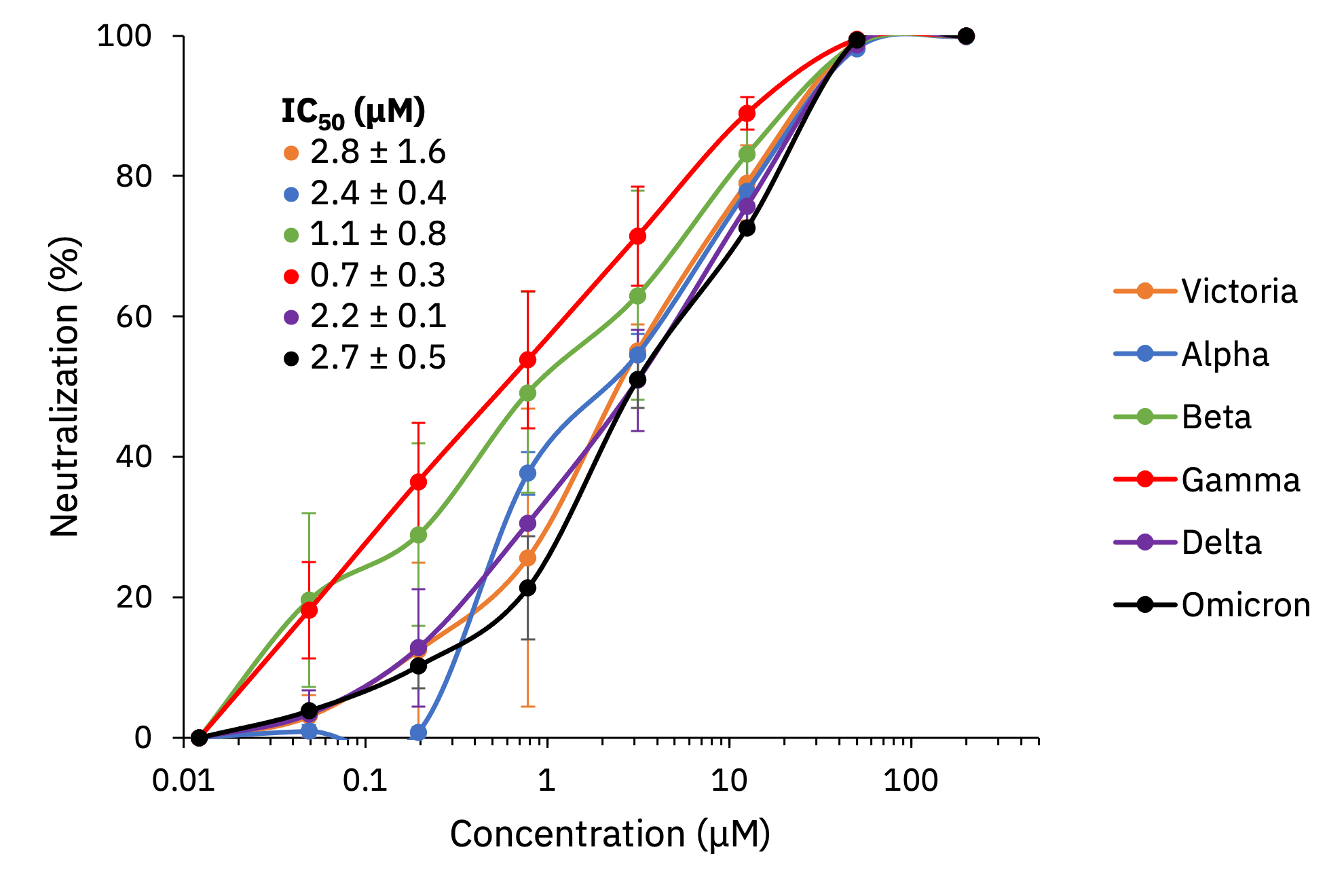}
        \caption{Pseudoviral neutralization --- VOCs}
        \label{fig:pseudo_variants}
    \end{subfigure}
    \begin{subfigure}[b]{0.45\textwidth}
        \centering
        \includegraphics[width=\textwidth]{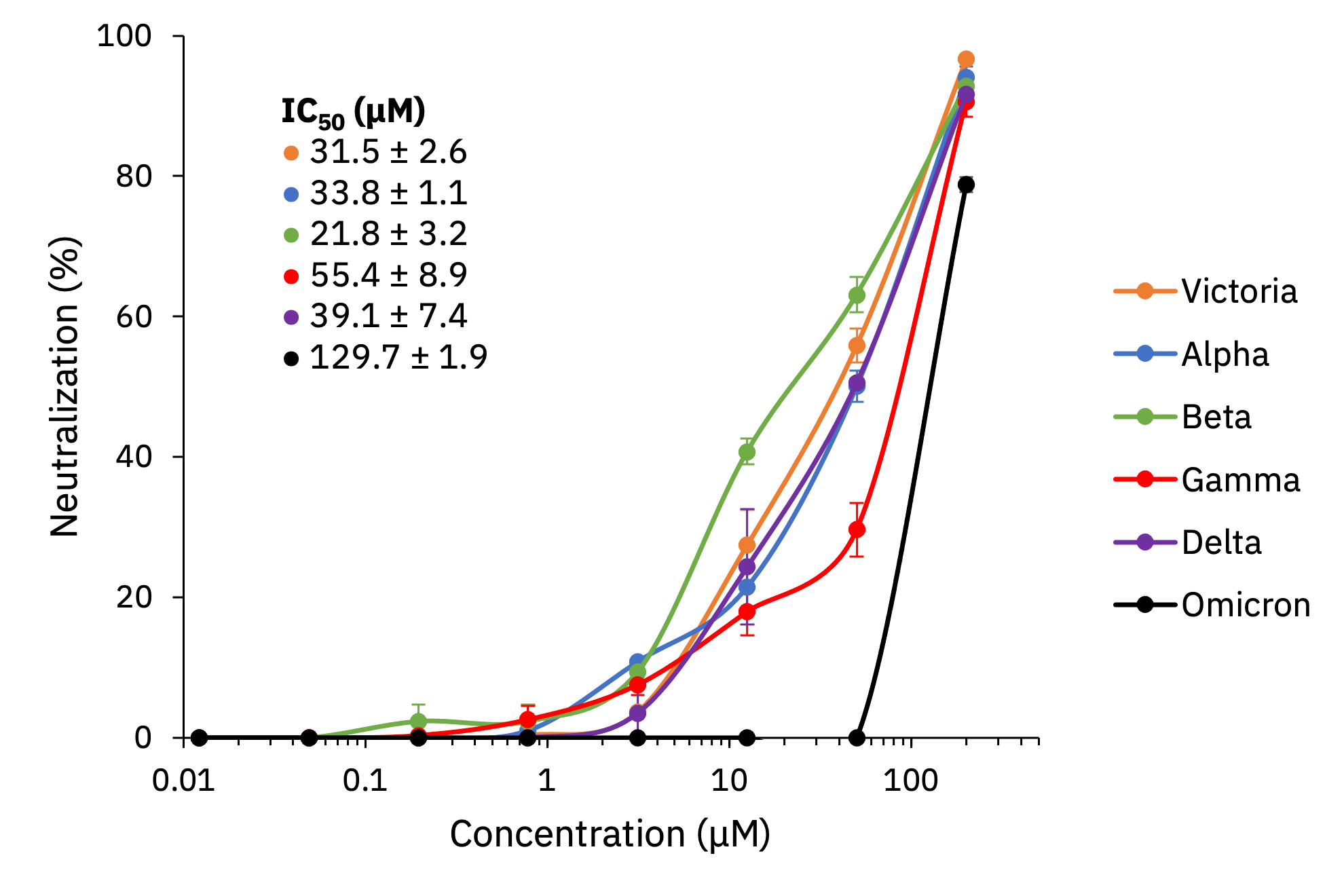}
        \caption{Live virus neutralization --- VOCs}
        \label{fig:live_variants}
    \end{subfigure}
    \caption{\textbf{SARS-CoV-2 spike neutralization assays.} Neutralization assay against SARS-CoV-2 pseudotyped lentivirus (\textbf{a}) and Victoria live virus (\textbf{b}) for four CogMol-generated compounds with DMSO as a control.  (\textbf{c}) The most effective compound, GEN727, was selected for a pseudoviral neutralization assay  against Victoria, Alpha, Beta, Gamma, Delta and Omicron variants of concern (VOCs), as well as (\textbf{d}) the live-virus neutralization assay. Error bars show the standard error of each measurement over two trials.}
    \label{fig:neutralization_assays}
\end{figure}

For the CogMol-designed compounds targeting the spike RBD, we  measured their neutralization ability using a  spike-containing
pseudotyped lentivirus  and a live viral isolate. These results are summarized in Figure \ref{fig:neutralization_assays}. Out of the four candidates, GEN725 and GEN727 showed IC$_{50}$ values less than \SI{50}{\micro\molar} (\SI{18.7}{\micro\molar} and \SI{2.8}{\micro\molar}, respectively), indicating discovery of novel hits with reasonable inhibition of the pseudovirus at a 50\% success rate (Figure \ref{fig:pseudo_compounds}). Importantly, GEN727 exhibited  live virus neutralization ability as well (Figure \ref{fig:live_compounds}).

We further checked if GEN727 is effective across different SARS-CoV-2 variants. We  compared  the  neutralization of viral variants of concern (VOCs) --- Alpha, Beta, Delta and Omicron ---
with neutralization of Victoria (SARS-CoV-2/human/AUS/VIC01/2020), a Wuhan-related strain isolated early in the
pandemic from Australia, in both pseudovirus and live virus. Figure \ref{fig:pseudo_variants} shows that GEN727  neutralizes spike-containing pseudovirus across all VOCs with an IC$_{50}$ value between \SIrange{0.7}{2.8}{\micro\molar}. Live virus data also shows inhibition with an IC$_{50}$ of less than \SI{50}{\micro\molar} for Victoria, Alpha, Beta and Delta (Figure~\ref{fig:live_variants}).

The virus neutralization results do not demonstrate direct interactions of GEN727 with the spike.   To probe this, we performed thermofluor measurements to determine if GEN727 affected the stability of the spike. The presence of the compound appeared to reduce the speed of the transition of the spike to a less stable form; after overnight incubation at pH 7.5, very little of the spike population remained in the more stable form with the higher $\mathrm{T_m}$ of \SI{65}{\degreeCelsius} (see SI Figure \ref{fig:thermafluor}).

\begin{figure}[]
\centering
\begin{subfigure}[b]{0.45\textwidth}
    \centering
    \begin{tabular}{cc}
        \parbox[c][8em]{0.45\textwidth}{\centering\includegraphics[scale=0.3]{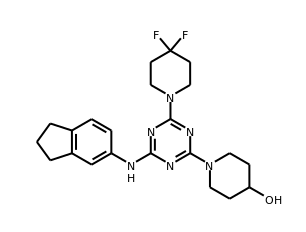}} &
        \parbox[c][8em]{0.45\textwidth}{\centering\includegraphics[scale=0.3]{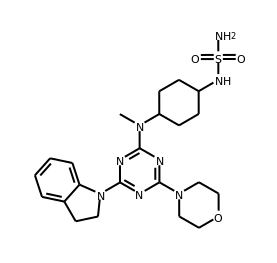}} \\
        GXA70 & GXA112 \\
        \addlinespace[1em]
        \parbox[c][8em]{0.45\textwidth}{\centering\includegraphics[scale=0.3]{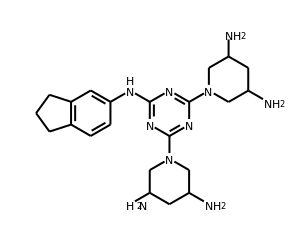}} &
        \parbox[c][8em]{0.45\textwidth}{\centering\includegraphics[scale=0.3]{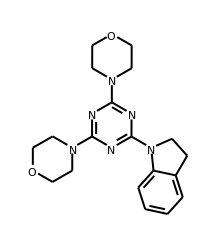}} \\
        CID 21104268 (0.544) & CID 42065574 (0.579) \\
    \end{tabular}
    \caption{M\textsuperscript{pro}}
    \label{fig:mpro_nov}
\end{subfigure}
\hfill
\begin{subfigure}[b]{0.45\textwidth}
    \centering
    \begin{tabular}{cc}
        \parbox[c][8em]{0.45\textwidth}{\centering\includegraphics[scale=0.3]{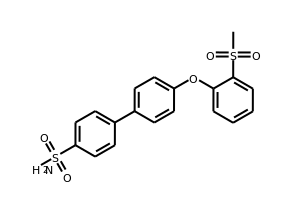}} &
        \parbox[c][8em]{0.45\textwidth}{\centering\includegraphics[scale=0.3]{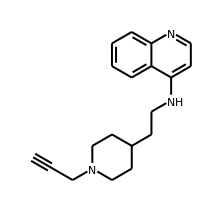}} \\
        GEN725 & GEN727 \\
        \addlinespace[1em]
        \parbox[c][8em]{0.45\textwidth}{\centering\includegraphics[scale=0.3]{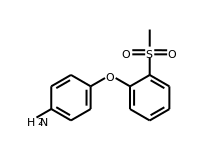}} &
        \parbox[c][8em]{0.45\textwidth}{\centering\includegraphics[scale=0.3]{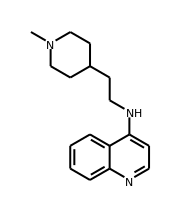}} \\
        CID 54908902 (0.658) & CID 114516038 (0.700) \\
    \end{tabular}
    \caption{Spike RBD}
    \label{fig:rbd_nov}
\end{subfigure}
\caption{\textbf{Molecular similarity with PubChem compounds.} Top: Validated  \textit{de novo} compounds  targeting (\textbf{a}) M\textsuperscript{pro} and (\textbf{b}) spike RBD. Bottom: Most similar molecules from PubChem. Values in parenthesis indicate Tanimoto similarity between the machine-designed and nearest Pubchem molecules.}

\label{fig:novelty}
\end{figure}

\begin{table}[h]
\centering
\begin{tabular}{@{}l@{}m{0.2\textwidth}|ccc@{}}
\toprule
               & & GXA70 & GXA112 & Z68337194 \\
 & & {\centering\includegraphics[scale=0.15]{img/Z4519300136_bw.png}} & {\centering\includegraphics[scale=0.15]{img/Z4519300867_bw.png}} & {\centering\includegraphics[scale=0.15]{img/Z68337194_bw.png}} \\
\midrule
TRY-UNI-714a760b-6\cite{achdout2020noncovalent} &\centering\includegraphics[scale=0.21]{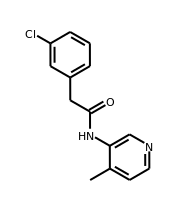}& 0.101 & 0.091  & 0.200     \\

X77 \cite{luttens2022ultralarge}         &  \centering\includegraphics[scale=0.21]{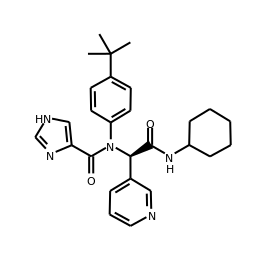}  & 0.116 & 0.150  & 0.115     \\
Ensitrelvir (S-217622) \cite{sasaki2022oral} & \centering\includegraphics[scale=0.21]{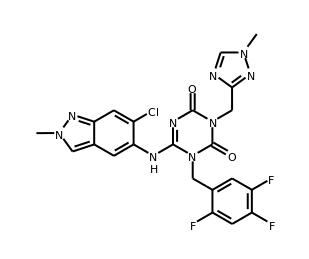}  & 0.093 & 0.075  & 0.128     \\

Nirmatrelvir (PF-07321332) \cite{Abdelnabi_pfizer} &  \centering\includegraphics[scale=0.21]{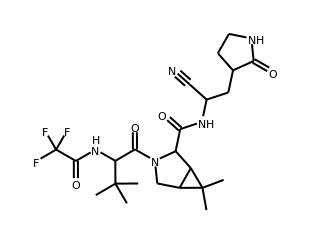}  & 0.109 & 0.100  & 0.051     \\
Compound 21 \cite{zhang2021potent}& \centering\includegraphics[scale=0.21]{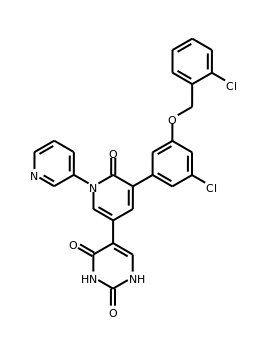} & 0.077 & 0.080  & 0.132     \\
Molnupiravir  \cite{Fischer_molnupiravir_2021} & \centering\includegraphics[scale=0.21]{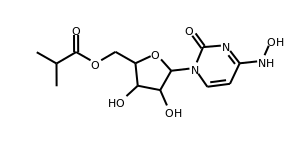} & 0.146 & 0.170  & 0.118    \\
\bottomrule
\end{tabular}
\caption{\textbf{Molecular similarity with existing inhibitors.} Tanimoto similarity of the validated machine-designed \textit{de novo} candidates to existing SARS-CoV-2 M\textsuperscript{pro} inhibitors.}
\label{tab:sim_known_inhibitors_similarity}
\end{table}

\subsection*{Novelty of  the \textit{de novo} designs and comparison with known SARS-CoV-2 inhibitors}
In order to characterize the novelty of the \textit{de novo} bioactive hits, we identified
the nearest compound from the PubChem database, 
in terms of their Tanimoto similarity~\cite{tanimoto1958elementary} estimated using Morgan fingerprints~\cite{Rogers_Hahn_2010_ecfp}. Figure \ref{fig:novelty} reveals that none of the \textit{de novo} molecules shares $\ge 0.7$ Tanimoto similarity with  PubChem molecules.
We further computed the Tanimoto similarity of the \textit{de novo} compounds to known SARS-CoV-2 M\textsuperscript{pro} inhibitors in literature. These results are shown in Table \ref{tab:sim_known_inhibitors_similarity}. In this category, specifically, we considered the following: an aminipyridine hit identified in the COVID-19 Moonshot initiative~\cite{achdout2020noncovalent}, X77 identified using ultralarge docking~\cite{luttens2022ultralarge}, the oral inhibitor S-217622 from reference~\cite{sasaki2022oral}
Nirmatrelvir in PAXLOVID~\cite{Abdelnabi_pfizer},
an $\alpha$-ketoamide inhibitor (Compound 21 from Zhang, et al.~\cite{zhang2021potent}), and
Molnupiravir~\cite{Fischer_molnupiravir_2021}. Consistently, the CogMol-designed inhibitors show high dissimilarity (as indicated by a low Tanimoto similarity around 0.1) to  existing SARS-CoV-2 M\textsuperscript{pro} inhibitors.

\begin{figure}[ht]
    \centering
    \begin{subfigure}[b]{0.4\textwidth}
        \centering
        \includegraphics[height=12em]{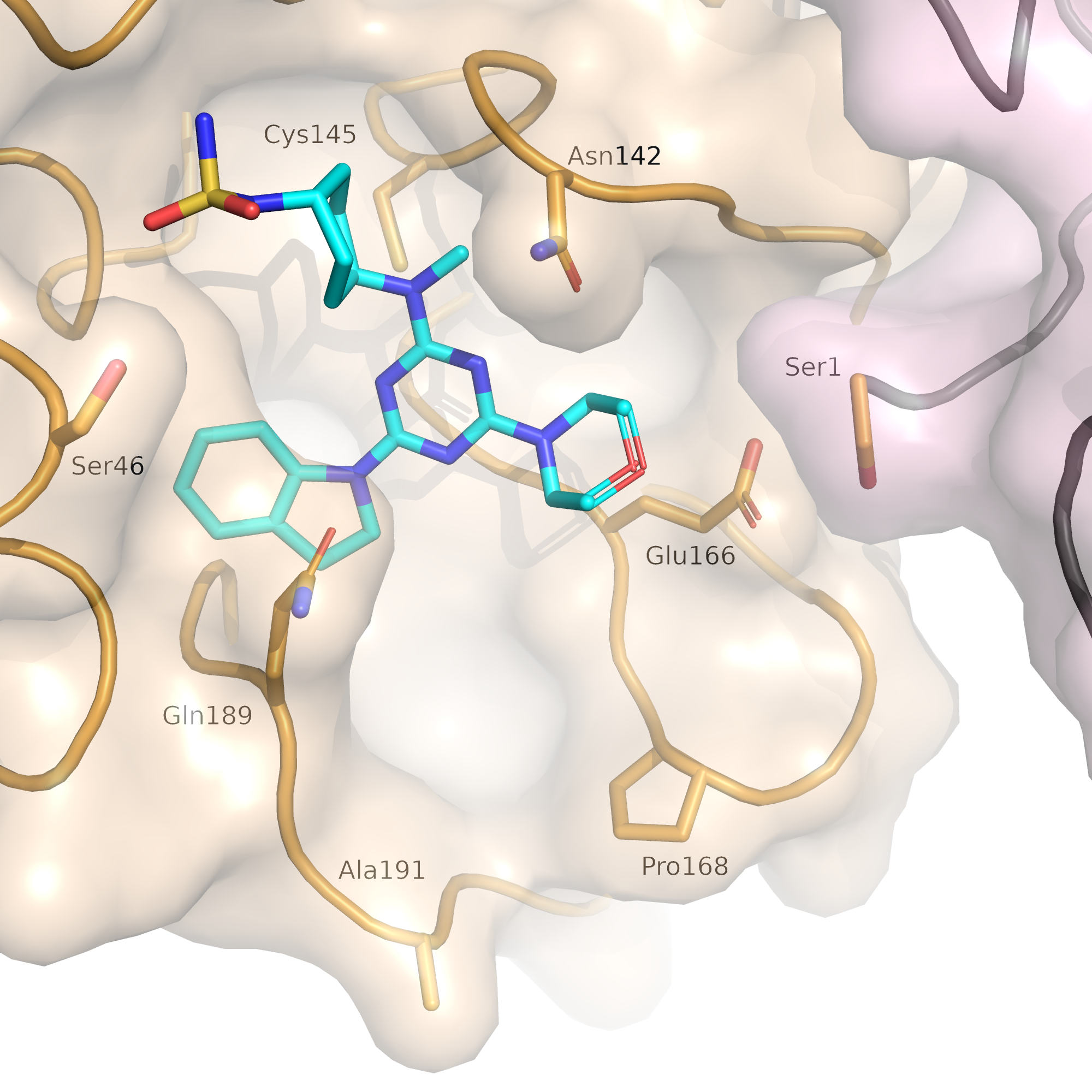}
        \caption{GXA112 --- Pocket view}
        \label{fig:docked_gxa112_pocket}
    \end{subfigure}
    \begin{subfigure}[b]{0.4\textwidth}
        \centering
        \includegraphics[height=12em]{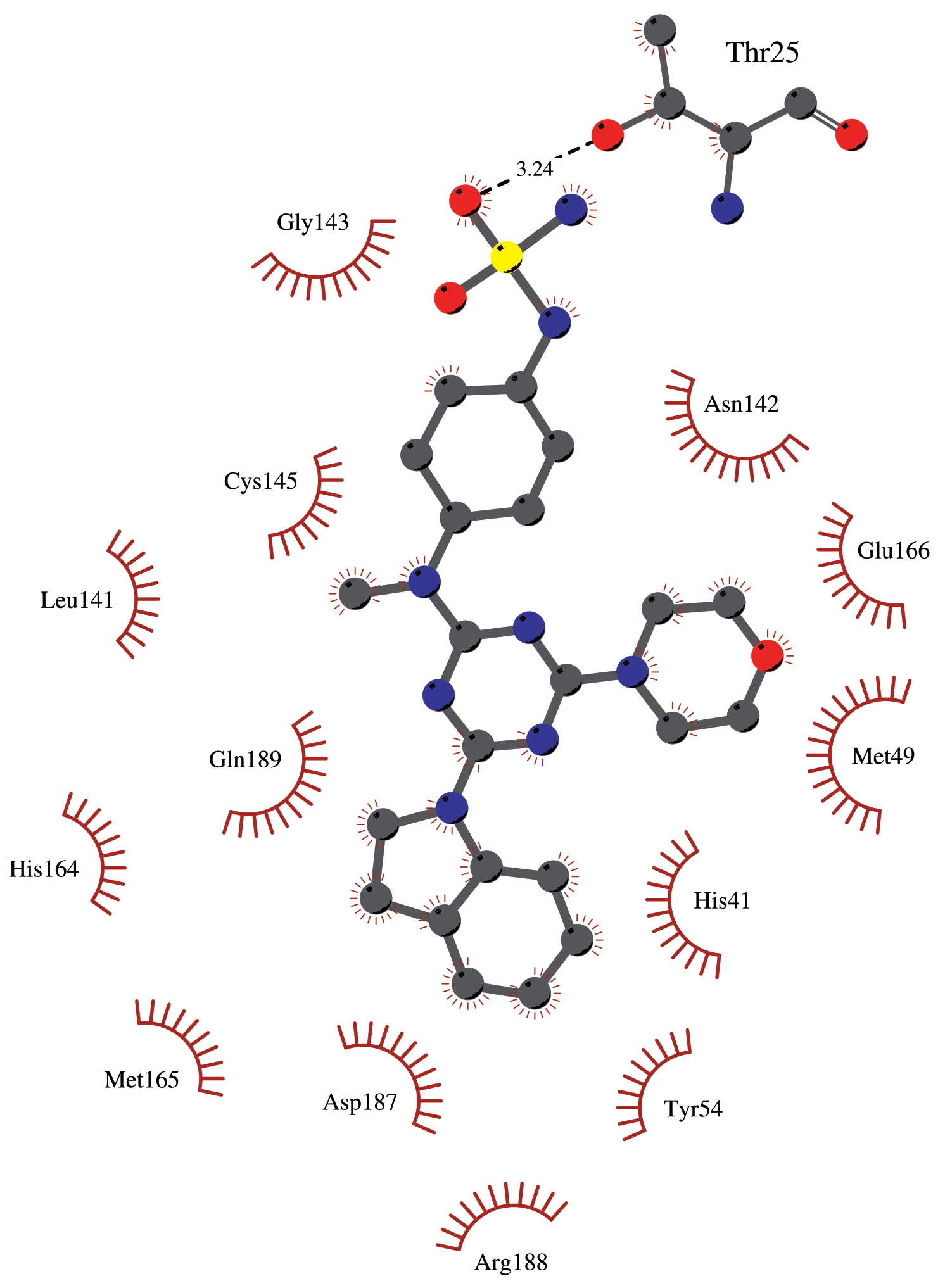}
        \caption{GXA112 --- Residue interaction map}
        \label{fig:docked_gxa112_ligplot}
    \end{subfigure}
    \begin{subfigure}[b]{0.4\textwidth}
        \centering
        \includegraphics[height=12em]{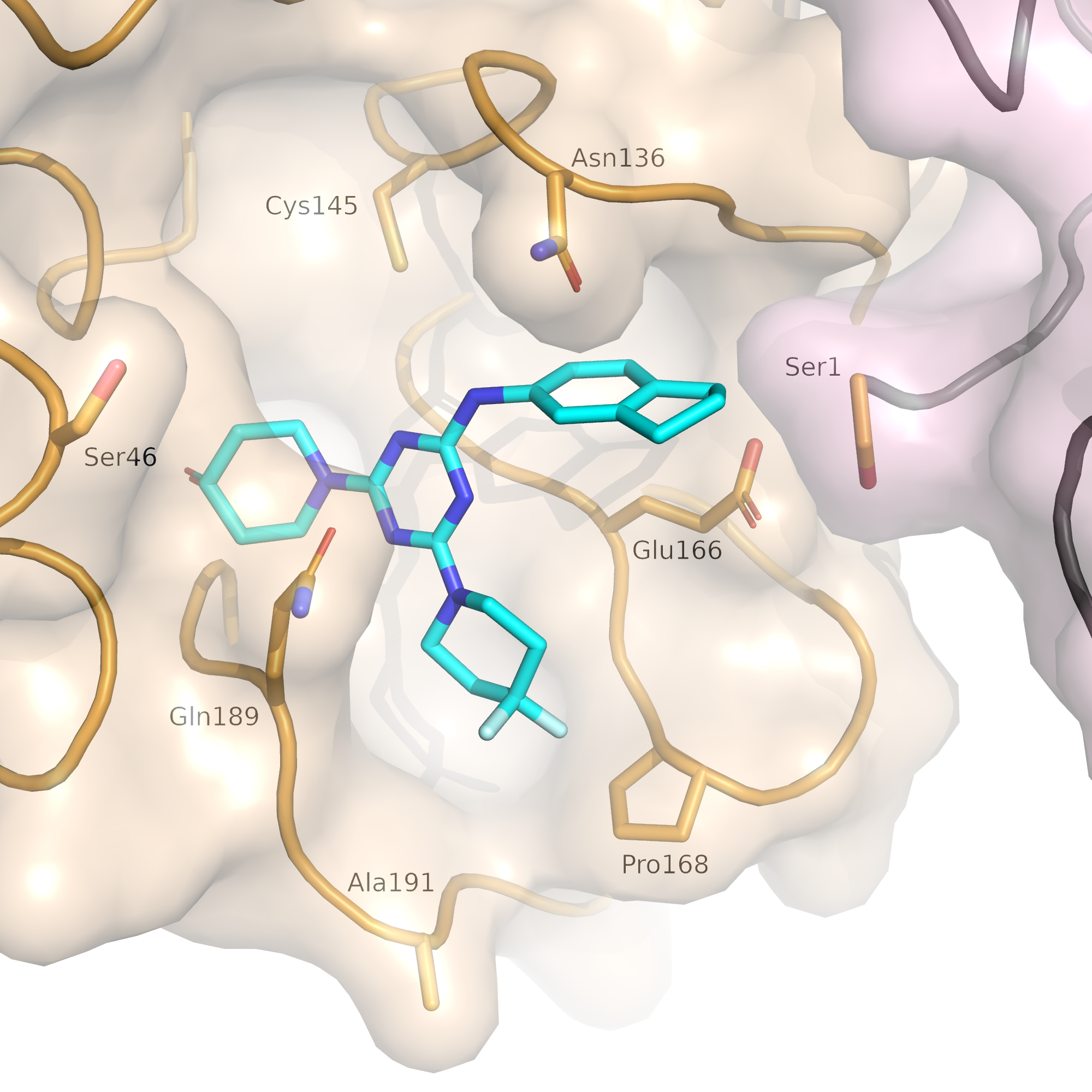}
        \caption{GXA70 --- Pocket view}
        \label{fig:docked_gxa70_pocket}
    \end{subfigure}
    \begin{subfigure}[b]{0.4\textwidth}
        \centering
        \includegraphics[height=12em]{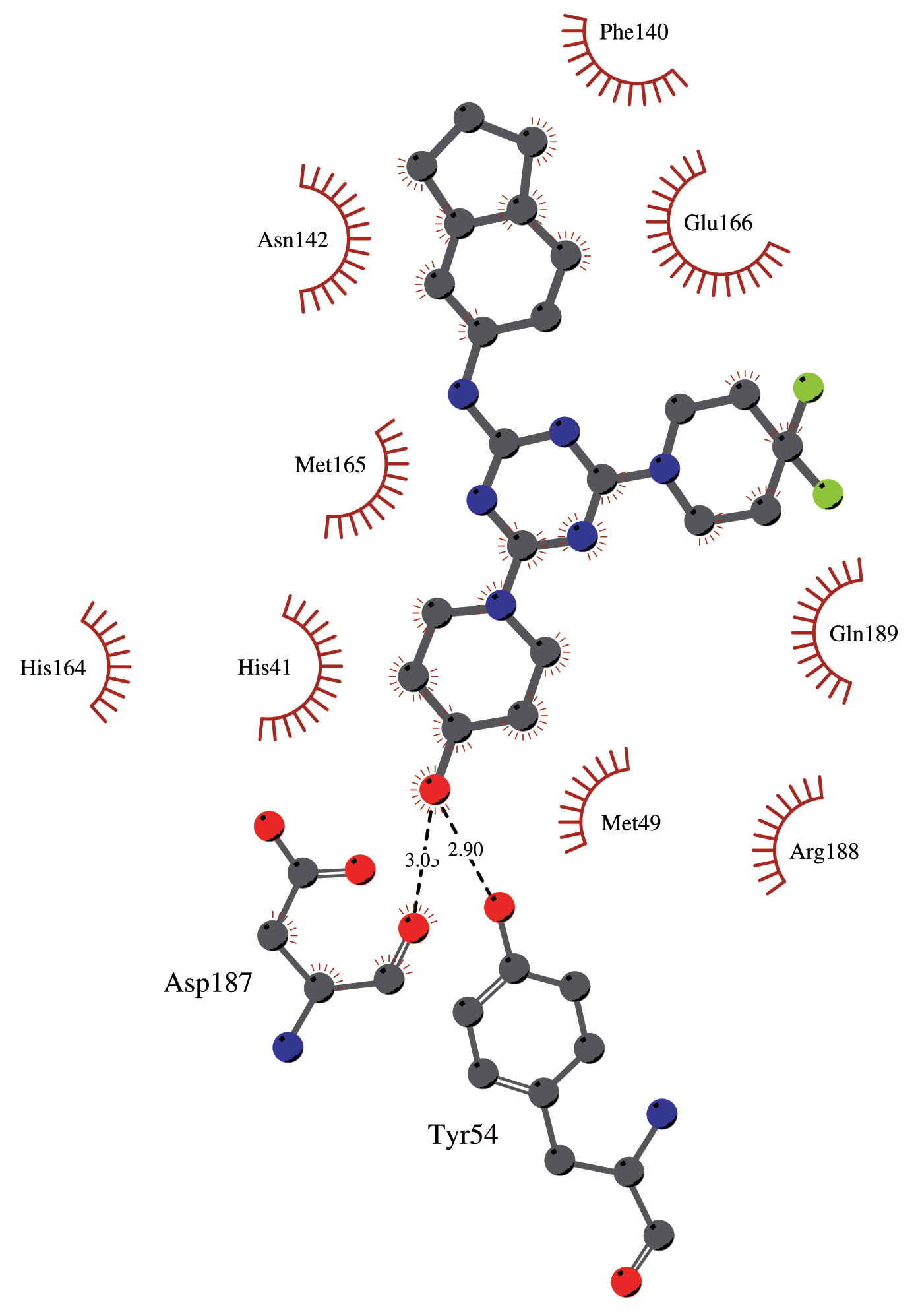}
        \caption{GXA70 --- Residue interaction map}
        \label{fig:docked_gxa70_ligplot}
    \end{subfigure}
    \caption{\textbf{Docked structures of SARS-CoV-2 M\textsuperscript{pro} with GXA112 and GXA70.} Surface representation depicting the overall ligand binding modes of (\textbf{a}) GXA112 and (\textbf{c}) GXA70 at the active site of M\textsuperscript{pro}.  Schematic representation of the ligand interactions with M\textsuperscript{pro} for (\textbf{b}) GXA112 and (\textbf{d}) GXA70.}
    \label{fig:docked_mpro_gxa}
\end{figure}

\begin{figure}[ht!]
    \centering
    \begin{subfigure}[b]{0.4\textwidth}
        \centering
        \includegraphics[height=12em]{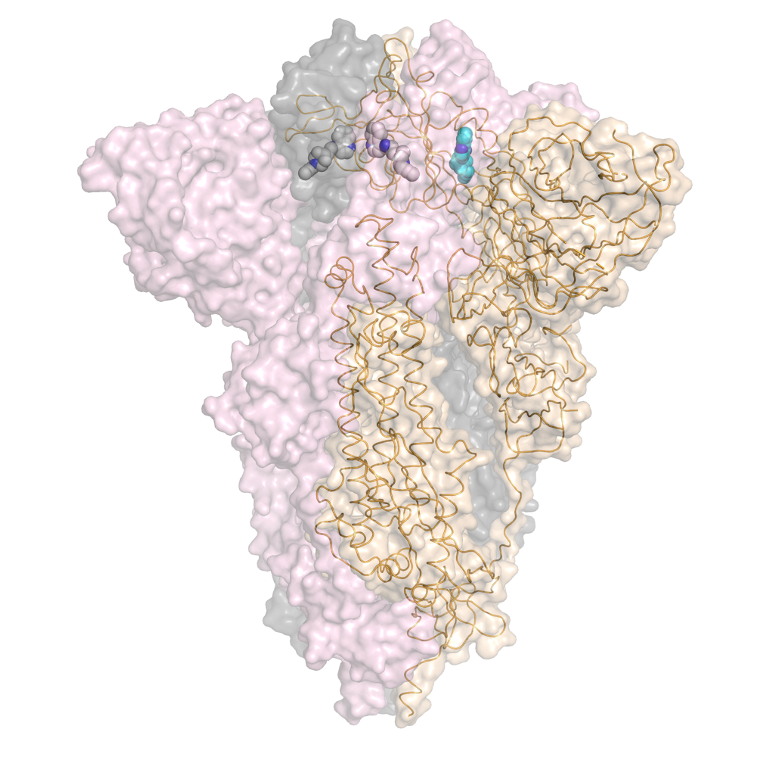}
        \caption{GEN727 --- Macro view}
        \label{fig:docked_gen727_macro}
    \end{subfigure}
    \begin{subfigure}[b]{0.4\textwidth}
        \centering
        \includegraphics[height=12em]{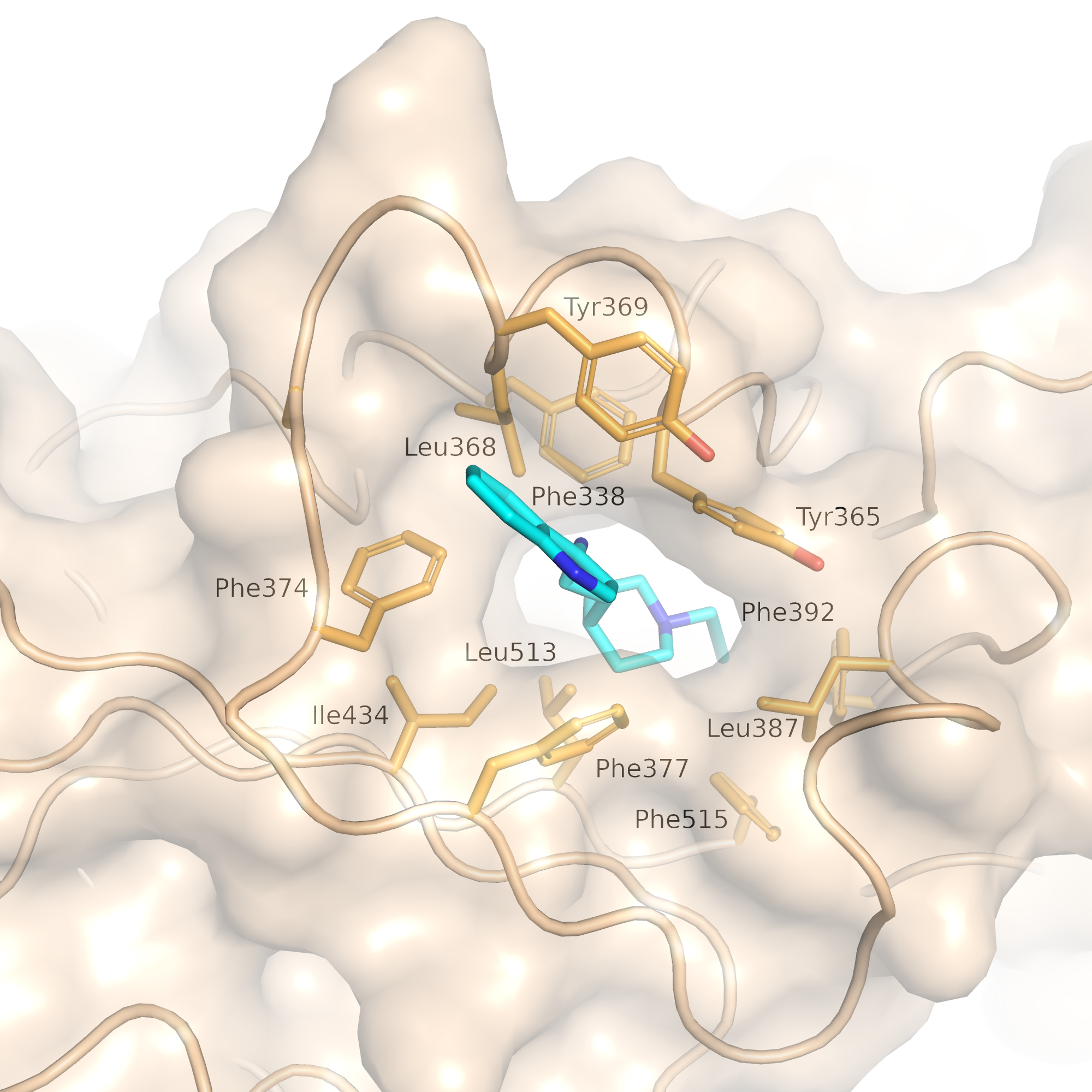}
        \caption{GEN727 --- Pocket view}
        \label{fig:docked_gen727_pocket}
    \end{subfigure}
    \begin{subfigure}[b]{0.4\textwidth}
        \centering
        \includegraphics[height=12em]{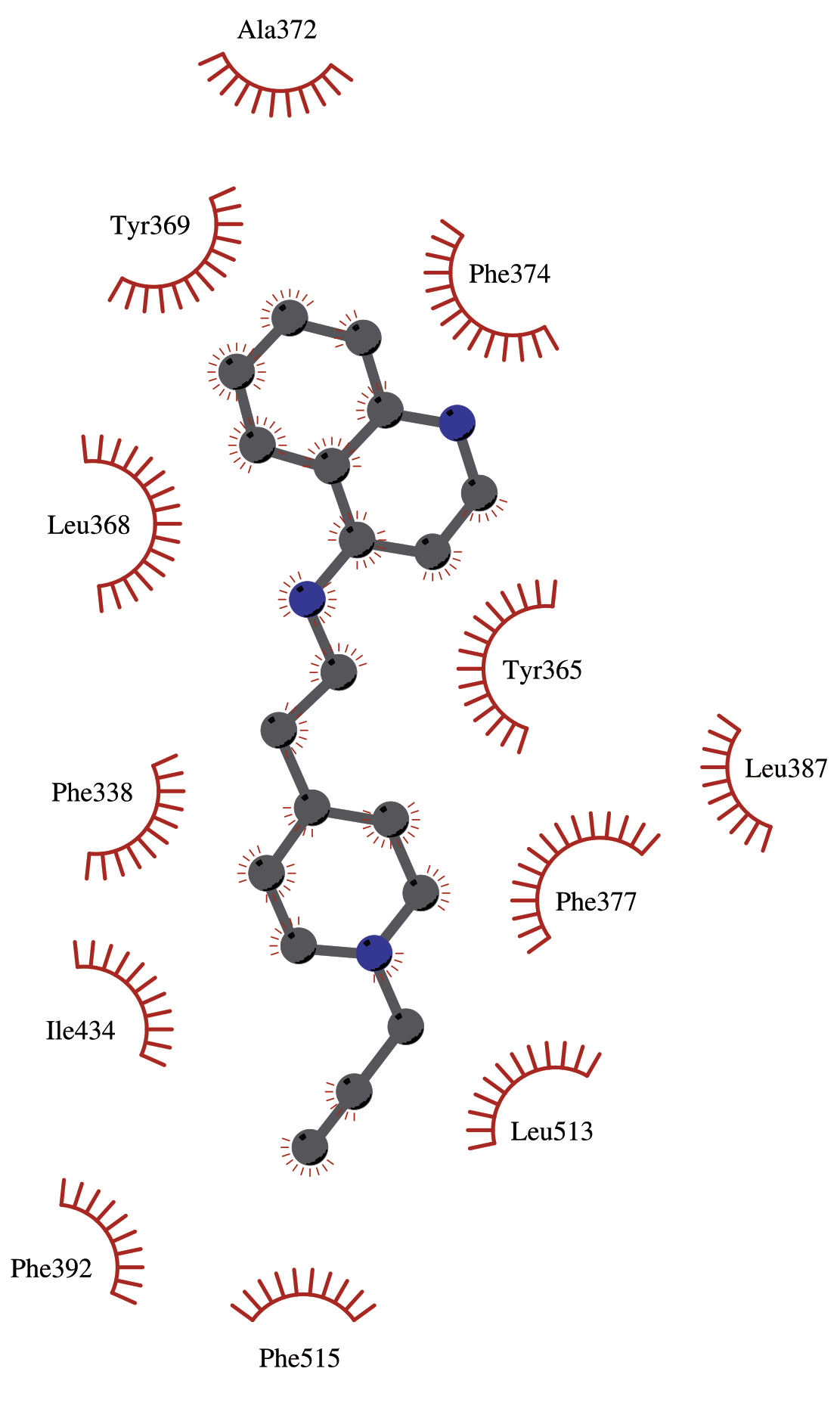}
        \caption{GEN727 --- Residue interaction map}
        \label{fig:docked_gen727_ligplot}
    \end{subfigure}
    \begin{subfigure}[b]{0.4\textwidth}
        \centering
        \includegraphics[height=12em]{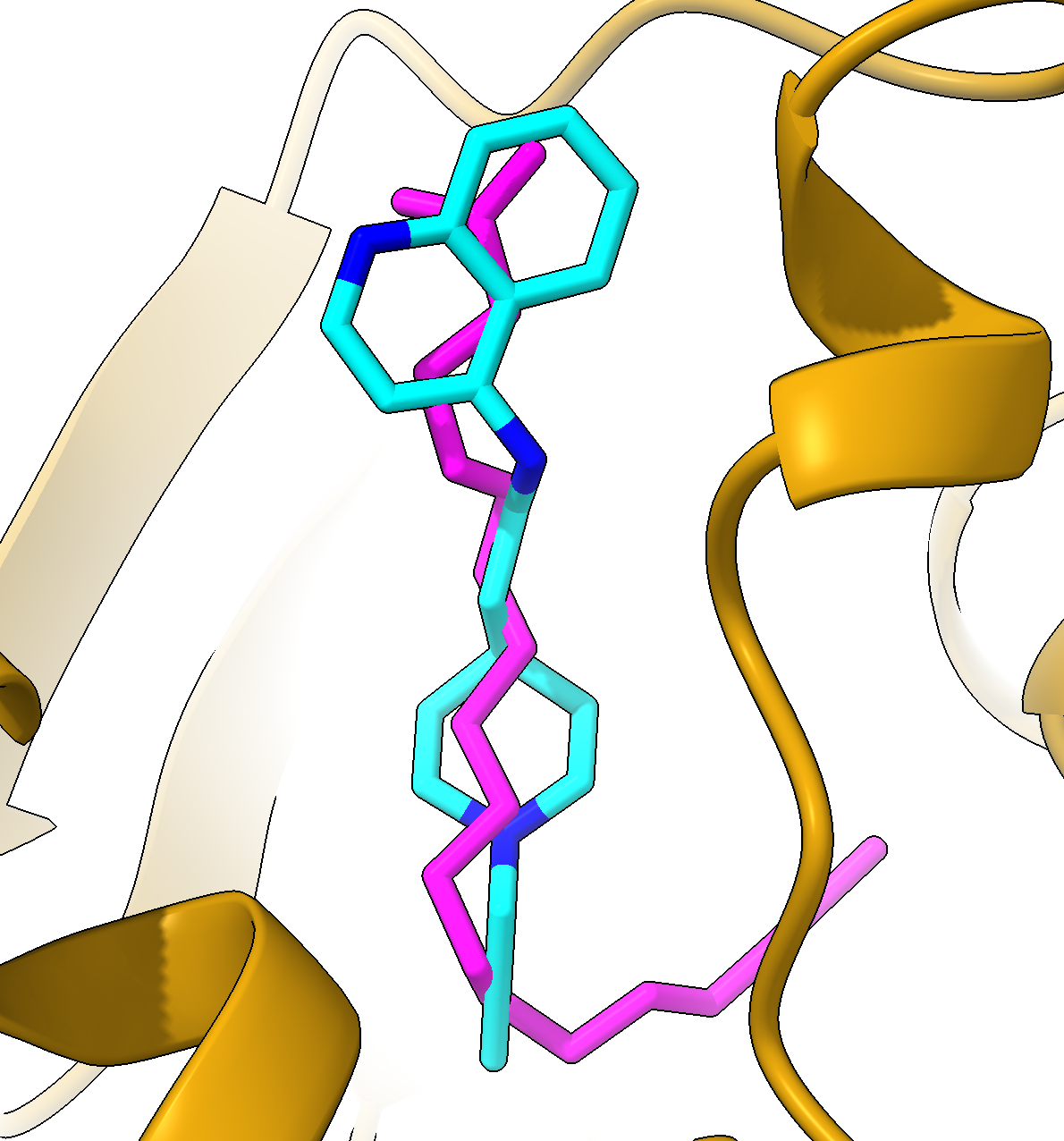}
        \caption{GEN727 (cyan) and stearic acid lipid (magenta)}
        \label{fig:docked_gen727_lipid}
    \end{subfigure}
    \begin{subfigure}[b]{0.6\textwidth}
        \centering
        \includegraphics[height=18em]{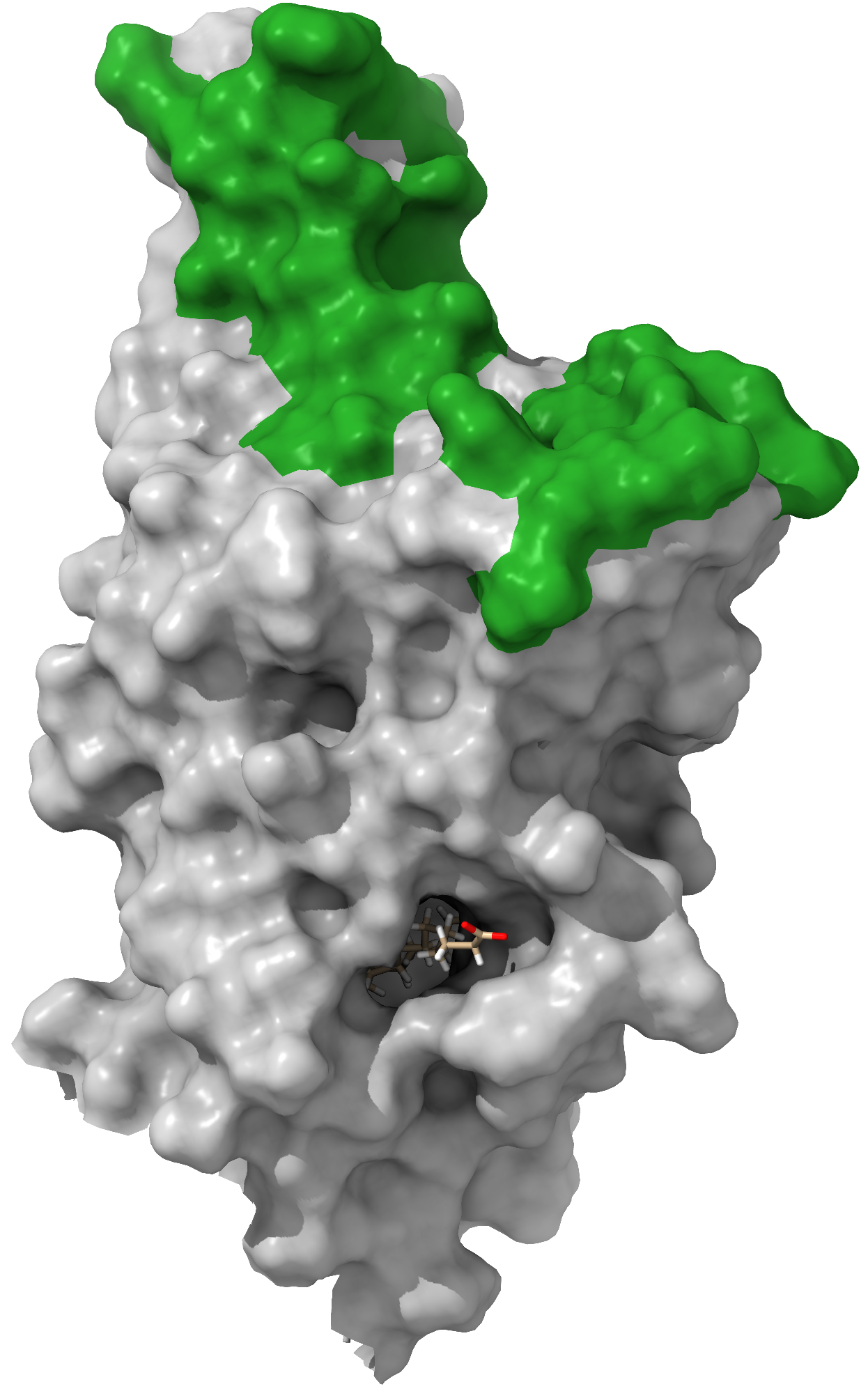}
        \caption{Lipid binding pocket vs. antibody binding hotspots}
        \label{fig:gen727_ab}
    \end{subfigure}
    \caption{\textbf{Docked structure of SARS-CoV-2 spike protein RBD in complex with GEN727.}  \textbf{(a)} illustrates the ribbon representation with transparent surface of the spike trimer. Wheat, gray, and light pink color is used to delineate each protomer. GEN727 (shown in stick representation) docked to a spike monomer structure is superimposed for reference. \textbf{(b)} Surface representation depicting the overall docking pose of GEN727  at the lipid binding site of the spike RBD.  \textbf{(c)} A schematic of GEN727 interacting  with the RBD.  \textbf{(d)} Docked GEN727 in reference to stearic acid lipid bound to the spike RBD. \textbf{(e)} Stearic acid binding pocket. Stearic acid (shown as sticks, almost completely buried) is distant from the sites of binding of most neutralising antibodies which attach much higher up the molecule, overlapping the site of attachment of ACE2 (the green surface) and thereby blocking attachment to the host cell.}
    \label{fig:docked_spike_gen}
\end{figure}

\subsection*{Insights into the binding mode of the  \textit{de novo} inhibitors via docking}
As experimental determinations of the  structure of either  M\textsuperscript{pro} or the spike protein in complex with the validated \textit{de novo} inhibitors were not fruitful, we used docking simulations to provide insight into the plausible binding modes. Docking simulations on the generated molecules were performed in the presence of their respective target structure --- PDB ID: 6LU7 \cite{Jin2020} for M\textsuperscript{pro} and  PDB ID: 7Z3Z for spike RBD (See  Methods for details). As shown in Figure \ref{fig:docked_mpro_gxa}, both machine-designed M\textsuperscript{pro} inhibitors, GXA112 and GXA70, revealed mainly hydrophobic contacts to the residues from the P1 and P2 subsites  which are the hotspots of interactions~\cite{Jin2020}.
The hydrogen bonding pattern revealed by the two molecules is, however, starkly different: GXA112 forms hydrogen bonding mainly with P1' site (T25), whereas GXA70 interacts with the P2 residues (D187 and Y54). The non-extensive and  diverse interaction pattern of the \textit{de novo} and commercially sourced M\textsuperscript{pro} inhibitors reported in this study is consistent with  reported observations for non-covalent inhibitors~\cite{durdagi2021near}.

For the validated \textit{de novo} spike inhibitor,  docking simulation (see Figure \ref{fig:docked_spike_gen}) reveals that GEN727 contacts with several hydrophobic residues, such as Tyr365, Tyr369, and Phe374, from RBD. Those   residues that constitute the lipid binding pocket of the spike RBD are conserved across seven coronaviruses that infect humans~\cite{toelzer2020free}. Also, the docking strikingly recapitulates the binding of the natural lipid (see Figure \ref{fig:docked_gen727_lipid}), suggesting that the lipid binding function maintains the conserved site targeted by GEN727. The lipid binding pocket is  distant and distinct from the sites of binding of the vast majority of neutralising antibodies, which cluster at the site of ACE binding (see Fig. \ref{fig:docked_spike_gen}e).
Therefore,  binding of GEN727 might stabilize the closed form of the spike, reducing receptor interactions \cite{toelzer2020free}. In line with that, the thermofluor results (SI Figure~\ref{fig:thermafluor}) showed an (albeit weak) indication that incubation of spike with GEN727 somewhat destabilized the spike, suggestive of a direct interaction underlying its broad-spectrum neutralization ability.
We also perform docking on the other \textit{de novo} designed spike inhibitor that showed pseudoviral neutralization, GEN725, which can be found in SI Figure~\ref{fig:docked_spike_gen725}, revealing an interaction pattern similar to that of GEN727.

\subsection*{Drug-likeness analysis of validated hits}
Finally, we have  predicted the drug-like nature and the medicinal chemistry friendliness of the experimentally validated hits found in this study, which are either \textit{de novo} designed or commercially available. For this purpose, the SwissADME~\cite{daina2017swissadme} software was used. SI Table \ref{tab:adme} provides a summary of those results (complete analysis reports can be found in SI Figures \ref{fig:adme1}--\ref{fig:adme9}), suggesting that the inhibitor hits satisfy several typical properties of drug-like, orally bioavailable compounds in terms of the drug-likeness scores and bioavailability. Further, the compounds rarely contain any medicinal chemistry alerts and fulfill most criteria for lead-likeness. Such favorable bioavailability and drug-likeness suggest potential of these  inhibitor hits as a starting point  for optimization toward compounds with more potency and better pharmacokinetic properties.

\section*{Discussion}

The discovery of therapeutic hits for diseases, including COVID-19, has been greatly advanced by the combined power of numerous \textit{in silico} approaches.
Nevertheless, even the most effective methods face broad challenges that are at the same time inherent to  general inverse molecular design tasks and specific to biological target-ligand binding chemistry. The first of these pertains to the vastness of the chemical space  being explored and its impact on the throughput and practical utility of the prevailing methods. For example, the use of docking or molecular simulation methods to screen on the order of $10^8$ to $10^9$ commercially available compounds, would incur a prohibitively high computational cost, estimated to reach 10 CPU years~\cite{luttens2022ultralarge} per target (as opposed to screening of less than a thousand machine-designed \textit{de novo} candidates via docking in the present study).

The second challenge is availability of critical information: while methods such as pharmacophore modeling and molecular docking and simulations have been used successfully in virtual screening or design of molecules~\cite{luttens2022ultralarge, achdout2020covid, zhang2021potent, morris2021discovery, unoh2022discovery, glaab2021pharmacophore}, such approaches generally rely upon initial design constructs obtained from available crystal structure(s) of a target protein bound to a candidate compound or fragment hits. For example, Glaab, et al.~\cite{glaab2021pharmacophore} has reported experimental validation of computationally screened M\textsuperscript{pro} inhibitors: out of 95 candidates tested \textit{in vitro}, two showed IC$_{50}$ values less than 50 \si{\micro\molar}. A variety of different computational approaches were used for screening: (i) searching for the nearest neighbors of a known M\textsuperscript{pro} inhibitor, (ii) M\textsuperscript{pro} structure-based screening using molecular docking followed by molecular simulations, and (iii) binding prediction using a machine-learning model trained on existing M\textsuperscript{pro} binders and non-binders.  Such knowledge of structures and known inhibitors is not guaranteed to be available for all drug targets of interest and may take months to derive experimentally,  and consequently these approaches are not broadly applicable to the case where target structures or inhibitors are unknown.  Recently, the field of structural biology  has been revolutionized by deep-learning based methods (e.g.,  AlphaFold \cite{ren2022alphafold}  and RoseTTAFold \cite{baek2021accurate}) for predicting the three-dimensional structure of a protein from its sequence. While they predict structures with often astonishing accuracy, the structural models derived from neural networks  are still relatively limited in aiding the understanding of natural protein function, in particular understanding the interactions with protein partners or small ligands. Therefore, the deduction of functional ligand and drug interaction  still remains predominantly reliant on resource-intensive experimental (bio)chemical techniques, e.g., assays, structural determination, and synthesis.

In general, reliance on privileged information (the target protein structure and/or known hits), confines the discovery space to the neighborhood of known chemical entities\cite{glaab2021pharmacophore}. This dependency therefore presents a practical challenge to expand the accessible chemical exploration space  and to devise more readily generalizable approaches to inhibitor design for multiple targets, the structure and binders of which may not be known.

Previous generative machine learning models that have been subject to experimental validation of \textit{de novo}-designed molecules were primarily either trained or fine-tuned on a target-specific ligand library \cite{merk2018novo, merk2018tuning, Zhavoronkov2019natbio, polykovskiy2018entangled, putin2018adversarial, tan2020automated, assmann2020novel}. This work establishes the basis for an alternative discovery paradigm, wherein a  generative model is used to discover novel inhibitor hits for different protein targets in an automated fashion.
To our knowledge, this is the first validated demonstration of a single generative model enabling successful and efficient discovery of drug-like inhibitor molecules for two very different target proteins, based only upon the protein sequence that is used during model inference. The generation of novel, drug-like, target-specific inhibitor molecules is automated, as the approach  performs attribute-controlled sampling on the learned abstract molecular representation space and does not rely on virtual screening of generated compounds that were designed using cumbersome rule-based fragmentation (e.g., as in \cite{morris2021discovery}. Moreover, to our knowledge, none of the earlier studies considers the challenging, but highly practical, scenario of designing and experimentally validating inhibitors for several distinct targets in parallel, without using the target structure and binder information, which resembles the scenario of relatively novel targets. Further, to our knowledge, evaluation of AI-generated retrosynthesis pathway predictions  against wet-lab compound production has not reported at this scale for AI-designed novel inhibitors.

The sequence information of new drug targets typically emerges at a much faster (days vs.\ months) pace than their detailed structural information, thanks to the latest advances in sequencing. The structural deduction of target-ligand interaction takes even longer. In contrast, as shown in Figure \ref{fig:pipeline_overview}, it took us less than a week to design and prioritize the set of candidate molecules to be synthesized and tested in wet lab for the two SARS-CoV-2 targets, as our approach does not reply on target structure or binder information. The information on SARS-CoV-2 sequences was made publicly available starting around January of 2020 and CogMol-designed candidates were open-sourced in the IBM COVID-19 Molecule Explorer platform in April 2020.  While the prioritized \textit{de novo} compounds were ordered in August 2020, and the first round of  wet lab validation was  completed in October 2020.  This rapid pace of  novel drug-like inhibitor discovery across two distinct drug targets, when the world was experiencing a pandemic, shows the potential of a  sequence-guided generative machine learning-based framework to help  with better pandemic preparedness and other global urgency. 

The overall success rate of hit discovery found here is 50\% for both targets, which required synthesizing and screening only four compounds per target. In addition, one of the three commercially sourced compounds  also showed M\textsuperscript{pro} inhibition. This result shows promise of the proposed approach, particularly when compared to a <10\% hit discovery  obtained typically using high-throughput screening~\cite{lloyd2020high, glaab2021pharmacophore}.  Additionally, the validated \textit{de novo} hits reported in this study appear to be novel, based on molecular similarity analyses with existing chemicals and SARS-CoV-2 inhibitors, indicating impressive creative ability of the generative framework, which is not possible when screening known compounds. The compounds also satisfy criteria of drug-likeliness and bioavailability.
The efficiency of hit discovery realized here and the demonstrated generalizability to distinctly dissimilar targets advocate for pre-training on a large volume of general data, e.g., chemical SMILES, protein sequences, and protein-ligand binding affinities. Conceptually this is a key feature of so-called foundation  models~\cite{bommasani2021opportunities, IBMblog}, which are trained on broad data at scale and can be easily adapted to newer tasks.
This perspective is also consistent with recent work, establishing the informative nature of a deep language model trained on large number of protein sequences, in terms of capturing fundamental properties~\cite{das2021accelerated, rives2021biological}. Thus, the  validation of the framework reported   here satisfies the generally accepted criteria of a foundation model, in the sense that it is trained on a broad set of unlabelled data, without a specific bias to a particular target, and is applicable without  little or no fine tuning to the general target-specific inhibitor discovery problem. The broad-spectrum efficacy across SARS-CoV-2 VOCs of the most potent spike hit observed is a further example of the foundational aspect of the model: the VOC sequences were never made available to the generative framework during training or inference. Moreover, to our knowledge, this is the first report of a novel spike-based non-covalent inhibitor that exhibits broad-spectrum antiviral activity. This contrasts with therapeutic monoclonal antibodies, the only drugs currently in use that target the spike protein, where rather few are effective across VOCs~\cite{dejnirattisai2022sars}. While the mutability of Spike is obvious because of the pressure to escape antibody neutralization, the widespread use of small molecule drugs will also apply a strong pressure --- as seen for instance in the rapid development of resistance to the first generation of anti-HIV-1 drugs. The choice of a binding site that is likely to be preserved to maintain a biological function, as seems to be the case with the RBD lipid pocket is probably about the best we can do in the early stages of drug discovery to build in some resilience.

Taken together, the results presented here establish the efficiency, generality, scalability, and readiness of a generative machine intelligence foundation model for rapid inhibitor discovery against existing and emerging targets.  Such a framework, particularly when combined with autonomous synthesis planning and robotic synthesis and testing~\cite{grisoni2021combining},  can further enhance preparedness for novel pandemics by enabling more efficient therapeutic design. The generality and efficiency of the mechanisms employed in CogMol for precisely controlling the attributes of generated molecules,  by plugging in property predictors post-hoc to a learned chemical representation, makes it suitable for broader applications in   advancing molecular and material discoveries. For example, a similar framework has already enabled  novel photoacid generator molecule design in a data-efficient manner for performant and sustainable semiconductor manufacturing, which has been validated by subject matter experts~\cite{hoffman2021sampleefficient}.

There remains significant scope for improving the discovery power of the framework: incorporation of the 3D structural information (when available) \cite{gebauer2022inverse, schiff2021augmenting} and further constraining the generations (e.g. solubility, number of hydrogen bonding donor/acceptor sites, structural diversity) are potential directions for further work. Iterative optimization methods~\cite{hoffman2022optimizing} can be adopted to improve initial hits by querying a set of molecular property evaluators along with a retrosynthesis predictor. Active learning paradigms can be also explored for improving process efficiency.

\section*{Materials and Methods}

\subsection*{CogMol overview}
\textbf{SMILES VAE as a molecule generator}: CogMol leverages a variational autoencoder~\cite{kingma2013auto, bowman2015generating} paradigm as the base generative model for molecules. The encoder in the VAE encodes molecules to a latent vector representation. The decoder maps latent vectors back to molecules. New molecules are  generated by sampling from the latent space.  Here, molecular SMILES is used as the input and output  to the encoder and the decoder, respectively. A bidirectional Gated Recurrent Unit (GRU) with a linear output layer was used as an encoder. The decoder contained a 3 layer GRU with a hidden dimension of 512 units and dropout layers with a dropout probability of 0.2.
The parameters for the encoder-decoder pair is learned by optimizing  a variational lower bound on the log-likelihood of the training data. The loss objective is comprised of a reconstruction loss and a Kullback-Leibler (KL) divergence (a measure of divergence between the fixed prior distribution $\rvp(\rvz)$, standard normal in this case,  and the learned distribution $\rvq_\phi(\rvz|\rvx)$) term: \[
     \mathcal{L}_{\text{VAE}}(\theta, \phi) = 
    \E_{q_\phi(z|x)}[\log p_\theta(x|z)]
    - D_{KL}(q_\phi(z|x) || p(z))
\]
This implies that new samples can be generated  from random points in the latent space, while  points close in the latent space will be decoded into chemically similar molecules.

The VAE was first trained for 40 epochs on 1.6M chemical molecules from the MOSES benchmarking dataset\cite{polykovskiy2018molecular}, which  was chosen from the larger ZINC Clean Leads \cite{irwin2005zinc} collection. Then, along with the  KL and reconstruction loss, the VAE was also jointly trained for another 15 epochs to predict the molecular attributes
QED and synthetic accessibility (SA) from the latent vectors $\rvz$. Two separate linear regression models were trained,
such that the VAE latent space becomes organized based on those  physical properties and thus serves as an approximation of the joint probability distribution of molecular structure and the chemical properties~\cite{gomez2018automatic}.The training was further continued for 50 epochs on around 211k ligand molecules from the BindingDB database\cite{gilson2015bindingdb}. This paradigm therefore served as a molecule generator that is unbiased toward any particular target.  The detailed evaluation of the final model is reported in \cite{chenthamarakshan2020cogmol}.

The final VAE generates SMILES strings by sampling from $q_\phi(\rvz|\rvx)$ that are 99\% unique and exhibit   greater than 90\% chemical validity, while  root-mean-square errors (RMSE) on the QED and SA prediction  are 0.0262  and 0.0175, respectively. The comparison of the unconditionally generated molecules from CogMol with five baseline generative models is reported in SI Table \ref{tab:vae_comp}, showing comparable performance in term of producing molecules that are valid, unique, diverse, and pass different  medicinal chemistry and other filters.

\noindent\textbf{Molecular attribute predictors for conditional generation}:
Two predictors trained on the latent $\rvz$ vectors were used for target-specific inhibitor molecule design, which are also drug-like. The QED regressor was comprised of 4 hidden layers with 50 units each and ReLU nonlinearity. 
Further, a target-chemical binder (strong/weak)  predictor was trained on the latent $\rvz$ vectors of chemicals  and the pretrained protein sequence embeddings~\cite{Alley_2019}, which used  the data released as part of the DeepAffinity \cite{karimi2018deepaffinity}. A pIC$_{50}$ value of $>6$ was used as a threshold to decide if a compound was a strong binder. The protein embeddings and the molecular embeddings were concatenated and passed through a single hidden layer with 2048  units and ReLU nonlinearity. The $\rvz$-based QED and pIC$_{50}$ predictors yield an RMSE of 0.0281 and 1.282, respectively.   These set of predictors were used for controlled sampling from the VAE model to design molecules with desired attributes.

\noindent\textbf{CLaSS sampling used for conditional generation in CogMol}:
We briefly describe Conditional Latent (attribute) Space Sampling (CLaSS)\cite{das2021accelerated} here.
CLaSS uses (i) a density model of the VAE latent representation, and (ii) a set of molecular attribute predictors trained on the VAE latent vectors,  to generate molecules in an attribute-controlled manner. For this purpose, a rejection sampling approach utilizing Bayes' theorem is used. To elaborate further, first  an explicit density model is learned on the latent embeddings of the training data to ensure sampling is uniformly random.
A Gaussian mixture model with 100 components and diagonal covariance matrices was used for this purpose. Assuming the attributes are all independent of each other and can be conditioned on the latent embeddings (i.e., the latent space encompasses all combinations of attributes),   Bayes' rule was then used to define the conditional probability of a sample, given certain properties in terms of the predictor models above. Finally, we employ this definition in a rejection sampling scheme, such that samples drawn from the density model are accepted according to the product of the attribute predictor scores. For more details on the algorithm, see SI Section \ref{sec:class_alg}. Generating the 875k samples for each target took around two days using an NVIDIA Tesla K80 GPU.
\label{sec:cond_sampling}

\subsection*{Ranking and prioritization}
The  filtering criteria  included   molecular weight (MW) less than 500 Da, QED greater than 0.5, SA less than 5, and octanol-water partition coefficient (logP) less than 3.5. MW, SA, logP, and QED were calculated using the RDKit toolkit~\cite{rdkit}. A pIC$_{50}$ predictor trained on DeepAffinity \cite{karimi2018deepaffinity} data was also used for ranking the designed molecules based on predicted affinity (AFF). A SMILES-based binding affinity  (pIC$_{50}$) predictor was used for this purpose. SMILES sequences were first embedded using long short-term memory units (LSTMs). Those SMILES embeddings were then concatenated with pre-trained protein embeddings~\cite{Alley_2019}, resulting in RMSE of 0.8426 on the test data.  A threshold for predicted pIC$_{50}$ affinity with the respective target sequence was set --- greater than 8 for molecules targeting M\textsuperscript{pro} and greater than 7 for molecules targeting the spike RBD. This affinity predictor was also used to estimate target selectivity (SEL)\cite{chenthamarakshan2020cogmol}, defined as the excess affinity to the target compared to a random set of proteins, lack of which is a known cause for drug candidate failure.

The molecules were also evaluated for predicted toxicity~\cite{lim2020explaining}  across a total of 12 \textit{in vitro}\cite{Tox21} and  one clinical end-points \cite{wu2018moleculenet}. Morgan fingerprints  were used as the input features for the toxicity prediction model. A multitask deep neural network   containing a total of four hidden layers was  used~\cite{lim2020explaining}: two layers were shared across all toxicity endpoints and two were specific to each of the endpoints.  A ReLU activation were used for all layers except for the last, for which a sigmoid activation was used.
Molecules that were predicted to have no toxicity to any of the toxicity endpoints were progressed in the workflow.

We then ran docking simulations on a prioritized set of designed molecules, less than 1000, with their respective target structures, as the docking energy can provide an indication of actual inhibition.  For M\textsuperscript{pro}, we used a monomer from the first structure determined and deposited with the Protein Data Bank for SARS-CoV-2 M\textsuperscript{pro} complexed with the covalent inhibitor N3  (PDB ID: 6LU7 \cite{Jin2020}) and set the search space to fully encompass the receptor. For spike, we used a lipid-bound conformation (PDB ID: 7Z3Z) and kept the protomer frozen during docking, as the goal is to find molecules that dock to the lipid-bound spike RBD. Our intent was to exploit the lipid binding pocket for developing inhibitors that can trap the spike protein in the closed conformation as this is known to have reduced interaction with the host ACE2 receptor \cite{toelzer2020free, carrique2020sars}. Docking was performed using AutoDock Vina \cite{trott2010autodock} run blindly over the entire  protein structure with an exhaustiveness of 8, and  repeated 5 times to find the optimal conformation.
Compounds with a  binding free energy given by docking of less than $-8.4$ kcal/mol with M\textsuperscript{pro} were selected.
For the generated spike compounds, we prioritized those that exhibited a binding free energy   less than $-7.5$ kcal/mol. Further, we only considered the compounds were docked less than 3.9~\AA~ from the lipid binding pocket in the final docked configurations. The surface and ribbon representations of ligands docked (or bound) to the target structure  were produced with PyMol\cite{PyMOL} and the protein-ligand interaction plots were produced with LigPlot+\cite{laskowski2011ligplot+}.

In contrast with large-scale screening, docking is only used to provide additional validation of the binding affinity predictor model and therefore can be run after filtering candidates based on the easily computed properties described above. After this filtering, we were left with fewer than 1000 molecules combined between the two targets on which to run docking. Each simulation takes only a few minutes and can be run independently in parallel which means the entire \textit{in silico} screening can be performed in less than a day when run on a compute cluster consisting of Intel Xeon E5-2600 v2 processors.

\subsection*{Retrosynthesis prediction}

 We assessed synthesis plausibility for the novel compounds, as a major challenge in driving successes in molecular discovery is to devise plausible and efficient synthesis-planning protocols. Here we applied the  recent advances made by  machine learning-based approaches to predict retrosynthetic routes from large reaction databases.
To estimate the ease of synthesizability  and facilitate synthesis  planning of the selected compounds, we predicted the retrosynthesis pathways for each candidate  using the IBM RXN platform \cite{schwaller2020predicting}. RXN combines a transformer neural network for forward reaction prediction  and graph exploration techniques to evaluate retrosynthesis paths,  scoring them according to probability. The path is terminated when all reagents are found to be commercially available. Candidates for which RXN was unable to determine a feasible retrosynthesis route or which terminated with non-commercially available compounds were removed from consideration.
For each prediction we used the following parameters: maximum single step reactions (depth), 6; minimum acceptance probability for a single step, 0.6; maximum number of pathways (beams), 10; number of steps between removal of low probability steps (pruning), 2; and maximum execution time, 1 hour. Commercial availability was determined by searching the eMolecules database\cite{emolecules} with a restriction on lead time of 4 weeks or less but no restriction on price.

In the next section, we provide a  detailed comparisons between predicted retrosynthesis and actual synthesis routes, which  is also summarized in SI Table \ref{tab:synth-comp}. We considered three main aspects in the comparison: number of reaction steps leading to the final product, overlap of the products in the intermediate reaction steps, and overlap of reactants used in the reactions. We chose the best path from the top six predicted for comparison by optimizing first for product overlap and then for reactant overlap. Overall,
the total number of actual reaction steps showed good agreement with predictions, generally only off by one or two steps. This was confirmed by the overlap of intermediate products, which showed that retrosynthesis often predicted the correct high-level path.
Product overlap is highly variable, though, since there are relatively few per route (often only two or three).
The actual synthesis routes even used many of the same reactants as predicted, although occasionally alternatives had to be found due to stock limitations.
In general, the retrosynthesis prediction was used as a starting point and any ``major'' deviations required were considered a failure.
Around 90-95\% of the top 100 generate compounds turned out to be synthesizable, based on the the retrosynthesis pathway predictions by IBM RXN \cite{chenthamarakshan2020cogmol} and human evaluation from subject matter experts (SME) at Enamine.  Design prioritization  to a small representative set was achieved by considering  time, reactant and reagent availability, and  amount of human effort.

\subsection*{Synthesis protocols}
\begin{figure}[t!]
    \centering
    \begin{subfigure}[b]{0.35\textwidth}
        \centering
        \includegraphics[width=\textwidth]{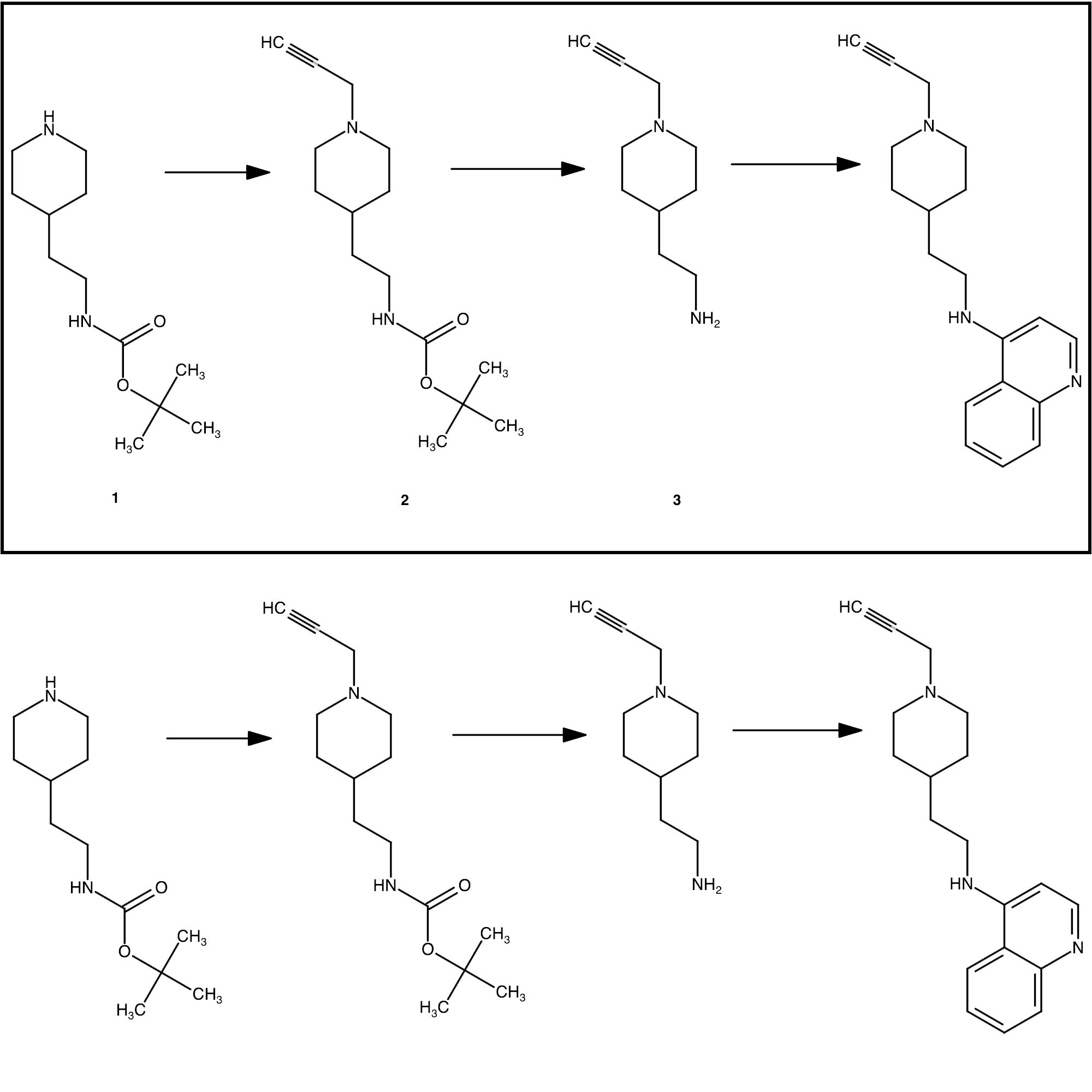}
        \caption{GEN727}
        \label{fig:gen727_route_comp}
    \end{subfigure}
    \hfill
    \begin{subfigure}[b]{0.55\textwidth}
        \centering
        \includegraphics[width=\textwidth,trim={0 150px 0 0},clip]{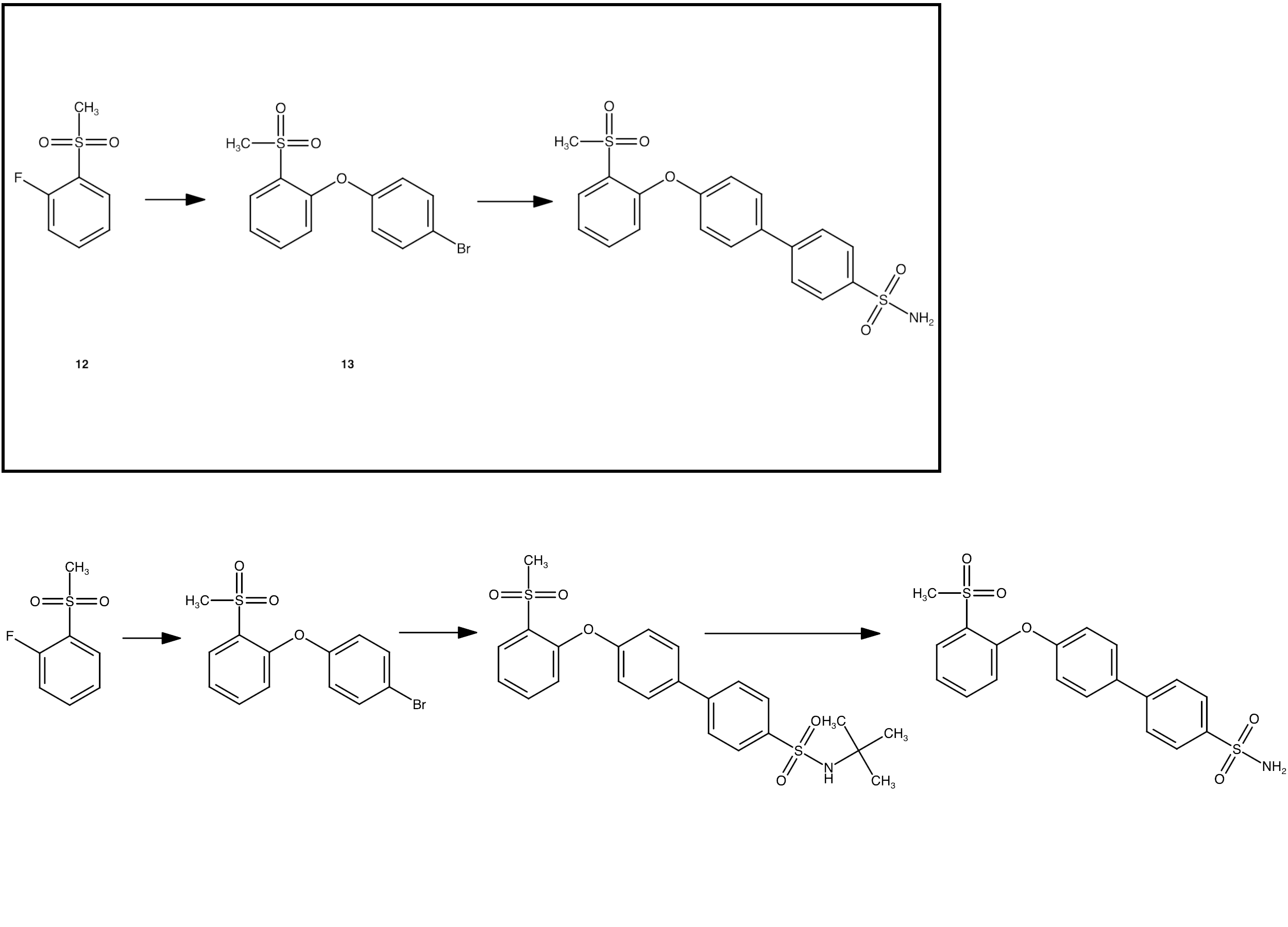}
        \caption{GEN725}
        \label{fig:gen725_route_comp}
    \end{subfigure}
    \begin{subfigure}[b]{0.45\textwidth}
        \centering
        \includegraphics[width=\textwidth,trim={0 50px 0 0},clip]{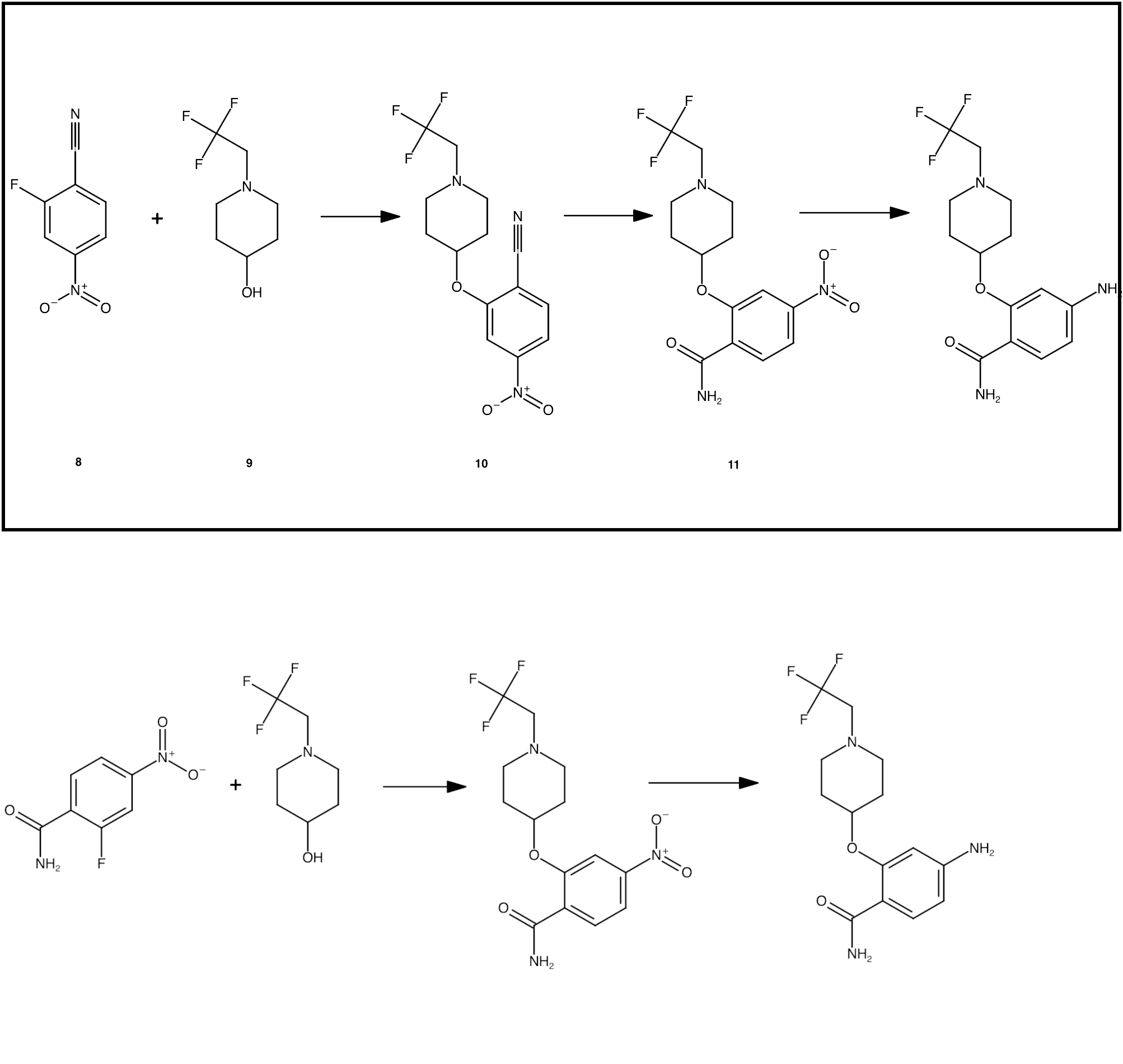}
        \caption{GEN626}
        \label{fig:gen626_route_comp}
    \end{subfigure}
    \hfill
    \begin{subfigure}[b]{0.5\textwidth}
        \centering
        \includegraphics[width=\textwidth]{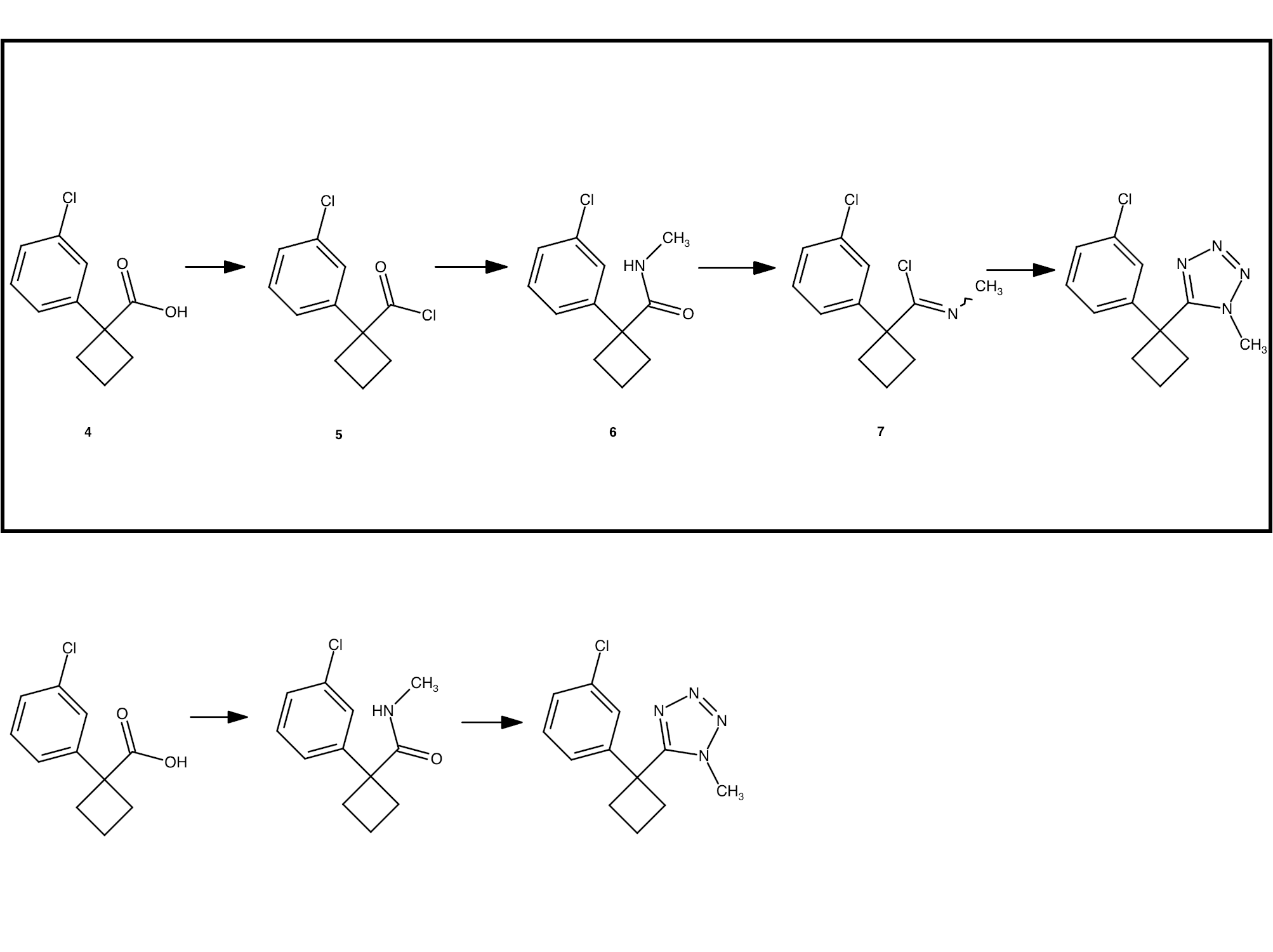}
        \caption{GEN777}
        \label{fig:gen777_route_comp}
    \end{subfigure}
    \caption{\textbf{Comparison between actual and predicted synthesis routes.} For each subfigure, the top reaction (enclosed in a box) is the actual synthesis procedure used in this study while the bottom reaction is the predicted retrosynthetic pathway.}
    \label{fig:synthesis_route_comparison_gen}
\end{figure}

In this section, we compare the retrosynthesis predictions to the actual routes used to synthesize the molecules:
GEN727 was synthesized according to the best RXN-predicted method (Figure~\ref{fig:gen727_route_comp}).
The synthesis of GEN725 was carried out by analogy to the best RXN strategy (Figure~\ref{fig:gen725_route_comp}). SNAr ester synthesis in DMF, gave intermediate compound \textbf{13} with high yield. Cross-coupling of \textbf{13} with sulfonamide- pinacolborane led to the final product with a moderate yield (see SI Section \ref{sec:synthesis_steps} steps K--L for full details of the synthesis procedure).
Several unsuccessful attempts were made to carry out the first step according to the retrosynthetic strategy for GEN626, which led to obtaining the desired intermediate with very low yield. As a result, the synthetic pathway was changed (Figure~\ref{fig:gen626_route_comp}). SNAr reaction was carried out with cyanide \textbf{8}, which was followed by hydrolysis of intermediate compound \textbf{10} (obtained with a moderate yield). Reduction of nitro-group of \textbf{11} led to GEN626 (see SI Section \ref{sec:synthesis_steps} steps H--J).
Unfortunately, following the pathway suggested by retrosynthesis for GEN777 didn't give good results and the synthetic strategy needed to be changed (Figure~\ref{fig:gen777_route_comp}). We synthesized acyl chloride \textbf{5}, which reacted with methyl amine on the next step. Thereafter, amide \textbf{6} was treated by $\mathrm{PCl}_5$ and the resulting intermediate was reacted in situ with azide-anion (see SI Section \ref{sec:synthesis_steps} steps D--G).


\begin{figure}[ht!]
    \ContinuedFloat
    \centering
    \begin{subfigure}[b]{0.55\textwidth}
        \centering
        \includegraphics[width=\textwidth,trim={0 150px 0 0},clip]{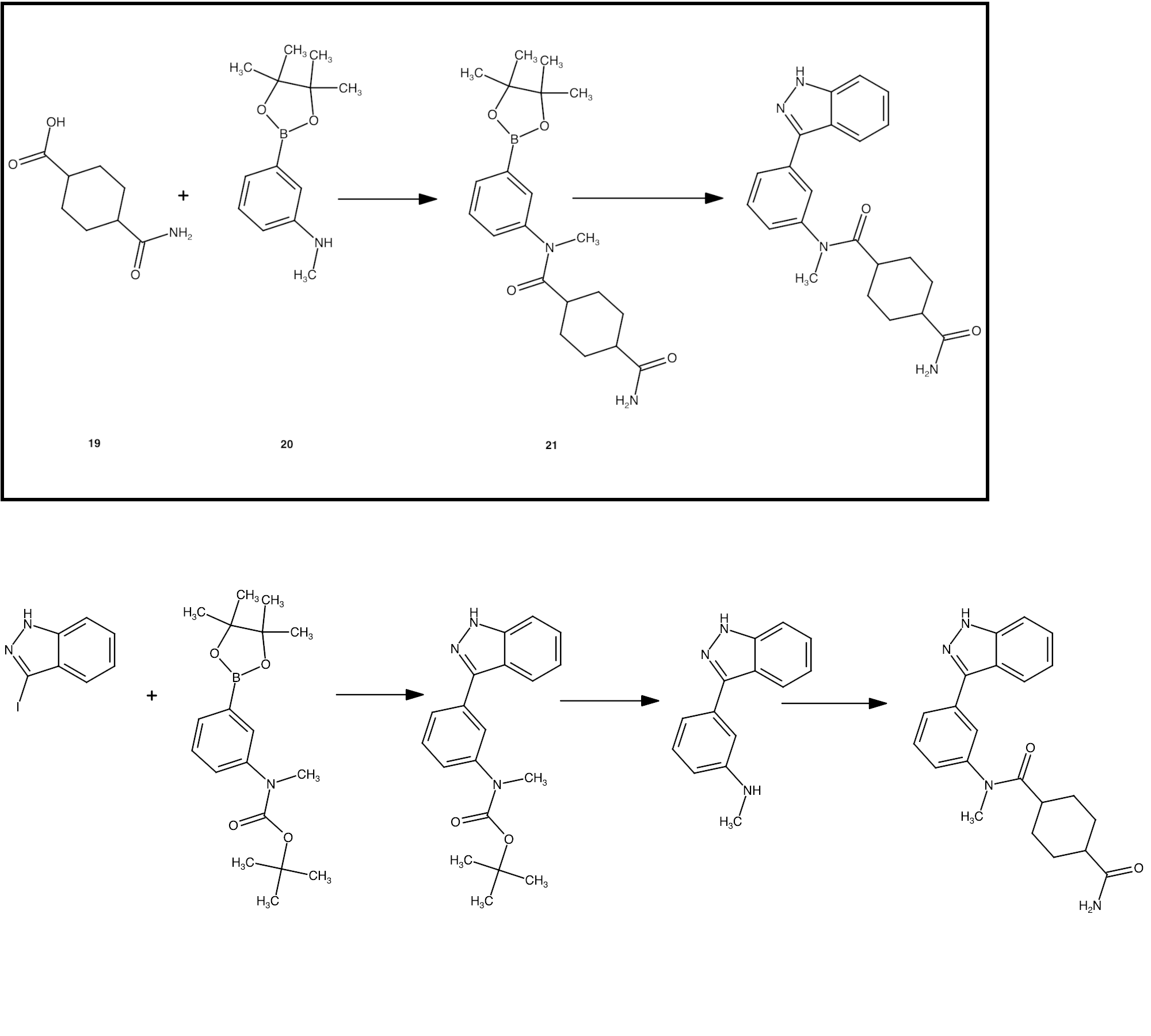}
        \caption{GXA104}
        \label{fig:gxa104_route_comp}
    \end{subfigure}
    \\
    \begin{subfigure}[b]{0.5\textwidth}
        \centering
        \includegraphics[width=\textwidth,trim={0 50px 0 0},clip]{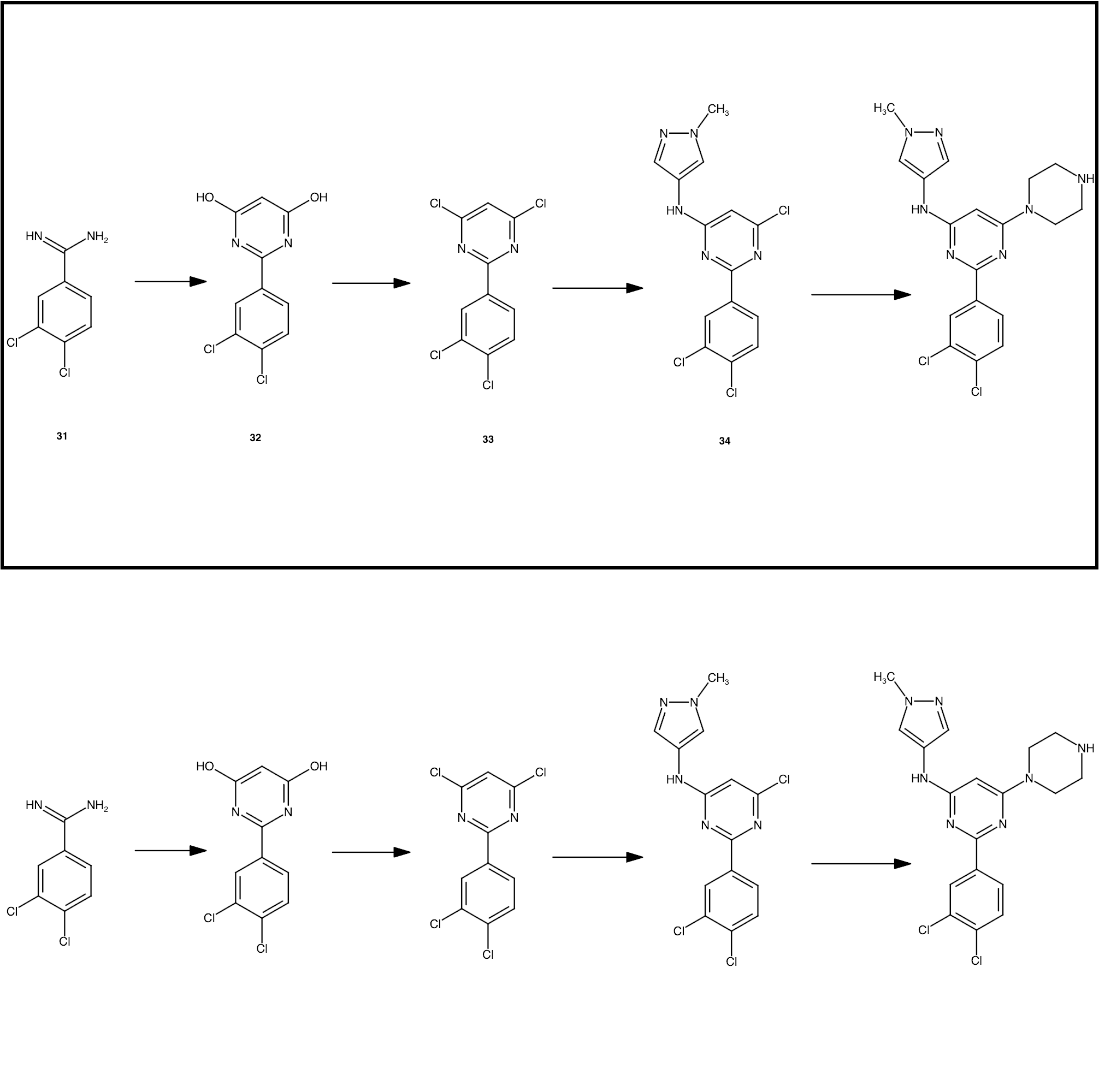}
        \caption{GXA56}
        \label{fig:gxa56_route_comp}
    \end{subfigure}
    \hfill
    \begin{subfigure}[b]{0.45\textwidth}
        \centering
        \includegraphics[width=\textwidth,trim={0 50px 0 0},clip]{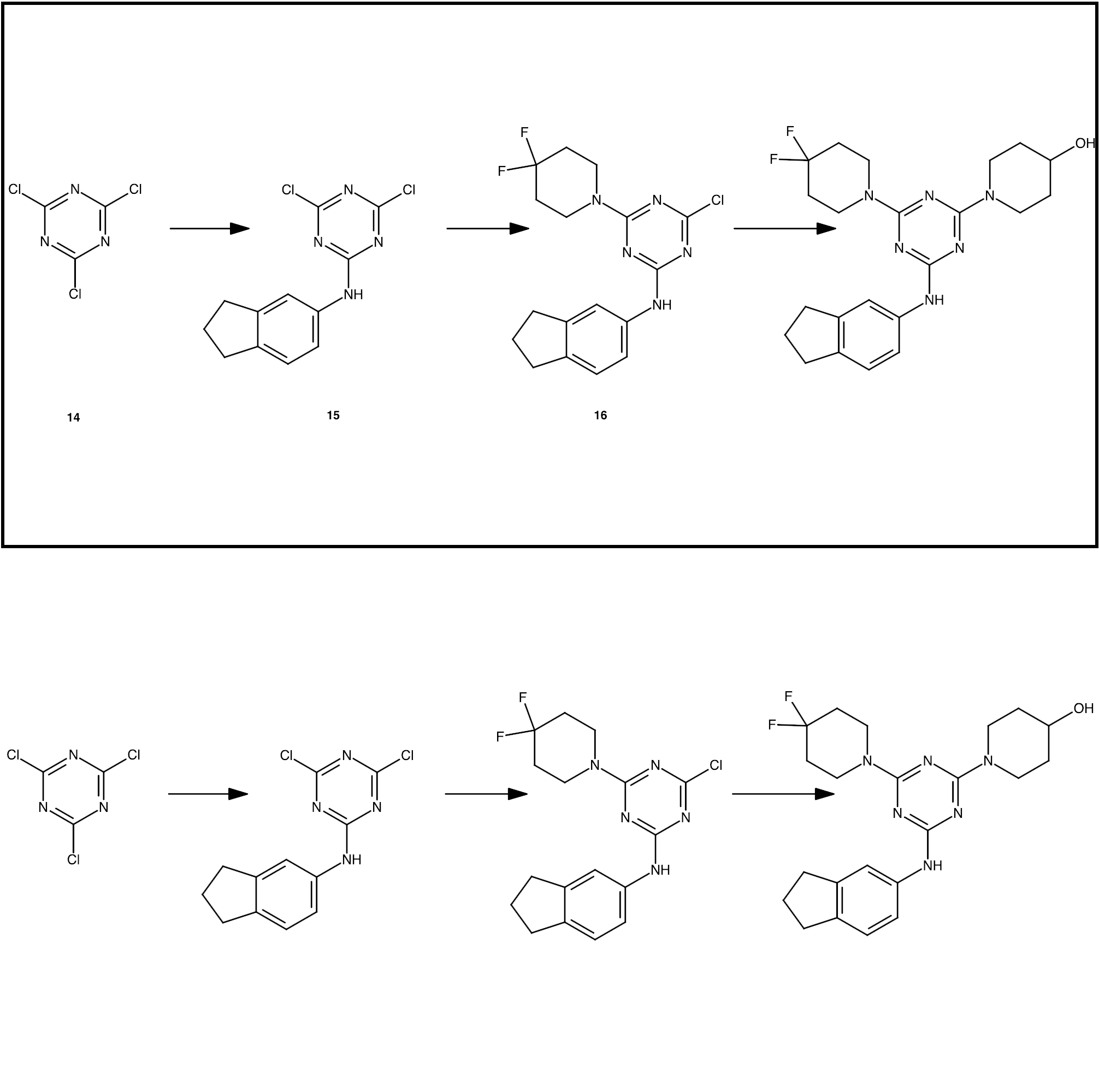}
        \caption{GXA70}
        \label{fig:gxa70_route_comp}
    \end{subfigure}
    \caption{\textbf{Comparison between actual and predicted synthesis routes.} For each subfigure, the top reaction (enclosed in a box) is the actual synthesis procedure used in this study while the bottom reaction is the predicted retrosynthetic pathway.}
    \label{fig:synthesis_route_comparison_gxa}
\end{figure}
Synthesis orders for designed compounds were placed with Enamine on August 4, 2020 (received by Enamine PO:8000109) and on September 4, 2020
(received by Enamine PO:8001023). Structures  were added to ACD commercial database as a part of regular auto-update of Enamine's catalog.
Enamine did not have boc-amino pinacolborane \textbf{20} in stock and could not follow the proposed retrosynthetic strategy for GXA104 (Figure~\ref{fig:gxa104_route_comp}). Unprotected amino-pinacolborane was available and so the strategy was changed, which made it possible to obtain GXA104 in fewer steps. At first, \textbf{20} was reacted with carboxylic acid \textbf{19}, which led to amide \textbf{21}. Cross-coupling of \textbf{21} with 3-iodo-1H-indazole led to GXA104 (see SI Section \ref{sec:synthesis_steps} steps P--Q).
GXA56 was synthesized according to the top RXN-predicted method (Figure~\ref{fig:gxa56_route_comp}).
GXA70 was synthesized by analogy to the best RXN-predicted method (Figure~\ref{fig:gxa70_route_comp}). Minor modifications were made to the synthetic steps, such as use of other bases and organic solvents (not significant for a whole scheme). The RXN strategy was chosen due to high reactivity of trichlorotriazine with amines and the need to substitute only one chlorine at the first stage (it is easier to be controlled with less nucleophilic aniline compared to more nucleophilic aliphatic secondary amines).

\begin{figure}[ht!]
    \ContinuedFloat
    \centering
    \begin{subfigure}[b]{0.65\textwidth}
        \centering
        \includegraphics[width=\textwidth,trim={0 250px 0 0},clip]{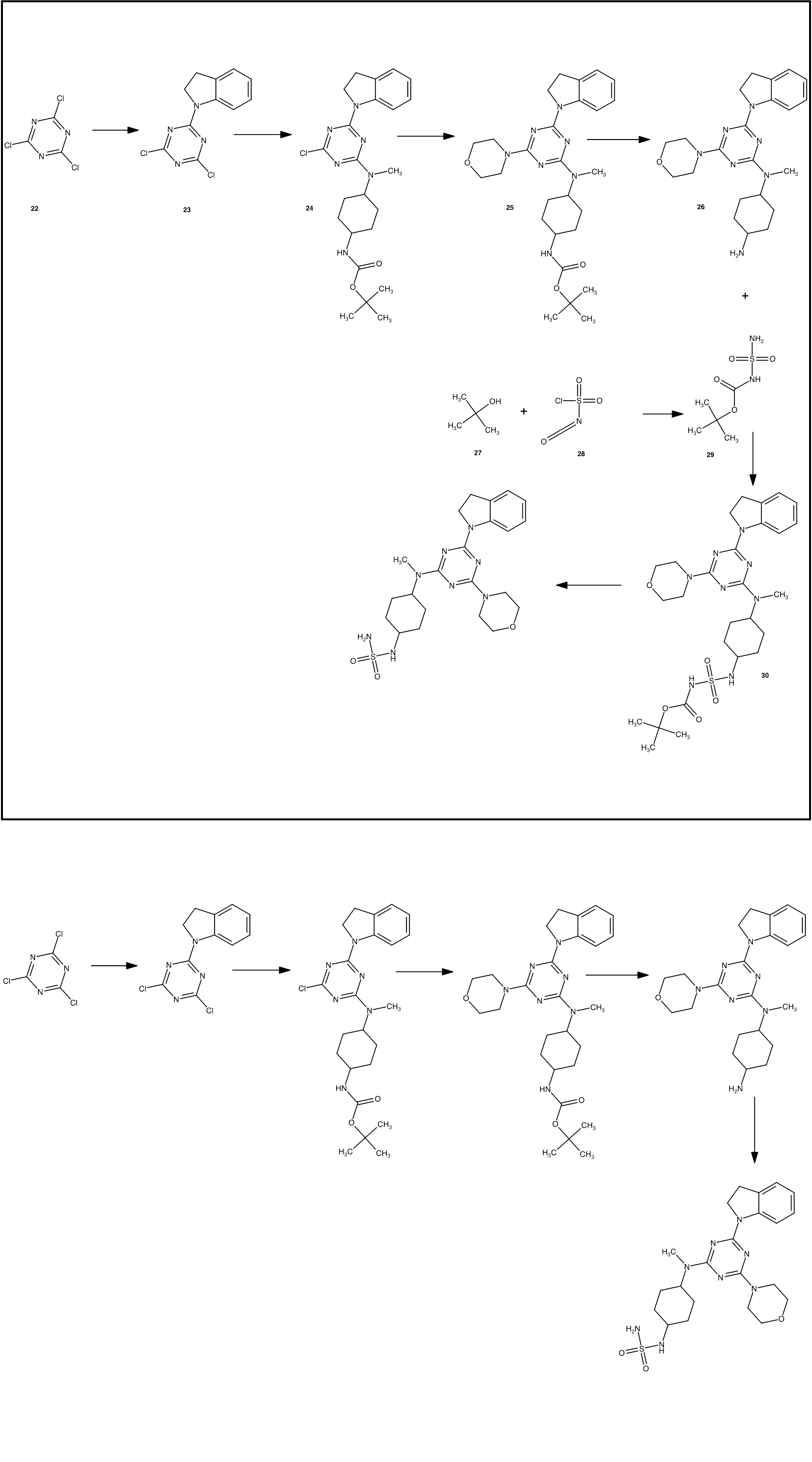}
        \caption{GXA112}
        \label{fig:gxa112_route_comp}
    \end{subfigure}
    \caption{\textbf{Comparison between actual and predicted synthesis routes.} For each subfigure, the top reaction (enclosed in a box) is the actual synthesis procedure used in this study while the bottom reaction is the predicted retrosynthetic pathway.}
    \label{fig:synthesis_route_comparison_gxa2}
\end{figure}

The RXN-predicted strategy for GXA112 was followed as closely as possible. The last synthetic step (reaction with $\mathrm{SO_2(NH_2)_2)}$ led to the final product with very low yield. To improve it, mono-Boc-protected $\mathrm{SO_2(NH_2)_2}$ was synthesized and reacted with \textbf{26}. Boc-protected final product \textbf{30} was obtained and readily deprotected via TFA cocktail (see SI Section \ref{sec:synthesis_steps} steps V--X). Spectroscopic  characterization of synthesized \textit{de novo} compounds can be found in SI Table \ref{tab:nmr}.

\subsection*{Cloning, protein production, and crystallization}
\textbf{M\textsuperscript{pro} production:} The M\textsuperscript{pro} coding sequence was codon optimised for expression in \textit{E.~coli} and synthesised by Integrated DNA technologies (IDT). The M\textsuperscript{pro} expression construct used for crystallization comprises an N-terminal GST region, an M\textsuperscript{pro} autocleavage site, the M\textsuperscript{pro} coding sequence, a hybrid cleavage site recognizable by 3C HRV protease and a C-terminal 6-Histidine tag \cite{xue2007production}. The overall construct was flanked by In-Fusion compatible ends for insertion into BamHI-XhoI cleaved pGEX-6P-1 (Sigma). Protein expression, purification and crystallisation was carried out in similar conditions to those previously described in Douangamath, et al.\cite{Douangamath2020}. Specifically, crystals were obtained from 0.1 M MES pH 6.5, 15 PEG4K, 5\% DMSO using drop ratios of \SI{0.15}{\ul} protein, \SI{0.3}{\ul} reservoir solution and \SI{0.05}{\ul} seed stock.

\noindent\textbf{Genetic constructs of spike ectodomain:} The gene encoding amino acids 1--1208 of the SARS-CoV-2 spike glycoprotein ectodomain, with mutations of RRAR > GSAS at residues 682--685 (the furin cleavage site) and KV > PP at residues 986--987, as well as inclusion of a T4 fibritin trimerisation domain, a HRV 3C cleavage site, a His-8 tag and a Twin-Strep-tag at the C-terminus, as reported by Wrapp, et al.~\cite{wrapp2020cryo}. All vectors were sequenced to confirm clones were correct.

\noindent\textbf{Spike protein production:} Recombinant spike ectodomain was expressed by transient transfection in HEK293S GnTI- cells (ATCC CRL-3022) for 9 days at \SI{30}{\degreeCelsius}. Conditioned media was dialysed against 2x phosphate buffered saline pH 7.4 buffer. The spike ectodomain was purified by immobilized metal affinity chromatography using Talon resin (Takara Bio) charged with cobalt followed by size exclusion chromatography using HiLoad 16/60 Superdex 200 column in 150 mM NaCl, 10 mM HEPES pH 8.0, 0.02\% NaN\textsubscript{3} at \SI{4}{\degreeCelsius}.

\subsection*{X-ray screening of M\textsuperscript{pro} binding compounds}
Compounds were dissolved in DMSO and directly added to the crystallization drops giving a final compound concentration of \SI{10}{\milli\molar} and DMSO concentration of 10\%. The crystals were left to soak in the presence of the compounds for 1--2 hours before being harvested and flash cooled in liquid nitrogen without the addition of further cryoprotectant.
X-ray diffraction data were collected on beamline I04-1 at Diamond Light Source and automatically processed using the Diamond automated processing pipelines \cite{Douangamath2021}.  Analysis was performed as outlined previously \cite{Douangamath2020}. Briefly, XChemExplorer\cite{Krojer2017} was used to analyse each processed dataset that was automatically selected and electron density maps were generated with Dimple\cite{Winn2011} Ligand-binding events were identified using PanDDA\cite{Pearce2017}, and ligands were modelled into PanDDA-calculated event maps using Coot\cite{Emsley2010}. Restraints were calculated with AceDRG\cite{Long2017} or GRADE\cite{grade}, structures were refined with Refmac\cite{Murshudov1997} and Buster\cite{buster} and models and quality annotations cross-reviewed. We have added PanDDA event maps in SI Figure \ref{fig:pandda} for structures of the protein-hit complexes obtained The PanDDA algorithm takes advantage of the large number of datasets collected during a fragment campaign to detect partial-occupancy ligands that are typically not readily detected in normal crystallographic maps and thus provides a better indication of bound compounds or hits than traditional omit maps.

\subsection*{Dose response assay for measuring M\textsuperscript{pro} inhibition} The solid phase extraction C4-cartridge coupled RapidFire 365 Mass Spectrometry (SPE RFMS) based high throughput dose response assay has been described~\cite{malla2021mass}. In brief, M\textsuperscript{pro} inhibitors were dry dispensed in an 11-point 3-fold dilution series using acoustic liquid transfer robot (Labcyte 550) in 384 well polypropylene plate (Greiner Bio-One). M\textsuperscript{pro} (\SI{0.3}{\micro\molar}) was dispensed across the well (\SI{25}{\uL}/well) using MultidropTM Combi (Thermo Scientific\texttrademark) and the reaction incubated at ambient temperature. Compounds were incubated with the protein for 15 minutes, following which an 11-mer substrate peptide TSAVLQ/SGFRK-NH\textsubscript{2} (\SI{4}{\micro\molar}) was dispensed (\SI{25}{\uL}/well) for probing inhibition activity.  Reaction was quenched by addition of 10\% aqueous formic acid (\SI{5}{\uL}/well) after 10 min incubation with the substrate at an ambient temperature. After addition of each reagent, the plates were centrifuged for 30s (Axygen Plate Spinner Centrifuge).
Samples were analysed by RapidFire (RF) 365 high-throughput sampling robot (Agilent) connected to an iFunnel Agilent 6550 accurate mass quadrupole time-of-flight (Q-TOF) mass spectrometer (operating parameters: capillary voltage (4000 V), nozzle voltage (1000 V), fragmentor voltage (365 V), drying gas temperature (\SI{280}{\degreeCelsius}), gas flow (13 L/min), sheath gas temperature (\SI{350}{\degreeCelsius}), sheath gas flow (12 L/min)). The peptide/protein sample was loaded onto a solid-phase extraction (SPE) C4-cartridge, AND washed with 0.1\% (v/v) aqueous formic acid to remove non-volatile buffer salts (5.5 s, 1.5 mL/min) prior to elution with aqueous 85\% (v/v) acetonitrile containing 0.1\% (v/v) formic acid (5.5 s, 1.25 mL/min). The cartridge was re-equilibrated with 0.1\% (v/v) aqueous formic acid (0.5 s, 1.25 mL/min) and sample aspirator washed with an aqueous, organic and aqueous wash before the injection of next protein: peptide mixture sample onto the SPE cartridge.

 Data were extracted with Rapid Fire integrator software (Agilent) and m/z (+1) was used for both N-terminal fragment TSAVLQ (681.34 Da), and the 11-mer substrate peptide (1191.68 Da). The percentage M\textsuperscript{pro} activity (N-terminal product peak integral/ (N-terminal product peak integral + substrate peak integral) *100) was calculated in Microsoft Excel and normalised data transferred to Prism 9 for non-linear regression curve analysis). IC$_{50}$-values are reported as the mean of technical duplicates (n = 2; mean $\pm$ SD). Signal to noise (S/N) and Z'-factor were calculated in Microsoft Excel (Z'> 0.8)~\cite{malla2021mass}.

\subsection*{Spike thermal shift-based binding assay}
Thermofluor (differential scanning fluorimetry, DSF) experiments were performed in triplicate in 96-well white PCR plates using a 1300-fold excess of small molecule (in DMSO) to \SI{1.5}{\ug} spike monomer in \SI{50}{\uL} buffer per well.  An Agilent MX3005p RT-PCR instrument ($\mathrm{\lambda_{ex}}$ 492 nm/$\mathrm{\lambda_{em}}$ 585 nm) was used to monitor the fluorescence change of a 3x final concentration of SYPRO Orange dye (Thermo) in an ``increasing-sawtooth'' temperature profile where the temperature was increased in \SI{1}{\degreeCelsius} increments from \SI{25}{\degreeCelsius} to \SI{98}{\degreeCelsius} with the fluorescence recorded at \SI{25}{\degreeCelsius}.
Four of the synthesised compounds were investigated using Thermofluor assay to assess effect upon stability.  Several conditions were tested: in 20 mM sodium acetate pH 4.6 150 mM NaCl, a storage buffer at which long term stability was observed to be much improved\cite{Zhou2020}; in 50 mM HEPES pH 7.5, 200 mM NaCl immediately after buffer exchange from the storage buffer; after incubation overnight at pH 7.5; and after incubation overnight at pH 7.5 in the presence of the compound.
Raw fluorescence data were analysed using Microsoft Excel and the JTSA software~\cite{jtsa}  using a 5-parameter model to produce melting temperature ($\mathrm{T_m}$) values.
Note that fresh spike protein exhibits a single melting transition which can be characterised as a melting point, $\mathrm{T_m}$, of \SI{65}{\degreeCelsius} in neutral pH buffer. At a reduced pH 4.6 the single melting transition is at \SI{62}{\degreeCelsius}.  As spike is incubated in pH 7.5 a second transition appears at a lower temperature with a $\mathrm{T_m}$ of \SI{50}{\degreeCelsius}. This transition increases as a proportion of the total melt until it is the only transition observed and correlates with a presumed conformational change of the spike trimer to a less stable form.

\subsection*{Focus reduction neutralization assay (FRNT) for measuring  SARS-CoV-2 live virus neutralization of spike RBD-targeting compounds}
Vero-CCL-81 cells (100,000 cells per well) were seeded in a 96-well, cell culture-treated, flat-bottom microplates for 48 hrs. Compounds were serially diluted and incubated with approximately 100 foci of SARS-CoV-2 for 1 hr at \SI{37}{\degreeCelsius}. The mixtures were added on cells and incubated for further 2 hrs at \SI{37}{\degreeCelsius} followed by the addition of 1.5\% semi-solid carboxymethyl cellulose (CMC) overlay medium to each well to limit virus diffusion. Twenty hours after infection, cells were fixed and permeabilized with 4\% paraformaldehyde and 2\% Triton-X 100, respectively.  The virus foci were stained with human anti-NP mAb (mAb206) and peroxidase-conjugated goat anti-human IgG (A0170; Sigma), and visualized by adding TrueBlue Peroxidase Substrate. Virus-infected cell foci were counted on the classic AID EliSpot reader using AID ELISpot software. The percentage of focus reduction was calculated by comparing the number of foci in treated wells with the number in untreated control wells and IC$_{50}$ was determined using the probit program from the SPSS package.

\subsection*{Pseudoviral neutralization assay for measuring inhibition of SARS-CoV-2 pseudovirus entry of spike RBD-targeting compounds}
Pseudotyped lentiviral particles expressing SARS-CoV-2 S protein were incubated with serial dilutions of compounds in white opaque 96-well plates for 1 hr at \SI{37}{\degreeCelsius}. The stable HEK293T/17 cells expressing human ACE2 were then added to the mixture at 15000 cells per well. Plates were spun at 500 RCF for 1 min and further incubated for 48 hrs. Finally, Culture supernatants were removed followed by the addition of Bright-GloTM Luciferase assay system (Promega, USA). The reaction was incubated at room temperature for 5 mins and the firefly luciferase activity was measured using CLARIOstar\textsuperscript{\textregistered} (BMG Labtech). The percentage of neutralization of compounds towards pseudotyped lentiviruses was calculated relative to the untreated control and IC$_{50}$ was determined using the probit program from the SPSS package.

\section*{Data availability}
Details of the generated molecules used in this study is available from Covid Molecule Explorer~\cite{cogmolexplorer}. Crystal structures of machine-identified small molecules bound to M\textsuperscript{pro} derived in this study have been deposited in Protein Data Bank (PDB ID: 5SML for M\textsuperscript{pro}-Z68337194,  5SMM for
M\textsuperscript{pro}-Z1633315555, and 5SMN for M\textsuperscript{pro}-
Z1365651030).

\bibliography{sample}

\begin{thebibliography}{10}
\urlstyle{rm}
\expandafter\ifx\csname url\endcsname\relax
  \def\url#1{\texttt{#1}}\fi
\expandafter\ifx\csname urlprefix\endcsname\relax\def\urlprefix{URL }\fi
\expandafter\ifx\csname doiprefix\endcsname\relax\def\doiprefix{DOI: }\fi
\providecommand{\bibinfo}[2]{#2}
\providecommand{\eprint}[2][]{\url{#2}}

\bibitem{lloyd2020high}
\bibinfo{author}{Lloyd, M.~D.}
\newblock \bibinfo{journal}{\bibinfo{title}{High-throughput screening for the
  discovery of enzyme inhibitors}}.
\newblock {\emph{\JournalTitle{Journal of Medicinal Chemistry}}}
  \textbf{\bibinfo{volume}{63}}, \bibinfo{pages}{10742--10772}
  (\bibinfo{year}{2020}).

\bibitem{polishchuk2013estimation}
\bibinfo{author}{Polishchuk, P.~G.}, \bibinfo{author}{Madzhidov, T.~I.} \&
  \bibinfo{author}{Varnek, A.}
\newblock \bibinfo{journal}{\bibinfo{title}{Estimation of the size of drug-like
  chemical space based on {GDB}-17 data}}.
\newblock {\emph{\JournalTitle{Journal of Computer-Aided Molecular Design}}}
  \textbf{\bibinfo{volume}{27}}, \bibinfo{pages}{675--679}
  (\bibinfo{year}{2013}).

\bibitem{dimasi2016innovation}
\bibinfo{author}{DiMasi, J.~A.}, \bibinfo{author}{Grabowski, H.~G.} \&
  \bibinfo{author}{Hansen, R.~W.}
\newblock \bibinfo{journal}{\bibinfo{title}{Innovation in the pharmaceutical
  industry: new estimates of {R\&D} costs}}.
\newblock {\emph{\JournalTitle{Journal of Health Economics}}}
  \textbf{\bibinfo{volume}{47}}, \bibinfo{pages}{20--33}
  (\bibinfo{year}{2016}).

\bibitem{zunger2018inverse}
\bibinfo{author}{Zunger, A.}
\newblock \bibinfo{journal}{\bibinfo{title}{Inverse design in search of
  materials with target functionalities}}.
\newblock {\emph{\JournalTitle{Nature Reviews Chemistry}}}
  \textbf{\bibinfo{volume}{2}}, \bibinfo{pages}{1--16} (\bibinfo{year}{2018}).

\bibitem{review1_sousa2021generative}
\bibinfo{author}{Sousa, T.}, \bibinfo{author}{Correia, J.},
  \bibinfo{author}{Pereira, V.} \& \bibinfo{author}{Rocha, M.}
\newblock \bibinfo{journal}{\bibinfo{title}{Generative deep learning for
  targeted compound design}}.
\newblock {\emph{\JournalTitle{Journal of Chemical Information and Modeling}}}
  \textbf{\bibinfo{volume}{61}}, \bibinfo{pages}{5343--5361}
  (\bibinfo{year}{2021}).

\bibitem{Zhavoronkov2019natbio}
\bibinfo{author}{Zhavoronkov, A.} \emph{et~al.}
\newblock \bibinfo{journal}{\bibinfo{title}{Deep learning enables rapid
  identification of potent {DDR1} kinase inhibitors}}.
\newblock {\emph{\JournalTitle{Nature Biotechnology}}}
  \textbf{\bibinfo{volume}{37}}, \bibinfo{pages}{1038--–1040}
  (\bibinfo{year}{2019}).

\bibitem{merk2018novo}
\bibinfo{author}{Merk, D.}, \bibinfo{author}{Friedrich, L.},
  \bibinfo{author}{Grisoni, F.} \& \bibinfo{author}{Schneider, G.}
\newblock \bibinfo{journal}{\bibinfo{title}{De novo design of bioactive small
  molecules by artificial intelligence}}.
\newblock {\emph{\JournalTitle{Molecular informatics}}}
  \textbf{\bibinfo{volume}{37}}, \bibinfo{pages}{1700153}
  (\bibinfo{year}{2018}).

\bibitem{grisoni2021combining}
\bibinfo{author}{Grisoni, F.} \emph{et~al.}
\newblock \bibinfo{journal}{\bibinfo{title}{Combining generative artificial
  intelligence and on-chip synthesis for de novo drug design}}.
\newblock {\emph{\JournalTitle{Science advances}}}
  \textbf{\bibinfo{volume}{7}}, \bibinfo{pages}{eabg3338}
  (\bibinfo{year}{2021}).

\bibitem{chenthamarakshan2020cogmol}
\bibinfo{author}{Chenthamarakshan, V.} \emph{et~al.}
\newblock \bibinfo{journal}{\bibinfo{title}{Cogmol: Target-specific and
  selective drug design for covid-19 using deep generative models}}.
\newblock {\emph{\JournalTitle{arXiv preprint arXiv:2004.01215}}}
  (\bibinfo{year}{2020}).

\bibitem{bommasani2021opportunities}
\bibinfo{author}{Bommasani, R.} \emph{et~al.}
\newblock \bibinfo{journal}{\bibinfo{title}{On the opportunities and risks of
  foundation models}}.
\newblock {\emph{\JournalTitle{arXiv preprint arXiv:2108.07258}}}
  (\bibinfo{year}{2021}).

\bibitem{IBMblog}
\bibinfo{author}{IBM}.
\newblock \emph{\bibinfo{title}{What are foundation models?}}
  (\bibinfo{year}{2022 (accessed May, 2022)}).

\bibitem{cogmolexplorer}
\bibinfo{author}{IBM}.
\newblock \bibinfo{title}{{CogMol Molecule Explorer}}.
\newblock \bibinfo{howpublished}{\url{https://covid19-mol.mybluemix.net/}}.
\newblock \bibinfo{note}{[Online; released April 2020; accessed
  07-March-2022]}.

\bibitem{kingma2013auto}
\bibinfo{author}{Kingma, D.~P.} \& \bibinfo{author}{Welling, M.}
\newblock \bibinfo{journal}{\bibinfo{title}{Auto-encoding variational
  {B}ayes}}.
\newblock {\emph{\JournalTitle{arXiv preprint arXiv:1312.6114}}}
  (\bibinfo{year}{2013}).

\bibitem{weininger1988smiles}
\bibinfo{author}{Weininger, D.}
\newblock \bibinfo{journal}{\bibinfo{title}{Smiles, a chemical language and
  information system. 1. introduction to methodology and encoding rules}}.
\newblock {\emph{\JournalTitle{Journal of chemical information and computer
  sciences}}} \textbf{\bibinfo{volume}{28}}, \bibinfo{pages}{31--36}
  (\bibinfo{year}{1988}).

\bibitem{Alley_2019}
\bibinfo{author}{Alley, E.~C.}, \bibinfo{author}{Khimulya, G.},
  \bibinfo{author}{Biswas, S.}, \bibinfo{author}{AlQuraishi, M.} \&
  \bibinfo{author}{Church, G.~M.}
\newblock \bibinfo{journal}{\bibinfo{title}{Unified rational protein
  engineering with sequence-based deep representation learning}}.
\newblock {\emph{\JournalTitle{Nature Methods}}} \textbf{\bibinfo{volume}{16}},
  \bibinfo{pages}{1315–1322}, \doiprefix\url{10.1038/s41592-019-0598-1}
  (\bibinfo{year}{2019}).

\bibitem{das2021accelerated}
\bibinfo{author}{Das, P.} \emph{et~al.}
\newblock \bibinfo{journal}{\bibinfo{title}{Accelerated antimicrobial discovery
  via deep generative models and molecular dynamics simulations}}.
\newblock {\emph{\JournalTitle{Nature Biomedical Engineering}}}
  \textbf{\bibinfo{volume}{5}}, \bibinfo{pages}{613--623}
  (\bibinfo{year}{2021}).

\bibitem{schwaller2020predicting}
\bibinfo{author}{Schwaller, P.} \emph{et~al.}
\newblock \bibinfo{journal}{\bibinfo{title}{Predicting retrosynthetic pathways
  using transformer-based models and a hyper-graph exploration strategy}}.
\newblock {\emph{\JournalTitle{Chemical science}}}
  \textbf{\bibinfo{volume}{11}}, \bibinfo{pages}{3316--3325}
  (\bibinfo{year}{2020}).

\bibitem{malla2021mass}
\bibinfo{author}{Malla, T.~R.} \emph{et~al.}
\newblock \bibinfo{journal}{\bibinfo{title}{Mass spectrometry reveals potential
  of $\beta$-lactams as sars-cov-2 m pro inhibitors}}.
\newblock {\emph{\JournalTitle{Chemical communications}}}
  \textbf{\bibinfo{volume}{57}}, \bibinfo{pages}{1430--1433}
  (\bibinfo{year}{2021}).

\bibitem{morris2021discovery}
\bibinfo{author}{Morris, A.} \emph{et~al.}
\newblock \bibinfo{journal}{\bibinfo{title}{Discovery of sars-cov-2 main
  protease inhibitors using a synthesis-directed de novo design model}}.
\newblock {\emph{\JournalTitle{Chemical Communications}}}
  \textbf{\bibinfo{volume}{57}}, \bibinfo{pages}{5909--5912}
  (\bibinfo{year}{2021}).

\bibitem{glaab2021pharmacophore}
\bibinfo{author}{Glaab, E.}, \bibinfo{author}{Manoharan, G.~B.} \&
  \bibinfo{author}{Abankwa, D.}
\newblock \bibinfo{journal}{\bibinfo{title}{Pharmacophore model for sars-cov-2
  3clpro small-molecule inhibitors and in vitro experimental validation of
  computationally screened inhibitors}}.
\newblock {\emph{\JournalTitle{Journal of Chemical Information and Modeling}}}
  \textbf{\bibinfo{volume}{61}}, \bibinfo{pages}{4082--4096}
  (\bibinfo{year}{2021}).

\bibitem{zhang2021potent}
\bibinfo{author}{Zhang, C.-H.} \emph{et~al.}
\newblock \bibinfo{journal}{\bibinfo{title}{Potent noncovalent inhibitors of
  the main protease of sars-cov-2 from molecular sculpting of the drug
  perampanel guided by free energy perturbation calculations}}.
\newblock {\emph{\JournalTitle{ACS central science}}}
  \textbf{\bibinfo{volume}{7}}, \bibinfo{pages}{467--475}
  (\bibinfo{year}{2021}).

\bibitem{enamineadvanced}
\bibinfo{author}{Enamine}.
\newblock \bibinfo{title}{{Enamine Advanced Collection}}.
\newblock
  \bibinfo{howpublished}{\url{https://enamine.net/compound-collections/screening-collection/advanced-collection}}
  (\bibinfo{year}{2022}).
\newblock \bibinfo{note}{[Online; accessed 07-March-2022]}.

\bibitem{Jin2020}
\bibinfo{author}{Jin, Z.} \emph{et~al.}
\newblock \bibinfo{journal}{\bibinfo{title}{Structure of mpro from
  {SARS}-{CoV}-2 and discovery of its inhibitors}}.
\newblock {\emph{\JournalTitle{Nature}}} \textbf{\bibinfo{volume}{582}},
  \bibinfo{pages}{289--293}, \doiprefix\url{10.1038/s41586-020-2223-y}
  (\bibinfo{year}{2020}).

\bibitem{Abdelnabi_pfizer}
\bibinfo{author}{Abdelnabi, R.} \emph{et~al.}
\newblock \bibinfo{journal}{\bibinfo{title}{The oral protease inhibitor
  (pf-07321332) protects syrian hamsters against infection with sars-cov-2
  variants of concern}}.
\newblock {\emph{\JournalTitle{Nature Communications}}}
  \textbf{\bibinfo{volume}{13}}, \bibinfo{pages}{719},
  \doiprefix\url{10.1038/s41467-022-28354-0} (\bibinfo{year}{2022}).

\bibitem{achdout2020noncovalent}
\bibinfo{author}{Consortium, T. C.~M.} \emph{et~al.}
\newblock \bibinfo{journal}{\bibinfo{title}{Open science discovery of oral
  non-covalent sars-cov-2 main protease inhibitor therapeutics}}.
\newblock {\emph{\JournalTitle{bioRxiv}}}
  \doiprefix\url{10.1101/2020.10.29.339317} (\bibinfo{year}{2021}).
\newblock
  \eprint{https://www.biorxiv.org/content/early/2021/10/18/2020.10.29.339317.full.pdf}.

\bibitem{luttens2022ultralarge}
\bibinfo{author}{Luttens, A.} \emph{et~al.}
\newblock \bibinfo{journal}{\bibinfo{title}{Ultralarge virtual screening
  identifies sars-cov-2 main protease inhibitors with broad-spectrum activity
  against coronaviruses}}.
\newblock {\emph{\JournalTitle{Journal of the American Chemical Society}}}
  (\bibinfo{year}{2022}).

\bibitem{sasaki2022oral}
\bibinfo{author}{Sasaki, M.} \emph{et~al.}
\newblock \bibinfo{journal}{\bibinfo{title}{Oral administration of s-217622, a
  sars-cov-2 main protease inhibitor, decreases viral load and accelerates
  recovery from clinical aspects of covid-19}}.
\newblock {\emph{\JournalTitle{bioRxiv}}}  (\bibinfo{year}{2022}).

\bibitem{Fischer_molnupiravir_2021}
\bibinfo{author}{Fischer, W.} \emph{et~al.}
\newblock \bibinfo{journal}{\bibinfo{title}{Molnupiravir, an oral antiviral
  treatment for covid-19}}.
\newblock {\emph{\JournalTitle{medRxiv}}} \bibinfo{pages}{2021.06.17.21258639},
  \doiprefix\url{10.1101/2021.06.17.21258639} (\bibinfo{year}{2021}).

\bibitem{tanimoto1958elementary}
\bibinfo{author}{Tanimoto, T.~T.}
\newblock \bibinfo{title}{Elementary mathematical theory of classification and
  prediction} (\bibinfo{year}{1958}).

\bibitem{Rogers_Hahn_2010_ecfp}
\bibinfo{author}{Rogers, D.} \& \bibinfo{author}{Hahn, M.}
\newblock \bibinfo{journal}{\bibinfo{title}{Extended-connectivity
  fingerprints}}.
\newblock {\emph{\JournalTitle{Journal of Chemical Information and Modeling}}}
  \textbf{\bibinfo{volume}{50}}, \bibinfo{pages}{742–754},
  \doiprefix\url{10.1021/ci100050t} (\bibinfo{year}{2010}).

\bibitem{durdagi2021near}
\bibinfo{author}{Durdagi, S.} \emph{et~al.}
\newblock \bibinfo{journal}{\bibinfo{title}{Near-physiological-temperature
  serial crystallography reveals conformations of sars-cov-2 main protease
  active site for improved drug repurposing}}.
\newblock {\emph{\JournalTitle{Structure}}} \textbf{\bibinfo{volume}{29}},
  \bibinfo{pages}{1382--1396} (\bibinfo{year}{2021}).

\bibitem{toelzer2020free}
\bibinfo{author}{Toelzer, C.} \emph{et~al.}
\newblock \bibinfo{journal}{\bibinfo{title}{Free fatty acid binding pocket in
  the locked structure of sars-cov-2 spike protein}}.
\newblock {\emph{\JournalTitle{Science}}} \textbf{\bibinfo{volume}{370}},
  \bibinfo{pages}{725--730} (\bibinfo{year}{2020}).

\bibitem{daina2017swissadme}
\bibinfo{author}{Daina, A.}, \bibinfo{author}{Michielin, O.} \&
  \bibinfo{author}{Zoete, V.}
\newblock \bibinfo{journal}{\bibinfo{title}{Swissadme: a free web tool to
  evaluate pharmacokinetics, drug-likeness and medicinal chemistry friendliness
  of small molecules}}.
\newblock {\emph{\JournalTitle{Scientific reports}}}
  \textbf{\bibinfo{volume}{7}}, \bibinfo{pages}{1--13} (\bibinfo{year}{2017}).

\bibitem{achdout2020covid}
\bibinfo{author}{Achdout, H.} \emph{et~al.}
\newblock \bibinfo{journal}{\bibinfo{title}{Covid moonshot: open science
  discovery of sars-cov-2 main protease inhibitors by combining crowdsourcing,
  high-throughput experiments, computational simulations, and machine
  learning}}.
\newblock {\emph{\JournalTitle{bioRxiv}}}  (\bibinfo{year}{2020}).

\bibitem{unoh2022discovery}
\bibinfo{author}{Unoh, Y.} \emph{et~al.}
\newblock \bibinfo{journal}{\bibinfo{title}{Discovery of s-217622, a
  non-covalent oral sars-cov-2 3cl protease inhibitor clinical candidate for
  treating covid-19}}.
\newblock {\emph{\JournalTitle{bioRxiv}}}  (\bibinfo{year}{2022}).

\bibitem{ren2022alphafold}
\bibinfo{author}{Ren, F.} \emph{et~al.}
\newblock \bibinfo{title}{Alphafold accelerates artificial intelligence powered
  drug discovery: Efficient discovery of a novel cyclin-dependent kinase 20
  (cdk20) small molecule inhibitor} (\bibinfo{year}{2022}).
\newblock \eprint{2201.09647}.

\bibitem{baek2021accurate}
\bibinfo{author}{Baek, M.} \emph{et~al.}
\newblock \bibinfo{journal}{\bibinfo{title}{Accurate prediction of protein
  structures and interactions using a three-track neural network}}.
\newblock {\emph{\JournalTitle{Science}}} \textbf{\bibinfo{volume}{373}},
  \bibinfo{pages}{871--876} (\bibinfo{year}{2021}).

\bibitem{merk2018tuning}
\bibinfo{author}{Merk, D.}, \bibinfo{author}{Grisoni, F.},
  \bibinfo{author}{Friedrich, L.} \& \bibinfo{author}{Schneider, G.}
\newblock \bibinfo{journal}{\bibinfo{title}{Tuning artificial intelligence on
  the de novo design of natural-product-inspired retinoid x receptor
  modulators}}.
\newblock {\emph{\JournalTitle{Communications Chemistry}}}
  \textbf{\bibinfo{volume}{1}}, \bibinfo{pages}{1--9} (\bibinfo{year}{2018}).

\bibitem{polykovskiy2018entangled}
\bibinfo{author}{Polykovskiy, D.} \emph{et~al.}
\newblock \bibinfo{journal}{\bibinfo{title}{Entangled conditional adversarial
  autoencoder for de novo drug discovery}}.
\newblock {\emph{\JournalTitle{Molecular pharmaceutics}}}
  \textbf{\bibinfo{volume}{15}}, \bibinfo{pages}{4398--4405}
  (\bibinfo{year}{2018}).

\bibitem{putin2018adversarial}
\bibinfo{author}{Putin, E.} \emph{et~al.}
\newblock \bibinfo{journal}{\bibinfo{title}{Adversarial threshold neural
  computer for molecular de novo design}}.
\newblock {\emph{\JournalTitle{Molecular pharmaceutics}}}
  \textbf{\bibinfo{volume}{15}}, \bibinfo{pages}{4386--4397}
  (\bibinfo{year}{2018}).

\bibitem{tan2020automated}
\bibinfo{author}{Tan, X.} \emph{et~al.}
\newblock \bibinfo{journal}{\bibinfo{title}{Automated design and optimization
  of multitarget schizophrenia drug candidates by deep learning}}.
\newblock {\emph{\JournalTitle{European Journal of Medicinal Chemistry}}}
  \textbf{\bibinfo{volume}{204}}, \bibinfo{pages}{112572}
  (\bibinfo{year}{2020}).

\bibitem{assmann2020novel}
\bibinfo{author}{Assmann, M.} \emph{et~al.}
\newblock \bibinfo{journal}{\bibinfo{title}{A novel machine learning approach
  uncovers new and distinctive inhibitors for cyclin-dependent kinase 9}}.
\newblock {\emph{\JournalTitle{bioRxiv}}}  (\bibinfo{year}{2020}).

\bibitem{rives2021biological}
\bibinfo{author}{Rives, A.} \emph{et~al.}
\newblock \bibinfo{journal}{\bibinfo{title}{Biological structure and function
  emerge from scaling unsupervised learning to 250 million protein sequences}}.
\newblock {\emph{\JournalTitle{Proceedings of the National Academy of
  Sciences}}} \textbf{\bibinfo{volume}{118}} (\bibinfo{year}{2021}).

\bibitem{dejnirattisai2022sars}
\bibinfo{author}{Dejnirattisai, W.} \emph{et~al.}
\newblock \bibinfo{journal}{\bibinfo{title}{Sars-cov-2 omicron-b. 1.1. 529
  leads to widespread escape from neutralizing antibody responses}}.
\newblock {\emph{\JournalTitle{Cell}}}  (\bibinfo{year}{2022}).

\bibitem{hoffman2021sampleefficient}
\bibinfo{author}{Hoffman, S.~C.}, \bibinfo{author}{Chenthamarakshan, V.},
  \bibinfo{author}{Zubarev, D.}, \bibinfo{author}{Sanders, D.~P.} \&
  \bibinfo{author}{Das, P.}
\newblock \bibinfo{title}{Sample-efficient generation of novel photo-acid
  generator molecules using a deep generative model}.
\newblock In \emph{\bibinfo{booktitle}{NeurIPS 2021 Workshop on Deep Generative
  Models and Downstream Applications}} (\bibinfo{year}{2021}).

\bibitem{gebauer2022inverse}
\bibinfo{author}{Gebauer, N.~W.}, \bibinfo{author}{Gastegger, M.},
  \bibinfo{author}{Hessmann, S.~S.}, \bibinfo{author}{M{\"u}ller, K.-R.} \&
  \bibinfo{author}{Sch{\"u}tt, K.~T.}
\newblock \bibinfo{journal}{\bibinfo{title}{Inverse design of 3d molecular
  structures with conditional generative neural networks}}.
\newblock {\emph{\JournalTitle{Nature Communications}}}
  \textbf{\bibinfo{volume}{13}}, \bibinfo{pages}{1--11} (\bibinfo{year}{2022}).

\bibitem{schiff2021augmenting}
\bibinfo{author}{Schiff, Y.}, \bibinfo{author}{Chenthamarakshan, V.},
  \bibinfo{author}{Hoffman, S.}, \bibinfo{author}{Ramamurthy, K.~N.} \&
  \bibinfo{author}{Das, P.}
\newblock \bibinfo{title}{Augmenting molecular deep generative models with
  topological data analysis representations}.
\newblock In \emph{\bibinfo{booktitle}{IEEE International Conference on
  Acoustics, Speech, and Signal Processing}} (\bibinfo{year}{2022}).

\bibitem{hoffman2022optimizing}
\bibinfo{author}{Hoffman, S.~C.}, \bibinfo{author}{Chenthamarakshan, V.},
  \bibinfo{author}{Wadhawan, K.}, \bibinfo{author}{Chen, P.-Y.} \&
  \bibinfo{author}{Das, P.}
\newblock \bibinfo{journal}{\bibinfo{title}{Optimizing molecules using
  efficient queries from property evaluations}}.
\newblock {\emph{\JournalTitle{Nature Machine Intelligence}}}
  \textbf{\bibinfo{volume}{4}}, \bibinfo{pages}{21--31} (\bibinfo{year}{2022}).

\bibitem{bowman2015generating}
\bibinfo{author}{Bowman, S.~R.} \emph{et~al.}
\newblock \bibinfo{journal}{\bibinfo{title}{Generating sentences from a
  continuous space}}.
\newblock {\emph{\JournalTitle{arXiv preprint arXiv:1511.06349}}}
  (\bibinfo{year}{2015}).

\bibitem{polykovskiy2018molecular}
\bibinfo{author}{Polykovskiy, D.} \emph{et~al.}
\newblock \bibinfo{journal}{\bibinfo{title}{Molecular sets ({MOSES}): A
  benchmarking platform for molecular generation models}}.
\newblock {\emph{\JournalTitle{arXiv preprint arXiv:1811.12823}}}
  (\bibinfo{year}{2018}).

\bibitem{irwin2005zinc}
\bibinfo{author}{Irwin, J.~J.} \& \bibinfo{author}{Shoichet, B.~K.}
\newblock \bibinfo{journal}{\bibinfo{title}{{ZINC}--a free database of
  commercially available compounds for virtual screening}}.
\newblock {\emph{\JournalTitle{Journal of Chemical Information and Modeling}}}
  \textbf{\bibinfo{volume}{45}}, \bibinfo{pages}{177--182}
  (\bibinfo{year}{2005}).

\bibitem{gomez2018automatic}
\bibinfo{author}{G{\'o}mez-Bombarelli, R.} \emph{et~al.}
\newblock \bibinfo{journal}{\bibinfo{title}{Automatic chemical design using a
  data-driven continuous representation of molecules}}.
\newblock {\emph{\JournalTitle{ACS Central Science}}}
  \textbf{\bibinfo{volume}{4}}, \bibinfo{pages}{268--276}
  (\bibinfo{year}{2018}).

\bibitem{gilson2015bindingdb}
\bibinfo{author}{Gilson, M.~K.} \emph{et~al.}
\newblock \bibinfo{journal}{\bibinfo{title}{{BindingDB} in 2015: a public
  database for medicinal chemistry, computational chemistry and systems
  pharmacology}}.
\newblock {\emph{\JournalTitle{Nucleic Acids Research}}}
  \textbf{\bibinfo{volume}{44}}, \bibinfo{pages}{D1045--D1053}
  (\bibinfo{year}{2015}).

\bibitem{karimi2018deepaffinity}
\bibinfo{author}{Karimi, M.}, \bibinfo{author}{Wu, D.}, \bibinfo{author}{Wang,
  Z.} \& \bibinfo{author}{Shen, Y.}
\newblock \bibinfo{journal}{\bibinfo{title}{Deepaffinity: interpretable deep
  learning of compound--protein affinity through unified recurrent and
  convolutional neural networks}}.
\newblock {\emph{\JournalTitle{Bioinformatics}}} \textbf{\bibinfo{volume}{35}},
  \bibinfo{pages}{3329--3338} (\bibinfo{year}{2019}).

\bibitem{rdkit}
\bibinfo{title}{{RDKit: Open-source cheminformatics}}.
\newblock \bibinfo{howpublished}{\url{http://www.rdkit.org}}.
\newblock \bibinfo{note}{[Online; accessed 07-March-2022]}.

\bibitem{lim2020explaining}
\bibinfo{author}{Lim, K.~W.}, \bibinfo{author}{Sharma, B.},
  \bibinfo{author}{Das, P.}, \bibinfo{author}{Chenthamarakshan, V.} \&
  \bibinfo{author}{Dordick, J.~S.}
\newblock \bibinfo{journal}{\bibinfo{title}{Explaining chemical toxicity using
  missing features}}.
\newblock {\emph{\JournalTitle{arXiv preprint arXiv:2009.12199}}}
  (\bibinfo{year}{2020}).

\bibitem{Tox21}
\bibinfo{author}{Huang, R.} \emph{et~al.}
\newblock \bibinfo{journal}{\bibinfo{title}{{Tox21} challenge to build
  predictive models of nuclear receptor and stress response pathways as
  mediated by exposure to environmental chemicals and drugs}}.
\newblock {\emph{\JournalTitle{Frontiers in Environmental Science}}}
  \textbf{\bibinfo{volume}{3}}, \bibinfo{pages}{85},
  \doiprefix\url{10.3389/fenvs.2015.00085} (\bibinfo{year}{2016}).

\bibitem{wu2018moleculenet}
\bibinfo{author}{Wu, Z.} \emph{et~al.}
\newblock \bibinfo{journal}{\bibinfo{title}{Moleculenet: a benchmark for
  molecular machine learning}}.
\newblock {\emph{\JournalTitle{Chemical science}}}
  \textbf{\bibinfo{volume}{9}}, \bibinfo{pages}{513--530}
  (\bibinfo{year}{2018}).

\bibitem{carrique2020sars}
\bibinfo{author}{Carrique, L.} \emph{et~al.}
\newblock \bibinfo{journal}{\bibinfo{title}{The sars-cov-2 spike harbours a
  lipid binding pocket which modulates stability of the prefusion trimer}}.
\newblock {\emph{\JournalTitle{Available at SSRN 3656643}}}
  (\bibinfo{year}{2020}).

\bibitem{trott2010autodock}
\bibinfo{author}{Trott, O.} \& \bibinfo{author}{Olson, A.~J.}
\newblock \bibinfo{journal}{\bibinfo{title}{Auto{D}ock {V}ina: improving the
  speed and accuracy of docking with a new scoring function, efficient
  optimization, and multithreading}}.
\newblock {\emph{\JournalTitle{Journal of Computational Chemistry}}}
  \textbf{\bibinfo{volume}{31}}, \bibinfo{pages}{455--461}
  (\bibinfo{year}{2010}).

\bibitem{PyMOL}
\bibinfo{author}{{Schr\"odinger, LLC}}.
\newblock \bibinfo{title}{The {PyMOL} molecular graphics system, version~2.4.1}
  (\bibinfo{year}{2020}).

\bibitem{laskowski2011ligplot+}
\bibinfo{author}{Laskowski, R.~A.} \& \bibinfo{author}{Swindells, M.~B.}
\newblock \bibinfo{journal}{\bibinfo{title}{Ligplot+: multiple ligand-protein
  interaction diagrams for drug discovery}}.
\newblock {\emph{\JournalTitle{Journal of chemical information and modeling}}}
  \textbf{\bibinfo{volume}{51}}, \bibinfo{pages}{2778--2786}
  (\bibinfo{year}{2011}).

\bibitem{emolecules}
\bibinfo{author}{eMolecules}.
\newblock \bibinfo{title}{{eMolecules} plus database}.
\newblock \bibinfo{howpublished}{\url{https://www.emolecules.com/}}
  (\bibinfo{year}{2020}).
\newblock \bibinfo{note}{[Online; accessed May 2020]}.

\bibitem{xue2007production}
\bibinfo{author}{Xue, X.} \emph{et~al.}
\newblock \bibinfo{journal}{\bibinfo{title}{Production of authentic sars-cov
  mpro with enhanced activity: application as a novel tag-cleavage
  endopeptidase for protein overproduction}}.
\newblock {\emph{\JournalTitle{Journal of molecular biology}}}
  \textbf{\bibinfo{volume}{366}}, \bibinfo{pages}{965--975}
  (\bibinfo{year}{2007}).

\bibitem{Douangamath2020}
\bibinfo{author}{Douangamath, A.} \emph{et~al.}
\newblock \bibinfo{journal}{\bibinfo{title}{Crystallographic and electrophilic
  fragment screening of the {SARS}-{CoV}-2 main protease}}.
\newblock {\emph{\JournalTitle{Nature Communications}}}
  \textbf{\bibinfo{volume}{11}}, \doiprefix\url{10.1038/s41467-020-18709-w}
  (\bibinfo{year}{2020}).

\bibitem{wrapp2020cryo}
\bibinfo{author}{Wrapp, D.} \emph{et~al.}
\newblock \bibinfo{journal}{\bibinfo{title}{Cryo-em structure of the 2019-ncov
  spike in the prefusion conformation}}.
\newblock {\emph{\JournalTitle{Science}}} \textbf{\bibinfo{volume}{367}},
  \bibinfo{pages}{1260--1263} (\bibinfo{year}{2020}).

\bibitem{Douangamath2021}
\bibinfo{author}{Douangamath, A.} \emph{et~al.}
\newblock \bibinfo{journal}{\bibinfo{title}{Achieving efficient fragment
  screening at {XChem} facility at diamond light source}}.
\newblock {\emph{\JournalTitle{Journal of Visualized Experiments}}}
  \doiprefix\url{10.3791/62414} (\bibinfo{year}{2021}).

\bibitem{Krojer2017}
\bibinfo{author}{Krojer, T.} \emph{et~al.}
\newblock \bibinfo{journal}{\bibinfo{title}{The {\it xchemexplorer} graphical
  workflow tool for routine or large-scale protein{\textendash}ligand structure
  determination}}.
\newblock {\emph{\JournalTitle{Acta Crystallographica Section D Structural
  Biology}}} \textbf{\bibinfo{volume}{73}}, \bibinfo{pages}{267--278},
  \doiprefix\url{10.1107/s2059798316020234} (\bibinfo{year}{2017}).

\bibitem{Winn2011}
\bibinfo{author}{Winn, M.~D.} \emph{et~al.}
\newblock \bibinfo{journal}{\bibinfo{title}{Overview of the {\it ccp}4 suite
  and current developments}}.
\newblock {\emph{\JournalTitle{Acta Crystallographica Section D Biological
  Crystallography}}} \textbf{\bibinfo{volume}{67}}, \bibinfo{pages}{235--242},
  \doiprefix\url{10.1107/s0907444910045749} (\bibinfo{year}{2011}).

\bibitem{Pearce2017}
\bibinfo{author}{Pearce, N.~M.} \emph{et~al.}
\newblock \bibinfo{journal}{\bibinfo{title}{A multi-crystal method for
  extracting obscured crystallographic states from conventionally
  uninterpretable electron density}}.
\newblock {\emph{\JournalTitle{Nature Communications}}}
  \textbf{\bibinfo{volume}{8}}, \doiprefix\url{10.1038/ncomms15123}
  (\bibinfo{year}{2017}).

\bibitem{Emsley2010}
\bibinfo{author}{Emsley, P.}, \bibinfo{author}{Lohkamp, B.},
  \bibinfo{author}{Scott, W.~G.} \& \bibinfo{author}{Cowtan, K.}
\newblock \bibinfo{journal}{\bibinfo{title}{Features and development of {\it
  coot}}}.
\newblock {\emph{\JournalTitle{Acta Crystallographica Section D Biological
  Crystallography}}} \textbf{\bibinfo{volume}{66}}, \bibinfo{pages}{486--501},
  \doiprefix\url{10.1107/s0907444910007493} (\bibinfo{year}{2010}).

\bibitem{Long2017}
\bibinfo{author}{Long, F.} \emph{et~al.}
\newblock \bibinfo{journal}{\bibinfo{title}{{AceDRG}: a stereochemical
  description generator for ligands}}.
\newblock {\emph{\JournalTitle{Acta Crystallographica Section D Structural
  Biology}}} \textbf{\bibinfo{volume}{73}}, \bibinfo{pages}{112--122},
  \doiprefix\url{10.1107/s2059798317000067} (\bibinfo{year}{2017}).

\bibitem{grade}
\bibinfo{author}{Smart, O.~S.} \emph{et~al.}
\newblock \bibinfo{title}{grade, version 1.2.20}.
\newblock \bibinfo{howpublished}{https://www.globalphasing.com/}
  (\bibinfo{year}{2021}).

\bibitem{Murshudov1997}
\bibinfo{author}{Murshudov, G.~N.}, \bibinfo{author}{Vagin, A.~A.} \&
  \bibinfo{author}{Dodson, E.~J.}
\newblock \bibinfo{journal}{\bibinfo{title}{Refinement of macromolecular
  structures by the maximum-likelihood method}}.
\newblock {\emph{\JournalTitle{Acta Crystallographica Section D Biological
  Crystallography}}} \textbf{\bibinfo{volume}{53}}, \bibinfo{pages}{240--255},
  \doiprefix\url{10.1107/s0907444996012255} (\bibinfo{year}{1997}).

\bibitem{buster}
\bibinfo{author}{Bricogne, G.} \emph{et~al.}
\newblock \bibinfo{title}{Buster, version 2.10.4}.
\newblock \bibinfo{howpublished}{Cambridge, United Kingdom: Global Phasing
  Ltd.} (\bibinfo{year}{2017}).

\bibitem{Zhou2020}
\bibinfo{author}{Zhou, T.} \emph{et~al.}
\newblock \bibinfo{journal}{\bibinfo{title}{Cryo-{EM} structures of
  {SARS}-{CoV}-2 spike without and with {ACE}2 reveal a {pH}-dependent switch
  to mediate endosomal positioning of receptor-binding domains}}.
\newblock {\emph{\JournalTitle{Cell Host {\&} Microbe}}}
  \textbf{\bibinfo{volume}{28}}, \bibinfo{pages}{867--879.e5},
  \doiprefix\url{10.1016/j.chom.2020.11.004} (\bibinfo{year}{2020}).

\bibitem{jtsa}
\bibinfo{author}{Bond, P.}
\newblock \bibinfo{title}{{JavaScript Thermal Shift Analysis Software}}.
\newblock \bibinfo{howpublished}{\url{https://paulsbond.co.uk/jtsa}}.
\newblock \bibinfo{note}{[Online; accessed 07-March-2022]}.

\bibitem{segler2017generating}
\bibinfo{author}{Segler, M.~H.}, \bibinfo{author}{Kogej, T.},
  \bibinfo{author}{Tyrchan, C.} \& \bibinfo{author}{Waller, M.~P.}
\newblock \bibinfo{journal}{\bibinfo{title}{Generating focused molecule
  libraries for drug discovery with recurrent neural networks}}.
\newblock {\emph{\JournalTitle{ACS Central Science}}}
  \textbf{\bibinfo{volume}{4}}, \bibinfo{pages}{120--131}
  (\bibinfo{year}{2017}).

\bibitem{kadurin2017cornucopia}
\bibinfo{author}{Kadurin, A.} \emph{et~al.}
\newblock \bibinfo{journal}{\bibinfo{title}{The cornucopia of meaningful leads:
  Applying deep adversarial autoencoders for new molecule development in
  oncology}}.
\newblock {\emph{\JournalTitle{Oncotarget}}} \textbf{\bibinfo{volume}{8}},
  \bibinfo{pages}{10883} (\bibinfo{year}{2017}).

\bibitem{jin2018junction}
\bibinfo{author}{Jin, W.}, \bibinfo{author}{Barzilay, R.} \&
  \bibinfo{author}{Jaakkola, T.}
\newblock \bibinfo{journal}{\bibinfo{title}{Junction tree variational
  autoencoder for molecular graph generation}}.
\newblock {\emph{\JournalTitle{arXiv preprint arXiv:1802.04364}}}
  (\bibinfo{year}{2018}).

\bibitem{prykhodko2019novo}
\bibinfo{author}{Prykhodko, O.} \emph{et~al.}
\newblock \bibinfo{journal}{\bibinfo{title}{A de novo molecular generation
  method using latent vector based generative adversarial network}}.
\newblock {\emph{\JournalTitle{Journal of Cheminformatics}}}
  \textbf{\bibinfo{volume}{11}}, \bibinfo{pages}{1--13} (\bibinfo{year}{2019}).

\bibitem{lipinski2004lead}
\bibinfo{author}{Lipinski, C.~A.}
\newblock \bibinfo{journal}{\bibinfo{title}{Lead-and drug-like compounds: the
  rule-of-five revolution}}.
\newblock {\emph{\JournalTitle{Drug discovery today: Technologies}}}
  \textbf{\bibinfo{volume}{1}}, \bibinfo{pages}{337--341}
  (\bibinfo{year}{2004}).

\bibitem{ghose1999knowledge}
\bibinfo{author}{Ghose, A.~K.}, \bibinfo{author}{Viswanadhan, V.~N.} \&
  \bibinfo{author}{Wendoloski, J.~J.}
\newblock \bibinfo{journal}{\bibinfo{title}{A knowledge-based approach in
  designing combinatorial or medicinal chemistry libraries for drug discovery.
  1. a qualitative and quantitative characterization of known drug databases}}.
\newblock {\emph{\JournalTitle{Journal of combinatorial chemistry}}}
  \textbf{\bibinfo{volume}{1}}, \bibinfo{pages}{55--68} (\bibinfo{year}{1999}).

\bibitem{veber2002molecular}
\bibinfo{author}{Veber, D.~F.} \emph{et~al.}
\newblock \bibinfo{journal}{\bibinfo{title}{Molecular properties that influence
  the oral bioavailability of drug candidates}}.
\newblock {\emph{\JournalTitle{Journal of medicinal chemistry}}}
  \textbf{\bibinfo{volume}{45}}, \bibinfo{pages}{2615--2623}
  (\bibinfo{year}{2002}).

\bibitem{egan2000prediction}
\bibinfo{author}{Egan, W.~J.}, \bibinfo{author}{Merz, K.~M.} \&
  \bibinfo{author}{Baldwin, J.~J.}
\newblock \bibinfo{journal}{\bibinfo{title}{Prediction of drug absorption using
  multivariate statistics}}.
\newblock {\emph{\JournalTitle{Journal of medicinal chemistry}}}
  \textbf{\bibinfo{volume}{43}}, \bibinfo{pages}{3867--3877}
  (\bibinfo{year}{2000}).

\bibitem{muegge2001simple}
\bibinfo{author}{Muegge, I.}, \bibinfo{author}{Heald, S.~L.} \&
  \bibinfo{author}{Brittelli, D.}
\newblock \bibinfo{journal}{\bibinfo{title}{Simple selection criteria for
  drug-like chemical matter}}.
\newblock {\emph{\JournalTitle{Journal of medicinal chemistry}}}
  \textbf{\bibinfo{volume}{44}}, \bibinfo{pages}{1841--1846}
  (\bibinfo{year}{2001}).

\bibitem{martin2005bioavailability}
\bibinfo{author}{Martin, Y.~C.}
\newblock \bibinfo{journal}{\bibinfo{title}{A bioavailability score}}.
\newblock {\emph{\JournalTitle{Journal of medicinal chemistry}}}
  \textbf{\bibinfo{volume}{48}}, \bibinfo{pages}{3164--3170}
  (\bibinfo{year}{2005}).

\bibitem{baell2010new}
\bibinfo{author}{Baell, J.~B.} \& \bibinfo{author}{Holloway, G.~A.}
\newblock \bibinfo{journal}{\bibinfo{title}{New substructure filters for
  removal of pan assay interference compounds (pains) from screening libraries
  and for their exclusion in bioassays}}.
\newblock {\emph{\JournalTitle{Journal of medicinal chemistry}}}
  \textbf{\bibinfo{volume}{53}}, \bibinfo{pages}{2719--2740}
  (\bibinfo{year}{2010}).

\bibitem{brenk2008lessons}
\bibinfo{author}{Brenk, R.} \emph{et~al.}
\newblock \bibinfo{journal}{\bibinfo{title}{Lessons learnt from assembling
  screening libraries for drug discovery for neglected diseases}}.
\newblock {\emph{\JournalTitle{ChemMedChem: Chemistry Enabling Drug
  Discovery}}} \textbf{\bibinfo{volume}{3}}, \bibinfo{pages}{435--444}
  (\bibinfo{year}{2008}).

\bibitem{teague1999design}
\bibinfo{author}{Teague, S.~J.}, \bibinfo{author}{Davis, A.~M.},
  \bibinfo{author}{Leeson, P.~D.} \& \bibinfo{author}{Oprea, T.}
\newblock \bibinfo{journal}{\bibinfo{title}{The design of leadlike
  combinatorial libraries}}.
\newblock {\emph{\JournalTitle{Angewandte Chemie International Edition}}}
  \textbf{\bibinfo{volume}{38}}, \bibinfo{pages}{3743--3748}
  (\bibinfo{year}{1999}).

\end{thebibliography}

\section*{Acknowledgements}
This work was supported by the IBM Science for Social Good program. D.I.S. is supported by the UKRI MRC (MR/N00065X/1) and is a Jenner Investigator. This is a contribution from the UK Instruct-ERIC Centre.   Authors thank the IBM COVID-19 Molecular Explorer team for help with open sourcing machine-designed inhibitor candidates,  the IBM RXN team, especially Matteo Manica, for
help with retrosynthesis predictions, Diamond for beamtime through the COVID-19 dedicated call, the Diamond MX group and the IBM-Oxford partnership for valuable support.  

\section*{Author contributions statement}

P.D, D.I.S, and M.A.W. conceived the study. V.C., S.C.H., and P.D. devised and performed the target-guided molecule generation and \textit{in silico} screening. V.C. trained and tested the CogMol generative model. S.C.H. performed and analyzed the retrosynthesis predictions and the docking simulations, with the help from P.D. V.C. assessed the novelty of the \textit{de novo} designed compounds. T.M. synthesized the \textit{de novo} designed compounds. S.C.H., P.D. and T.M. compared predicted and actual retrosynthesis routes. P.D. performed ADME prediction. M.A.W. and C.J.S. conceptualized M\textsuperscript{pro} experiments. P.L. and C.S-d. performed M\textsuperscript{pro} protein production and purification and analysis of associated data. C.D.O. and D.F. performed crystallization and crystallography. C.D.O., D.F., P.L., and M.A.W. analysed crystallographic data.
T.R.M. and A.T. performed functional analysis of M\textsuperscript{pro} by Rapid MS.
D.I.S. and G.R.S. designed the spike-related wet lab experiments.
T.S.W. performed the thermofluor analysis.
L.C. and H.M.E.D. analysed the lipid bound spike structure.
W.D. performed the neutralisation studies.
All authors wrote and reviewed the manuscript.

\clearpage
\appendix
\appendixpage

\section{Supplementary Tables}
\subsection{Target sequences used in generation pipeline}

\begin{table}[h]
    \centering
    \begin{tabular}{@{}lp{0.8\textwidth}@{}}
        \toprule
        Target & Sequence \\
        \midrule
        M\textsuperscript{pro} & \texttt{SGFRKMAFPSGKVEGCMVQVTCGTTTLNGLWLDDVVYCPRHVICTSEDMLNPNYEDLLIRKSNHNFL\newline
        VQAGNVQLRVIGHSMQNCVLKLKVDTANPKTPKYKFVRIQPGQTFSVLACYNGSPSGVYQCAMRPNF\newline
        TIKGSFLNGSCGSVGFNIDYDCVSFCYMHHMELPTGVHAGTDLEGNFYGPFVDRQTAQAAGTDTTIT\newline
        VNVLAWLYAAVINGDRWFLNRFTTTLNDFNLVAMKYNYEPLTQDHVDILGPLSAQTGIAVLDMCASL\newline
        KELLQNGMNGRTILGSALLEDEFTPFDVVRQCSGVTFQ} \\
        \addlinespace[1em]
        Chimeric RBD           & \texttt{RVVPSGDVVRFPNITNLCPFGEVFNATKFPSVYAWERKKISNCVADYSVLYNSTFFSTFKCYGVSAT\newline
        KLNDLCFSNVYADSFVVKGDDVRQIAPGQTGVIADYNYKLPDDFMGCVLAWNTRNIDATSTGNYNYK\newline
        YRLFRKSNLKPFERDISTEIYQAGSTPCNGVEGFNCYFPLQSYGFQPTNGVGYQPYRVVVLSFELLN\newline
        APATVCGPKLSTDLIK} \\
        \bottomrule
    \end{tabular}
    \caption{\textbf{SARS-CoV-2 target protein sequences.} The amino acid sequences of the protein targets used in the generation pipeline}
    \label{tab:fasta}
\end{table}
\subsection{Baseline Comparison of the generative model}
\begin{table}[!ht]
\centering
\scalebox{0.85}{
\begin{tabular}{
    r|
    S[table-format=1.4]
    S[table-format=1.1]
    S[table-format=1.4]
    S[table-format=1.4]
    S[table-format=1.4]
    S[table-format=1.4]
}
\toprule
Model & {Valid} & {Unique@1k} & {Unique@10k} & {IntDiv1} & {IntDiv2} & {Filters}  \\
\midrule

CogMol \cite{chenthamarakshan2020cogmol} & 0.95 & 1.0 & 0.999 & 0.8578 & 0.8521 & 0.9888  \\
CharRNN \cite{segler2017generating} & 0.809 & 1.0 & 1.0 & 0.855 & 0.849 &  0.975 \\
AAE \cite{kadurin2017cornucopia} & 0.997 & 1.0 & 0.995  & 0.857 & 0.85  &  0.997 \\
VAE \cite{gomez2018automatic} & 0.969 & 1.0 & 0.999 & 0.856 & 0.851 &  0.996  \\
JT-VAE \cite{jin2018junction} & 1.0 & 1.0 & 0.999 & 0.851 & 0.845 & 0.978  \\
LatentGan \cite{prykhodko2019novo} & 0.8966 & 1.0 & 0.9968 & 0.8565 & 0.8505 & 0.9735 \\
Training & 1.0 & 1.0 & 1.0 & 0.857 & 0.851 & 1.0  \\
\bottomrule
\end{tabular}}
\caption{\textbf{Comparison of generated molecules in terms of fraction of valid, unique (out of 1,000 and 10,000 generated), internal diversity, and passing filters (medicinal chemistry filters, PAINS, ring sizes, charges, atom type).} All generative models were trained and tested on MOSES benchmark \cite{polykovskiy2018molecular}. Performances of baseline models are from Polykovskiy, et al.~\cite{polykovskiy2018molecular}.}
\label{tab:vae_comp}
\end{table}

\clearpage
\subsection{Predicted properties of validated \textit{de novo} hits}
\begin{table}[h]
\centering
\begin{tabular}{lccccccccc}
\toprule
ID          & AFF & SEL      & TOX & QED        & SA         & logP    & MW   & docking & dist.\ to pocket \\
   & (pIC$_{50}$) & (pIC$_{50}$) &  &  &  &  &  (Da) & (kcal/mol) & (\AA) \\
\midrule
GXA56 & 8.050  & 0.646 & 0   & 0.695 & 2.562 & 3.337 & 404.305 & $-9.2$ & 3.88 \\
GXA70 & 8.162 & 0.744 & 0   & 0.771 & 2.774 & 3.301 & 430.503 & $-9.1$ & 6.77 \\
GXA104 & 8.16 & 1.112 & 0   & 0.730  & 2.417 & 3.484 & 376.460 & $-8.9$ & 6.65 \\
GXA112 & 8.280 & 0.721 & 0   & 0.610  & 2.934 & 0.943 & 488.618 & $-8.8$ & 4.97 \\
\bottomrule
\end{tabular}
\caption{\textbf{Predicted and estimated properties of \textit{de novo} compounds targeting M\textsuperscript{pro}.} See the Ranking and prioritization section for explanations of the column headers.}
\label{tab:mpro-denovo}
\end{table}

\begin{table}[h]
\centering
\begin{tabular}{lccccccccc}
\toprule
ID          & AFF & SEL      & TOX & QED        & SA         & logP    & MW   & docking & dist.\ to pocket \\
   & (pIC$_{50}$) & (pIC$_{50}$) &  &  &  &  &  (Da) & (kcal/mol) & (\AA) \\
\midrule
GEN626 & 7.077  & 0.754 & 0   & 0.829 & 2.392 & 1.773  & 317.311 & $-7.6$   & 1.93    \\
GEN725 & 8.140  & 0.752 & 0   & 0.704 & 1.951 & 3.197  & 403.481 & $-8.8$   & 2.06   \\
GEN727 & 7.920  & 0.826 & 0   & 0.857  & 2.322  & 3.382   & 293.414 & $-8.1$   & 2.69    \\
GEN777 & 7.513  & 0.834 & 0   & 0.819 & 2.603  & 2.333  & 248.717 & $-7.9$   & 3.36   \\
\bottomrule
\end{tabular}
\caption{\textbf{Predicted and estimated properties of \textit{de novo} compounds targeting spike RBD.} See the Ranking and prioritization section for explanations of the column headers.}
\label{tab:rbd-denovo}
\end{table}

\newpage
\subsection{Comparison of predicted and actual synthesis routes}
\begin{table}[h!]
\centering
\begin{tabular}{@{}lm{0.2\textwidth}lllrrcp{0.2\textwidth}@{}}
\toprule
            & & path & conf. & steps  & products  & reactants & success & comments \\ \midrule
GEN626 & \centering\includegraphics[scale=0.21]{img/Z4620186698_bw.png} & 5    & 1.0   & 150\% & 50.0\% & 62.5\% & \xmark & \\
GEN725 & \centering\includegraphics[scale=0.21]{img/Z4620186699_bw.png} & 5    & 1.0   & 66.7\%  & 50.0\%  & 50.0\% & \cmark & minor changes; moderate yield \\
GEN727 & \centering\includegraphics[scale=0.21]{img/Z4620186700_bw.png} & 5    & 1.0   & 100\%  & 100\%  & 70.0\% & \cmark &  followed top prediction\\
GEN777 & \centering\includegraphics[scale=0.21]{img/Z4620186701_bw.png} & 3    & 1.0   & 200\% & 33.3\% & 75.0\% & \xmark & \\
\addlinespace[2em] 
GXA56 & \centering\includegraphics[scale=0.21]{img/Z4519299683_bw.png} & 0    & 1.0   & 100\%  & 100\%  & 52.9\% & \cmark & followed top prediction\\
GXA70 & \centering\includegraphics[scale=0.21]{img/Z4519300136_bw.png} & 2    & 1.0   & 100\%  & 100\%  & 38.5\% & \cmark & minor changes to top prediction\\
GXA104 & \centering\includegraphics[scale=0.21]{img/Z4519300813_bw.png} & 0    & 0.88  & 66.7\%  & 0\%    & 35.7\% & \textbf{--} & reactant unavailable \\
GXA112 & \centering\includegraphics[scale=0.21]{img/Z4519300867_bw.png} & 4    & 1.0   & 140\% & 75.0\% & 62.5\% & \cmark & low yield \\
\bottomrule
\end{tabular}
\caption{\textbf{Consolidated results comparing predicted and actual synthesis paths.} The top 6 predicted retrosynthesis paths (by confidence) are considered and the path with the best agreement is shown. ``Steps'' is simply the number of reaction steps actual / predicted number of reaction steps. ``Products'' shows the intermediate (not including the final molecule) reaction products overlap in terms of recall (with respect to the predicted path) while ``reactants'' similarly shows the overlap of reactants from all steps in terms of recall. The ``success'' column shows whether the given predicted path was successfully synthesized as is or with minor changes or failed (but still synthesized via an alternative method devised by Enamine).}
\label{tab:synth-comp}
\end{table}

\clearpage

\subsection{Crystallography table}
\begin{table}[h]
\centering
\begin{tabular}{@{}llll@{}}
\toprule
                                             & Z68337194               & Z1633315555           & Z1365651030                  \\
                                             & 5SML               & 5SMM           & 5SMN                  \\
Data Collection                              &                         &                       &                              \\ \midrule
Wavelength (\AA)                             & 0.9126                  & 0.9126                & 0.9126                       \\
Resolution range (\AA)                       & 47.57-1.53 (1.585-1.53) & 47.8-1.58 (1.64-1.58) & 47.36-1.36 (1.41-1.36)       \\
Space group                                  & C2                      & C2                    & C2                           \\
Unit cell                                    &                         &                       &                              \\
\hspace*{1em}$a,b,c$ (\AA)                   & 112.12, 52.83, 44.46    & 113.12, 53.04, 44.38  & 111.93, 52.57, 44.59         \\
\hspace*{1em}$\alpha,\beta,\gamma$ (\degree) & 90.00, 102.99, 90.00    & 90.00, 102.90, 90.00  & 90.00, 102.94, 90.00         \\
Total   reflections                          & 119085 (10469)          & 112151 (10502)        & 158637 (10806)               \\
Unique   reflections                         & 38187 (3690)            & 35130 (3385)          & 53606 (4976)                 \\
Multiplicity                                 & 3.1 (2.7)               & 3.2 (3.0)             & 3.0 (2.1)                    \\
Completeness   (\%)                          & 98.88 (95.80)           & 99.16 (96.18)         & 98.60 (92.03)                \\
Mean $I/\sigma I$                            & 11.16 (0.81)            & 12.16 (0.76)          & 14.56 (0.77)                 \\
$R_\mathrm{merge}$                           & 0.088 (0.937)           & 0.097 (1.38)          & 0.068 (1.02)                 \\
$R_\mathrm{meas}$                            & 0.106 (1.164)           & 0.117 (1.68)          & 0.082 (1.32)                 \\
CC1/2                                        & 0.995 (0.342)           & 0.997 (0.336)         & 0.998 (0.347)                \\ \addlinespace[1em]
Refinement                                   &                         &                       &                              \\ \midrule
Reflections used in refinement               & 37923 (3674)            & 34946 (3378)          & 53514 (4976)                 \\
$R_\mathrm{work}$                            & 0.1962 (0.3414)         & 0.1966 (0.3680)       & 0.1934 (0.3734)              \\
$R_\mathrm{free}$                            & 0.2250 (0.3324)         & 0.2322 (0.3891)       & 0.2181 (0.3577)              \\
Number of non-hydrogen atoms                 & 4084                    & 3818                  & 3304                         \\
\hspace*{1em}Protein                         & 3598                    & 3412                  & 2935                         \\
\hspace*{1em}Ligands                         & 54                      & 76                    & 54                           \\
\hspace*{1em}Solvent                         & 432                    & 330                   & 315                          \\
RMSD bond   lengths (\AA)                    & 0.013                   & 0.013                 & 0.014                        \\
RMSD bond   angles (\degree)                 & 1.73                    & 1.77                  & 1.81                         \\
Ramachandran   favored (\%)                  & 97.35                 & 97.68                 & 97.68                        \\
Ramachandran   allowed (\%)                  & 2.32                    & 1.99                  & 1.99                         \\
Ramachandran   outliers (\%)                 & 0.33                    & 0.33                  & 0.33                         \\
Rotamer   outliers (\%)                      & 1.00                    & 2.08                  & 0.31                         \\
Clashscore                                   & 5.4                    & 3.91                  & 3.74                         \\
Average $B$-factors (\AA$^2$)                &                         &                       &                              \\
\hspace*{1em}All                             & 23.19                   & 23.55                 & 18.92                        \\
\hspace*{1em}Protein                         & 22.21                  & 22.55                 & 17.82                        \\
\hspace*{1em}Solvent                         & 31.33                   & 32.25                 & 27.53                        \\ \bottomrule
\end{tabular}
\caption{\textbf{Crystallographic data collection and refinement statistics.} Values in parentheses refer to the highest resolution shell.}
\label{tab:crystal_data}
\end{table}

\newpage
\subsection{SwissADME Analysis}
\begin{table}[!ht]
    \centering
    \begin{tabular}{lccccccccc}
    \toprule
        ID & Lipinski  & Ghose  & Veber & Egan  & Muegge  & Bioavailability  & PAINS & BRENK   & Leadlikeness  \\ \midrule
        Z68337194 & 0 & 0 & 0 & 0 & 0 & Medium & 0 & 0 & 0 \\
         GXA70 & 0 & 0 & 0 & 0 & 0 & Medium & 0 &  0 & 2 \\
         GXA56 & 0 & 0 & 0 & 0 & 0 & Medium & 0 & 0 &  1 \\
        GEN725 & 0 & 0 & 0 & 0 & 0 & Medium & 0 & 0 &  1 \\
         GEN727 & 0 & 0 & 0 & 0 & 0 & Medium & 0 & 1 & 0 \\
      \bottomrule
    \end{tabular}
    \caption{\textbf{ADME properties of validated hits.} Drug-likeness (as estimated using number of violations according to Lipinski's~\cite{lipinski2004lead}, Ghose's~\cite{ghose1999knowledge}, Veber's~\cite{veber2002molecular}, Egan's~\cite{egan2000prediction}, and Muegge's~\cite{muegge2001simple} criteria), bioavailability~\cite{martin2005bioavailability} (Low below 0.25, Medium between 0.25 and 0.75, and High above 0.75), number of medicinal chemistry (PAINS~\cite{baell2010new} and BRENK~\cite{brenk2008lessons}) alerts and Leadlikeness~\cite{teague1999design} (number of violations:  250 g/mol $\leq$ molecular weight $\leq$ 400 g/mol, xlogP $\leq$ 3.5, number of  rotatable bonds $\leq$ 7)  are  estimated using SwissADME software~\cite{daina2017swissadme}. }
\label{tab:adme}
\end{table}

\newpage
\subsection{Spectroscopic data}
\begin{table}[h]
\centering
\begin{tabular}{@{}lp{0.8\textwidth}@{}}
\toprule
GEN727 & $\mathrm{^1H}$ NMR (400 MHz, dmso) $\delta$ 8.37 (d, $J$ = 5.3 Hz, 1H), 8.21 (d, $J$ = 8.4 Hz, 1H), 7.76 (d, $J$ = 8.4 Hz, 1H), 7.59 (t, $J$ = 7.6, 7.6 Hz, 1H), 7.40 (t, $J$ = 7.6, 7.6 Hz, 1H), 7.08 (t, $J$ = 5.4, 5.4 Hz, 1H), 6.42 (d, $J$ = 5.3 Hz, 1H), 3.29 (q, $J$ = 6.7, 6.7, 6.4 Hz, 2H), 3.22 (d, $J$ = 2.5 Hz, 2H), 3.10 (t, $J$ = 2.5, 2.5 Hz, 1H), 2.76 (dt, $J$ = 11.8, 3.4, 3.4 Hz, 2H), 2.08 (td, $J$ = 11.5, 11.4, 2.6 Hz, 2H), 1.72 (m, 2H), 1.60 (q, $J$ = 7.1, 7.1, 7.1 Hz, 2H), 1.37 (m, 1H), 1.20 (qd, $J$ = 12.0, 11.8, 11.8, 3.8 Hz, 2H).
\newline
HPLC-MS m/z [M+H]+ = 294.2 , purity 100\% \\\addlinespace[0.5em]
GEN777 & $\mathrm{^1H}$ NMR (400 MHz, dmso) $\delta$ 7.41 (m, 1H), 7.37 (m, 2H), 7.19 (d, $J$ = 7.5 Hz, 1H), 3.64 (s, 3H), 2.94 (m, 2H), 2.75 (m, 2H), 1.98 (m, 2H).
\newline
HPLC-MS m/z [M+H]+ = 249.2 , purity 100\% \\\addlinespace[0.5em]
GEN626 & $\mathrm{^1H}$ NMR (400 MHz, dmso) $\delta$ 7.63 (d, $J$ = 8.6 Hz, 1H), 7.28 (br s, 1H), 7.08 (br s, 1H), 6.28 (d, $J$ = 2.0 Hz, 1H), 6.18 (dd, $J$ = 8.5, 2.0 Hz, 1H), 5.65 (m, 2H), 4.42 (m, 1H), 3.20 (q, $J$ = 10.3, 10.2, 10.2 Hz, 2H), 2.83 (m, 2H), 2.60 (m, 2H), 1.98 (m, 2H), 1.74 (m, 2H).
\newline
HPLC-MS m/z [M+H]+ = 318.2 , purity 100\% \\\addlinespace[0.5em]
GEN725 & $\mathrm{^1H}$ NMR (400 MHz, dmso) $\delta$ 7.97 (dd, $J$ = 7.9, 1.8 Hz, 1H), 7.90 (m, 4H), 7.82 (d, $J$ = 8.8 Hz, 2H), 7.73 (td, $J$ = 7.9, 7.8, 1.8 Hz, 1H), 7.41 (m, 3H), 7.26 (d, $J$ = 8.7 Hz, 2H), 7.14 (d, $J$ = 8.3 Hz, 1H), 3.38 (s, 3H).
\newline
HPLC-MS m/z [M+H]+ = 404.2 , purity 98.72\% \\\addlinespace[0.5em]
\midrule\addlinespace[0.5em]
GXA104 & $\mathrm{^1H}$ NMR (400 MHz, dmso) $\delta$ 13.34 (s, 1H), 8.08 (d, $J$ = 8.3 Hz, 1H), 8.01 (d, $J$ = 8.3 Hz, 1H), 7.87 (s, 1H), 7.61 (d, $J$ = 8.3 Hz, 2H), 7.42 (m, 1H), 7.36 (d, $J$ = 7.7 Hz, 1H), 7.23 (m, 1H), 7.11 (br s, 1H), 6.59 (br s, 1H), 3.22 (s, 3H), 2.23 (m, 1H), 2.01 (m, 1H), 1.69 (m, 4H), 1.42 (m, 2H), 1.00 (m, 2H).
\newline
HPLC-MS m/z [M+H]+ =  377.2 , purity 100\% \\\addlinespace[0.5em]
GXA112 & $\mathrm{^1H}$ NMR (500 MHz, dmso) $\delta$ 8.18 (s, 1H), 7.18 (d, $J$ = 7.1 Hz, 1H), 7.13 (t, $J$ = 7.7, 7.7 Hz, 1H), 6.86 (m, 1H), 6.64 (m, 1H), 6.46 (m, 2H), 4.44 (m, 1H), 4.09 (m, 2H), 3.69 (m, 4H), 3.62 (m, 4H), 3.50 (m, 1H), 3.07 (m, 2H), 2.98 (m, 2H), 2.94 (m, 1H), 2.07 (m, 1H), 1.90 (m, 2H), 1.67 (m, 1H), 1.59 (m, 2H), 1.36 (m, 2H).
\newline
HPLC-MS m/z [M+H]+ = 489.2 , purity 100\% \\\addlinespace[0.5em]
GXA70 & $\mathrm{^1H}$ NMR (400 MHz, dmso) $\delta$ 8.96 (s, 1H), 7.53 (s, 1H), 7.40 (dd, $J$ = 8.0, 2.1 Hz, 1H), 7.08 (d, $J$ = 8.1 Hz, 1H), 4.71 (d, $J$ = 4.2 Hz, 1H), 4.23 (m, 2H), 3.84 (m, 4H), 3.71 (m, 1H), 3.22 (m, 2H), 2.79 (m, 4H), 1.97 (m, 6H), 1.75 (m, 2H), 1.32 (m, 2H).
\newline
HPLC-MS m/z [M+H]+ = 431.2 , purity 100\% \\\addlinespace[0.5em]
GXA56 & $\mathrm{^1H}$ NMR (400 MHz, dmso) $\delta$ 8.96 (s, 1H), 7.53 (s, 1H), 7.40 (dd, $J$ = 8.0, 2.1 Hz, 1H), 7.08 (d, $J$ = 8.1 Hz, 1H), 4.71 (d, $J$ = 4.2 Hz, 1H), 4.23 (m, 2H), 3.84 (m, 4H), 3.71 (m, 1H), 3.22 (m, 2H), 2.79 (m, 4H), 1.97 (m, 6H), 1.75 (m, 2H), 1.48 (m, 2H).
\newline
HPLC-MS m/z [M+H]+ = 404.2 , purity 100\% \\\addlinespace[0.5em]
\bottomrule
\end{tabular}
\caption{\textbf{Compound characterization.} Nuclear magnetic resonance (NMR) and high pressure liquid chromatography–mass spectrometry (HPLC-MS).}
\label{tab:nmr}
\end{table}

\newpage
\section{Supplementary Figures}
\subsection{Thermofluor assay results}

\begin{figure}[h]
    \centering
    \includegraphics[width=0.7\textwidth]{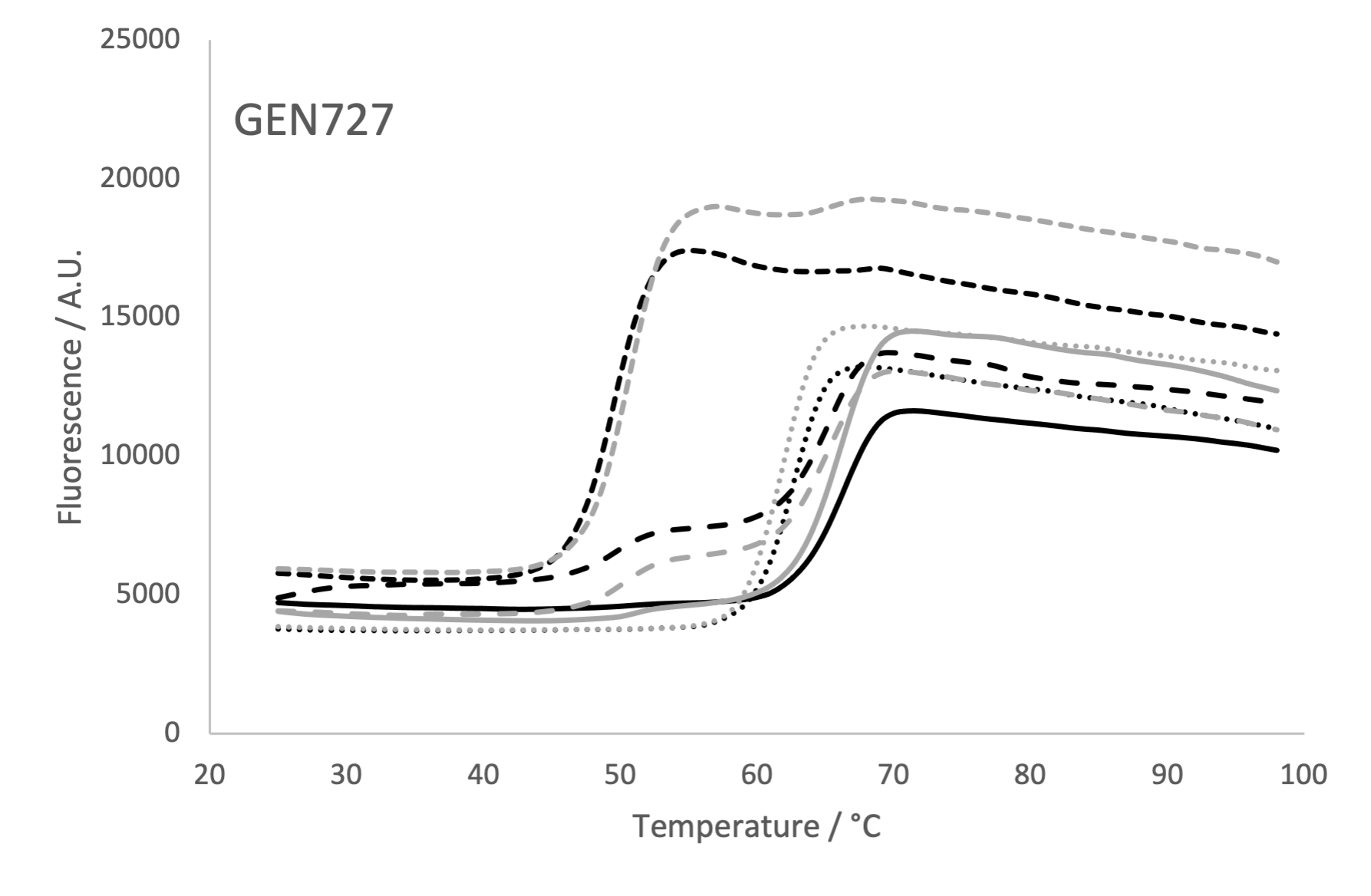}
    \caption{\textbf{Thermofluor assay results.} Thermofluor raw fluorescence data for experiments with AI-designed compound GEN727 (black) and a DMSO control (grey). Data were recorded using protein that was used immediately after dilution into neutral buffer (solid lines), incubated overnight in neutral buffer (long-dashed lines), or incubated overnight with the compound in neutral buffer (short-dashed lines).  For comparison, data from protein in pH 4.6 buffer is also shown (dotted lines).}
    \label{fig:thermafluor}
\end{figure}

\clearpage
\subsection{Additional structures}

\begin{figure}[h]
    \centering
    \includegraphics[height=16em]{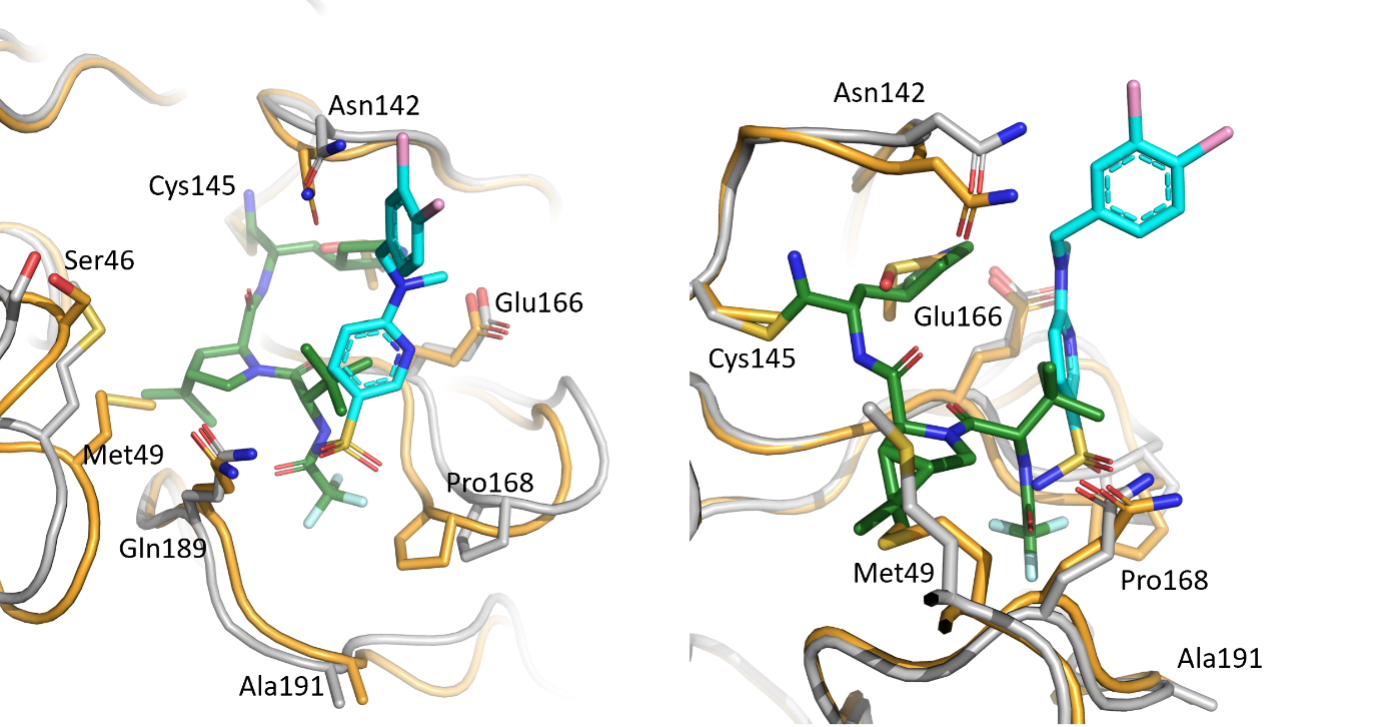}
    \caption{\textbf{Comparison of crystal structures of Z68337194 and nirmatrelvir.} SARS-CoV-2 M\textsuperscript{pro} in complex with Z68337194 (orange protein chain with ligand in cyan), aligned to SARS-CoV-2 M\textsuperscript{pro} in complex with nirmatrelvir (7TE0, grey protein with ligand in green). Images are related by a \SI{90}{\degree} rotation around the z-axis.}
    \label{fig:nirmatrelvir}
\end{figure}

\begin{figure}[h]
    \centering
    \begin{subfigure}[b]{0.4\textwidth}
        \centering
        \includegraphics[height=16em]{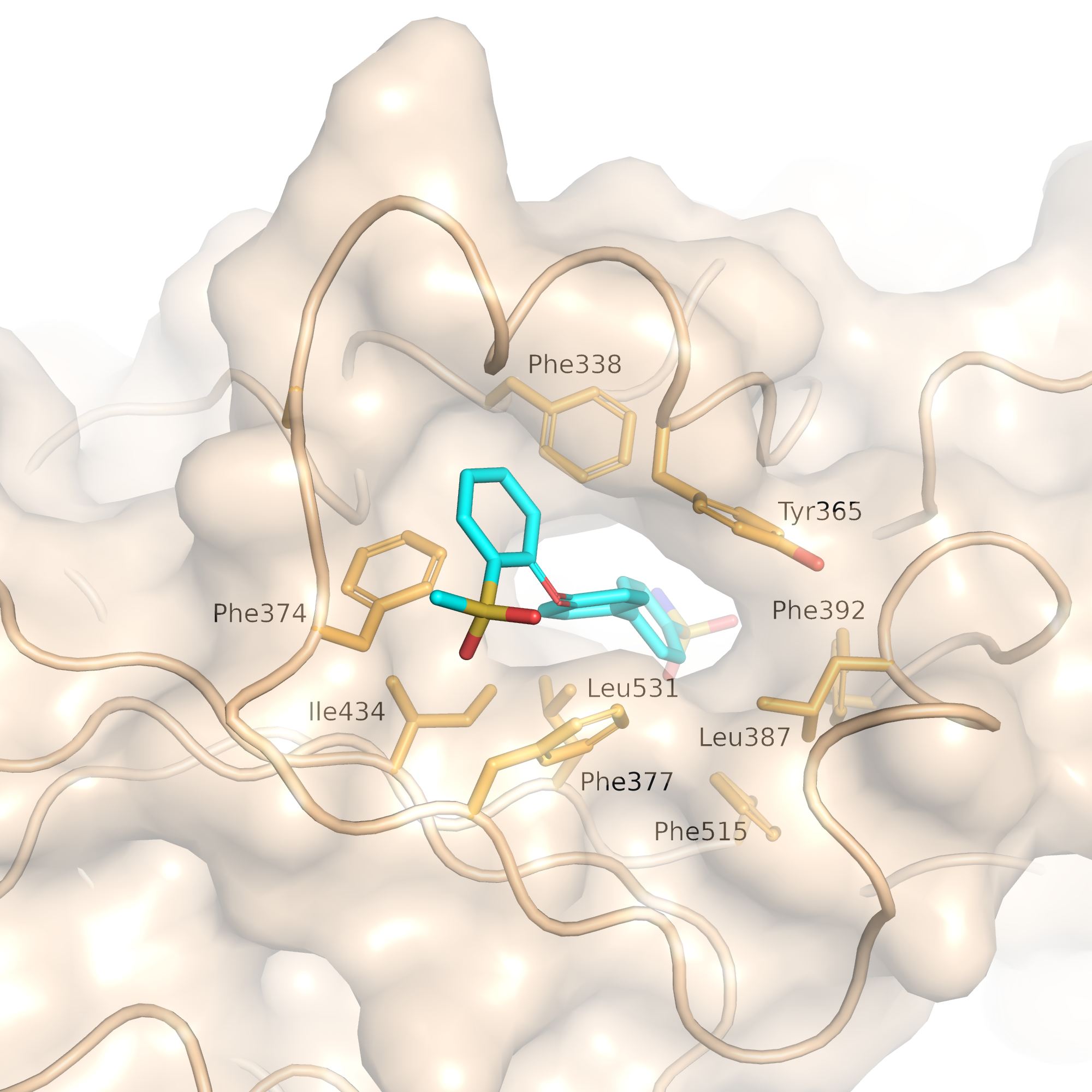}
        \caption{GEN725 --- Pocket view}
        \label{fig:docked_gen725_pocket}
    \end{subfigure}
    \begin{subfigure}[b]{0.4\textwidth}
        \centering
        \includegraphics[height=16em]{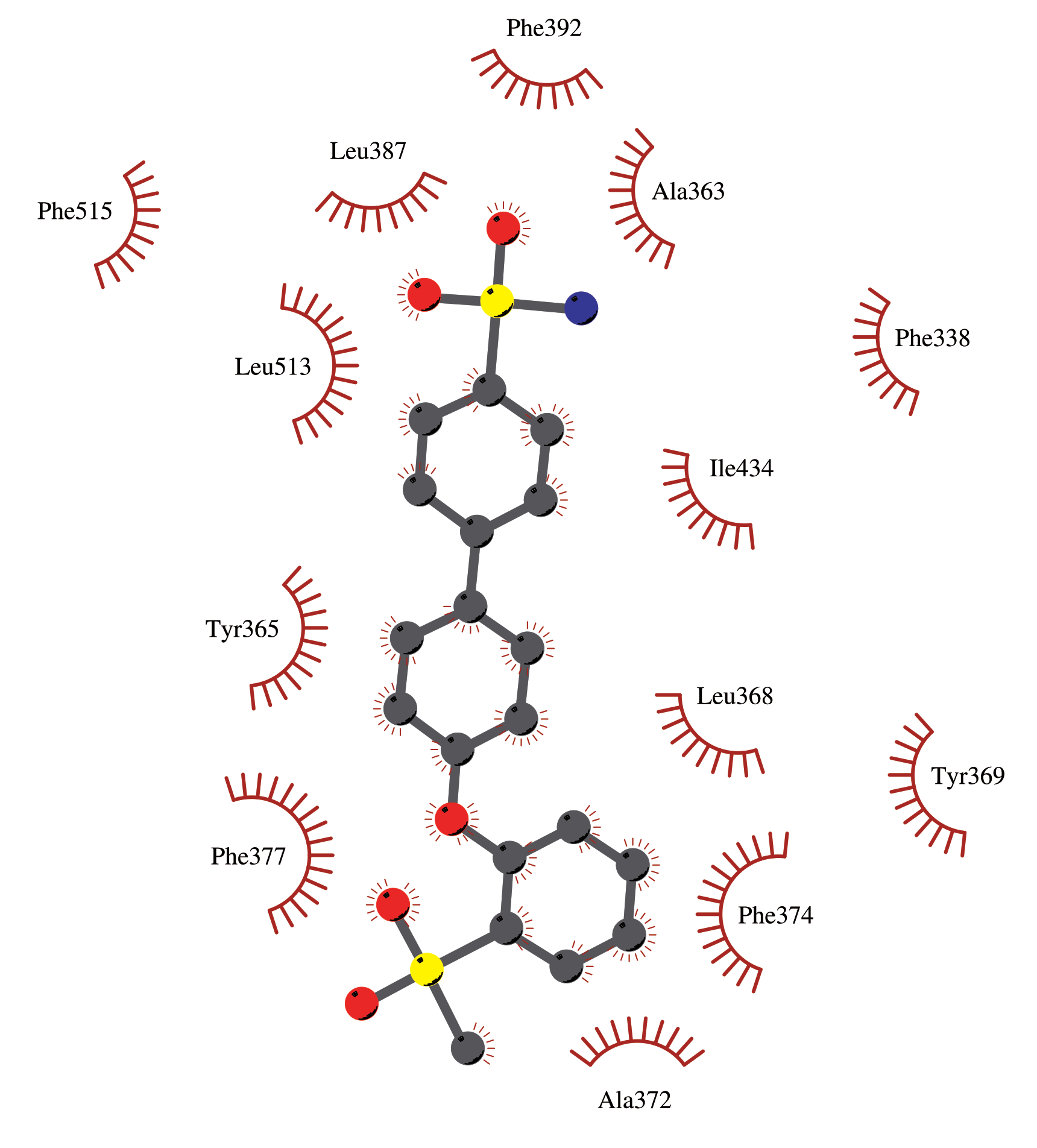}
        \caption{GEN725 --- Residue interaction map}
        \label{fig:docked_gen725_ligplot}
    \end{subfigure}
    \caption{\textbf{Docked structure of SARS-CoV-2 spike protein RBD in complex with GEN725.} (\textbf{a}) Surface representation depicting the overall ligand binding modes of GEN725 at the lipid binding site of the RBD. (\textbf{b}) Schematic representation of the ligand interactions with the spike RBD.}
    \label{fig:docked_spike_gen725}
\end{figure}

\clearpage
\subsection{PanDDA event maps}
\begin{figure}[h]
    \centering
    \begin{subfigure}[b]{0.3\textwidth}
        \centering
        \includegraphics[width=\textwidth,trim={0 0 0 15px},clip]{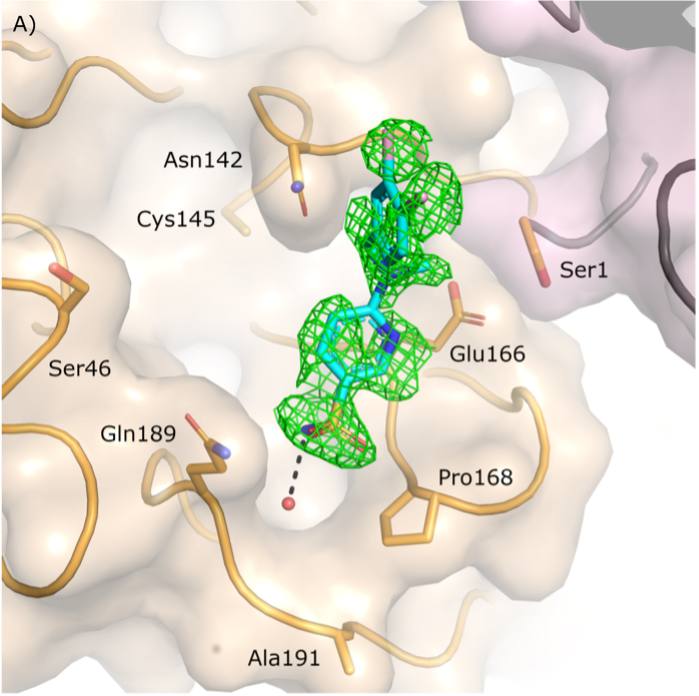}
        \caption{Z6833714}
    \end{subfigure}
    \begin{subfigure}[b]{0.3\textwidth}
        \centering
        \includegraphics[width=\textwidth,trim={0 0 0 15px},clip]{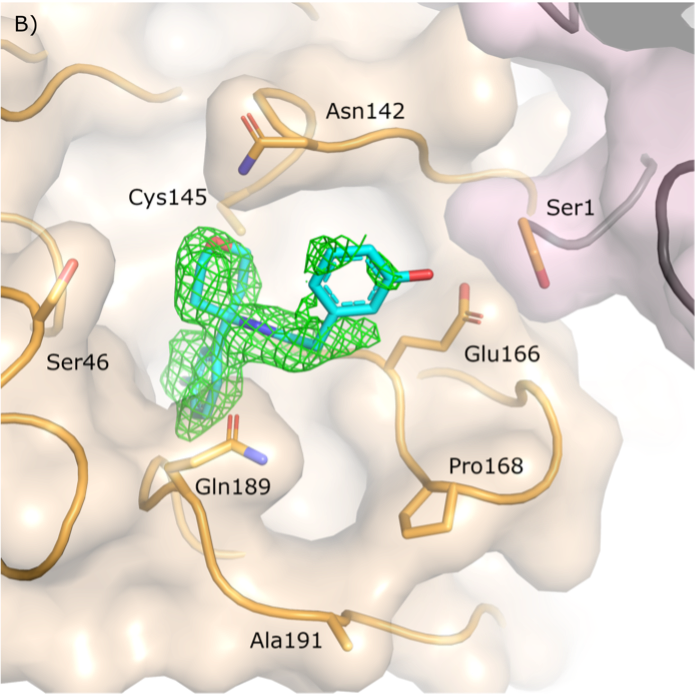}
        \caption{Z1633315555}
    \end{subfigure}
    \begin{subfigure}[b]{0.3\textwidth}
        \centering
        \includegraphics[width=\textwidth,trim={0 0 0 15px},clip]{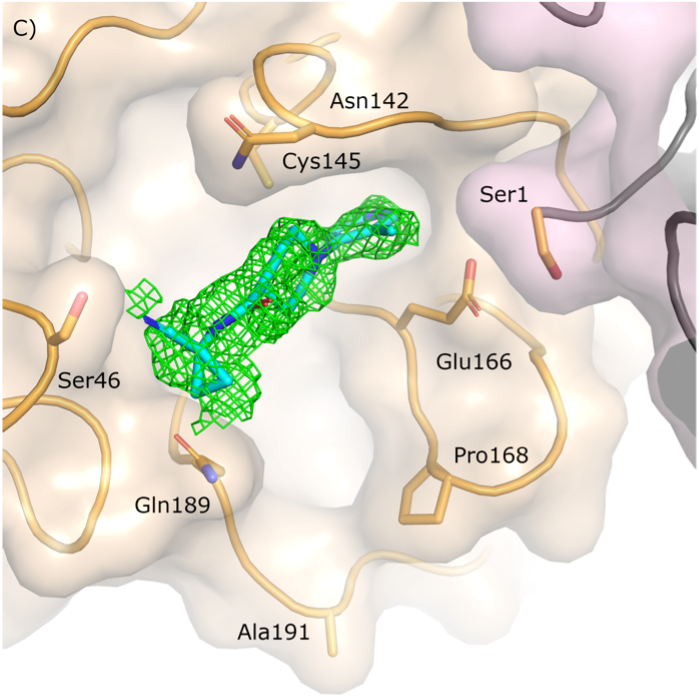}
        \caption{Z1365651030}
    \end{subfigure}
    \caption{\textbf{Pan Dataset Density Analysis (PanDDA) event maps.} PanDDA\cite{Pearce2017} event maps for crystal structures of the SARS-CoV-2 M\textsuperscript{pro} in complex with (\textbf{a}) Z6833714, (\textbf{b}) Z1633315555, and (\textbf{c}) Z1365651030. All event maps are contoured at the 1$\sigma$ level. The PanDDA algorithm facilitates identification of weakly bound ligands as described previously\cite{Douangamath2020}.}
    \label{fig:pandda}
\end{figure}

\clearpage
\subsection{SwissADME evaluations}
\begin{figure}[h]
    \centering
    \includegraphics[width=0.9\textwidth]{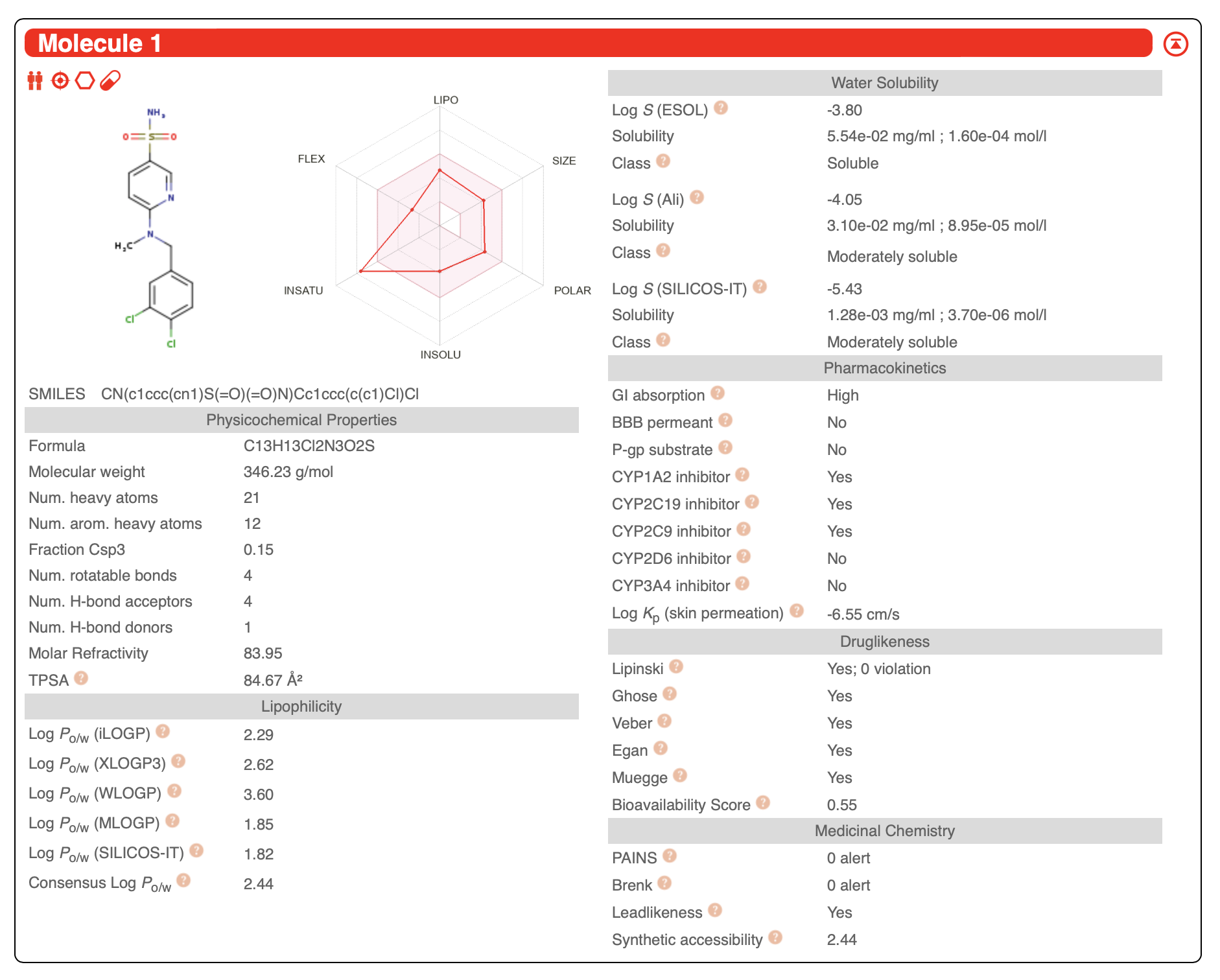}
    \caption{\textbf{SwissADME evaluation of Z68337194.}}
    \label{fig:adme1}
\end{figure}

\newpage

\begin{figure}[h]
    \centering
    \includegraphics[width=0.9\textwidth]{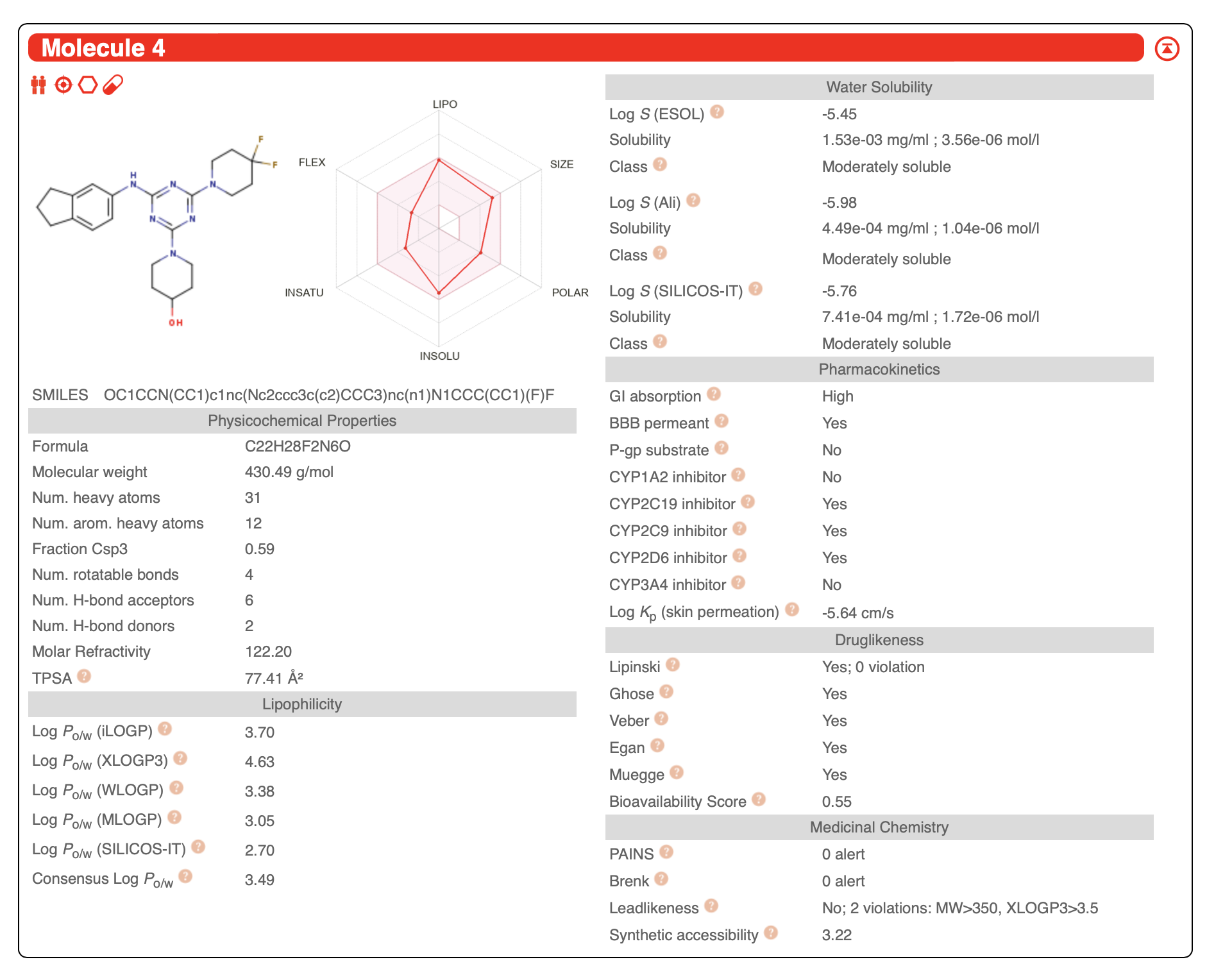}
    \caption{\textbf{SwissADME evaluation of GXA70.}}
    \label{fig:adme4}
\end{figure}

\newpage
\begin{figure}[h]
    \centering
    \includegraphics[width=0.9\textwidth]{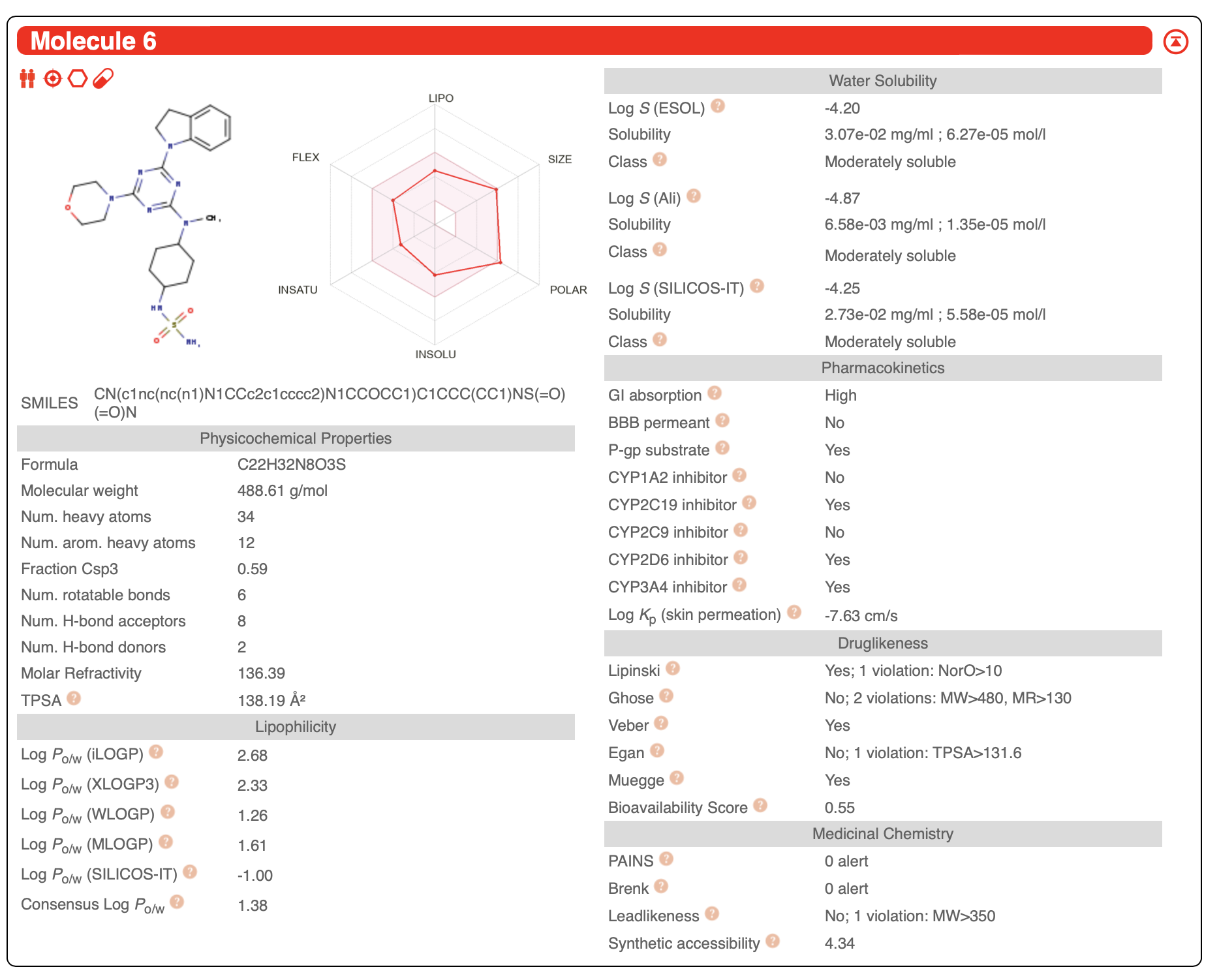}
    \caption{\textbf{SwissADME evaluation of GXA112.}}
    \label{fig:adme6}
\end{figure}
\newpage
\begin{figure}[h]
    \centering
    \includegraphics[width=0.9\textwidth]{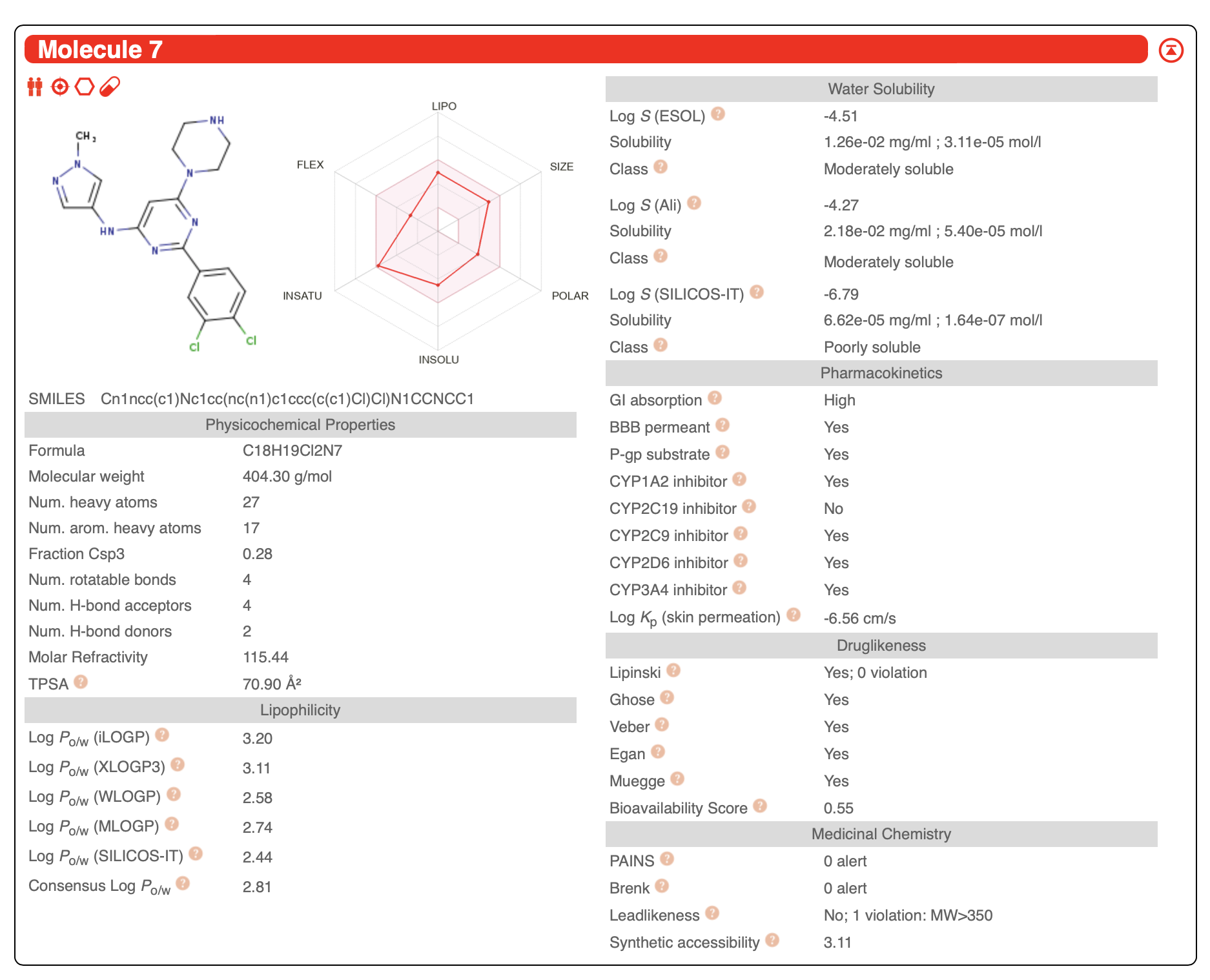}
    \caption{\textbf{SwissADME evaluation of GXA56.}}
    \label{fig:adme7}
\end{figure}
\newpage
\begin{figure}[h]
    \centering
    \includegraphics[width=0.9\textwidth]{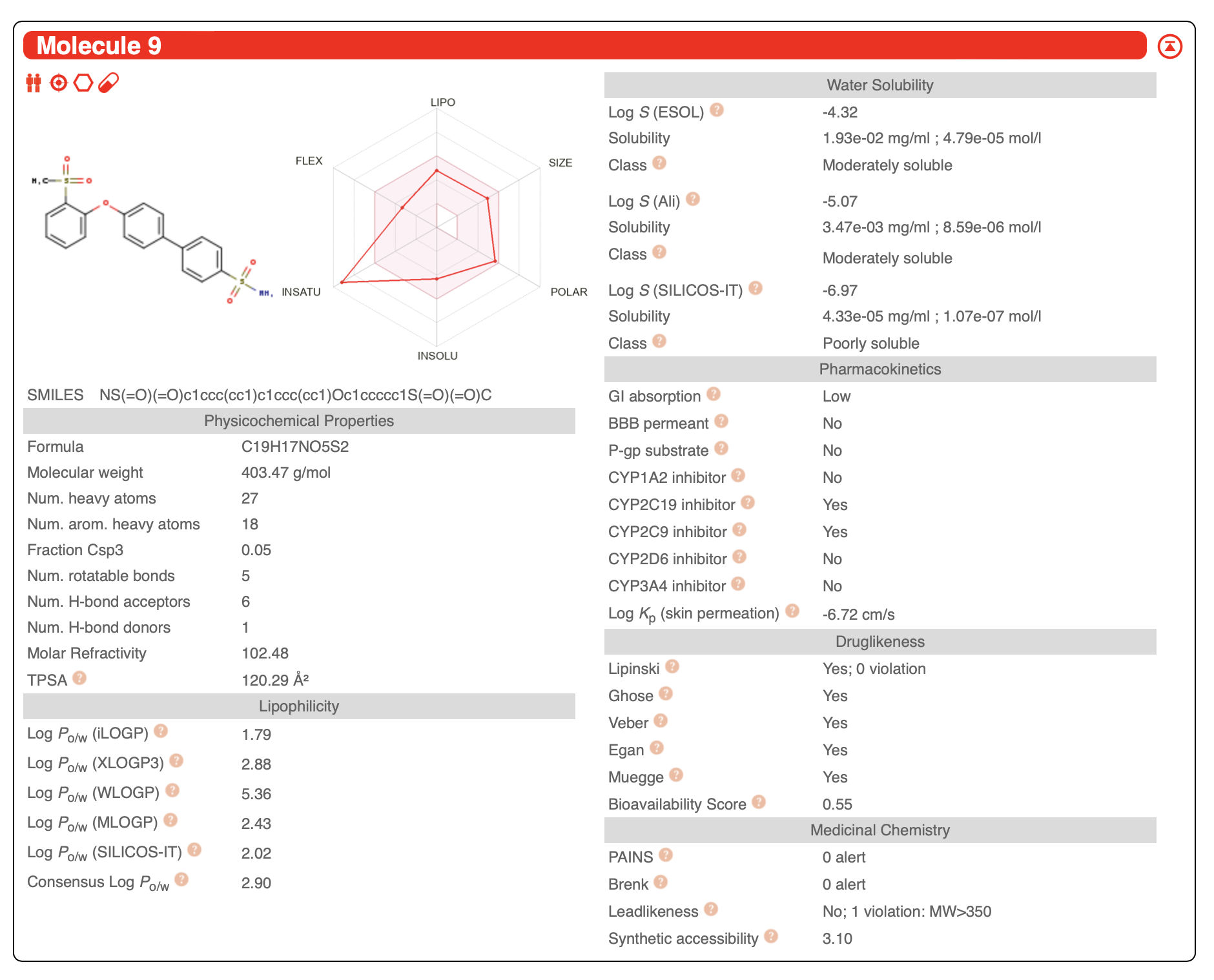}
    \caption{\textbf{SwissADME evaluation of GEN725.}}
    \label{fig:adme9}
\end{figure}
\newpage
\begin{figure}[h]
    \centering
    \includegraphics[width=0.9\textwidth]{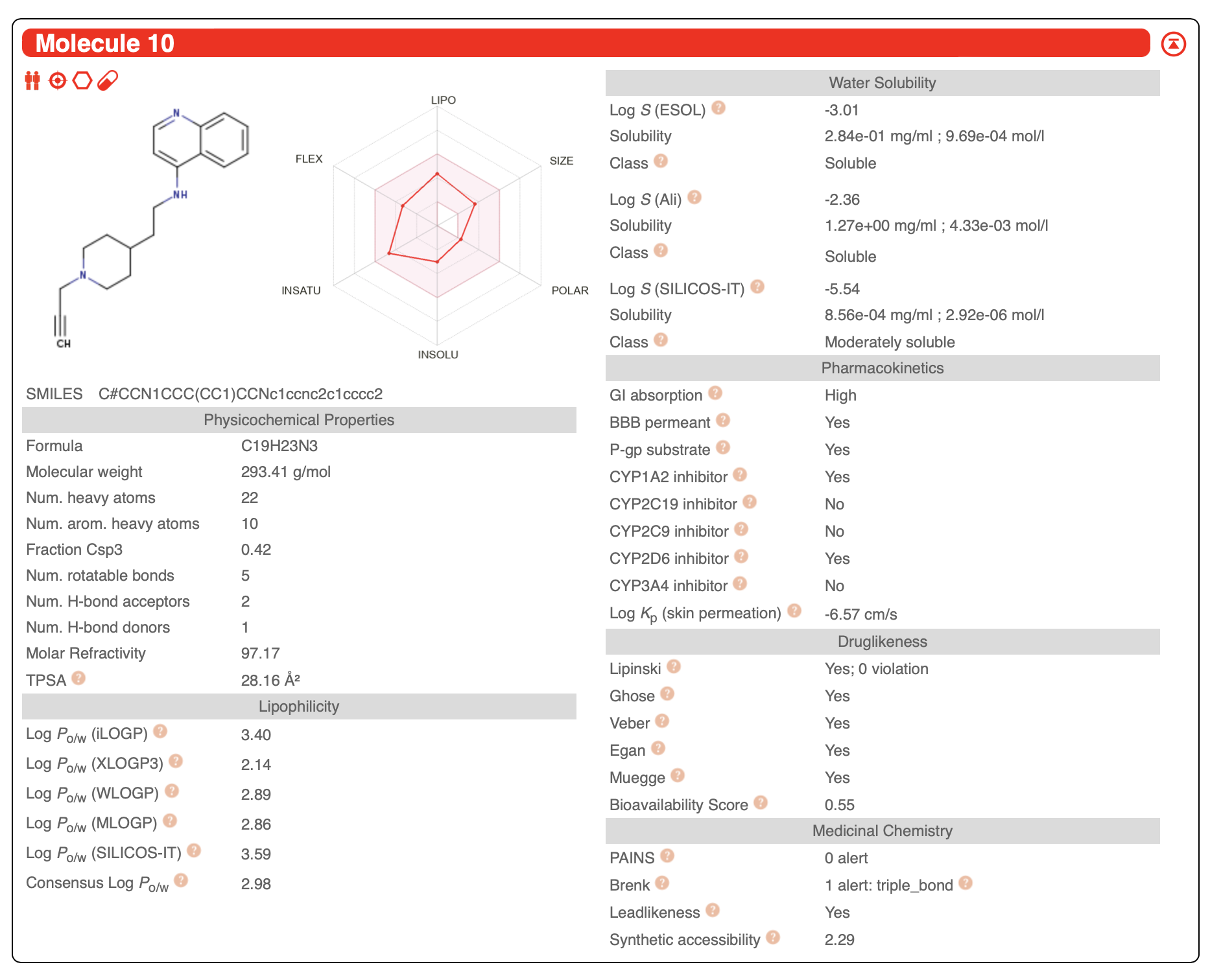}
    \caption{\textbf{SwissADME evaluation of GEN727.}}
    \label{fig:adme10}
\end{figure}
\newpage

\section{Supplementary Details}
\subsection{CLaSS algorithm details}
\label{sec:class_alg}

\begin{algorithm}[h!]
\caption{Conditional Latent (attribute) Space Sampling (CLaSS)}
\label{alg:class}
\begin{algorithmic}[1]
\Require Trained latent variable model (e.g.~VAE), samples $\rvz_j$ drawn from domain of interest, labeled samples for each attribute $a_i$.
\State Encode training data $\rvx_j$ in latent space: $\rvz_{j,k} \sim q_{\phi}(\rvz|\rvx_j)$ for $k=1,...,K$
\State Use $\rvz_{j,k}$ to fit explicit density model $Q_{\xi}(\rvz)$ to approximate marginal posterior $q_{\phi}(\rvz)$
\State Train classifier models $q_{\xi}(a_i|\rvz)$ using labeled samples for each attribute $a_i$ to approximate probability $p(a_i|\rvx)$
\State Assuming attributes $a_i$ are conditionally independent given $\rvz$, then
    $$\hat{p_{\xi}}(\rvz|\rva) = \frac{Q_{\xi}(\rvz)\prod_i{q_{\xi}(a_i|\rvz)}}{q_{\xi}(\rvz)}$$
via Bayes' rule.
\State Let $g(\rvz) = Q_{\xi}(\rvz)$ and $M = \frac{1}{q_{\xi}(\rva)}$
\Repeat
    \State Sample from $Q_{\xi}(\rvz)$
    \State Accept with probability $\frac{f(\rvz)}{Mg(\rvz)} = \prod_i{q_{\xi}(a_i|\rvz)} \leq 1$
    \If{Accepted}
        \State Decode sample from latent and save: $\rvx \sim p_\theta(\rvx|\rvz)$
    \EndIf
\Until{Desired number of samples attained}
\State \Return Accepted samples
\end{algorithmic}
\end{algorithm}

\newpage
\subsection{Detailed synthesis steps}
\label{sec:synthesis_steps}
\begin{figure}[h]
    \centering
    \includegraphics[width=0.6\textwidth]{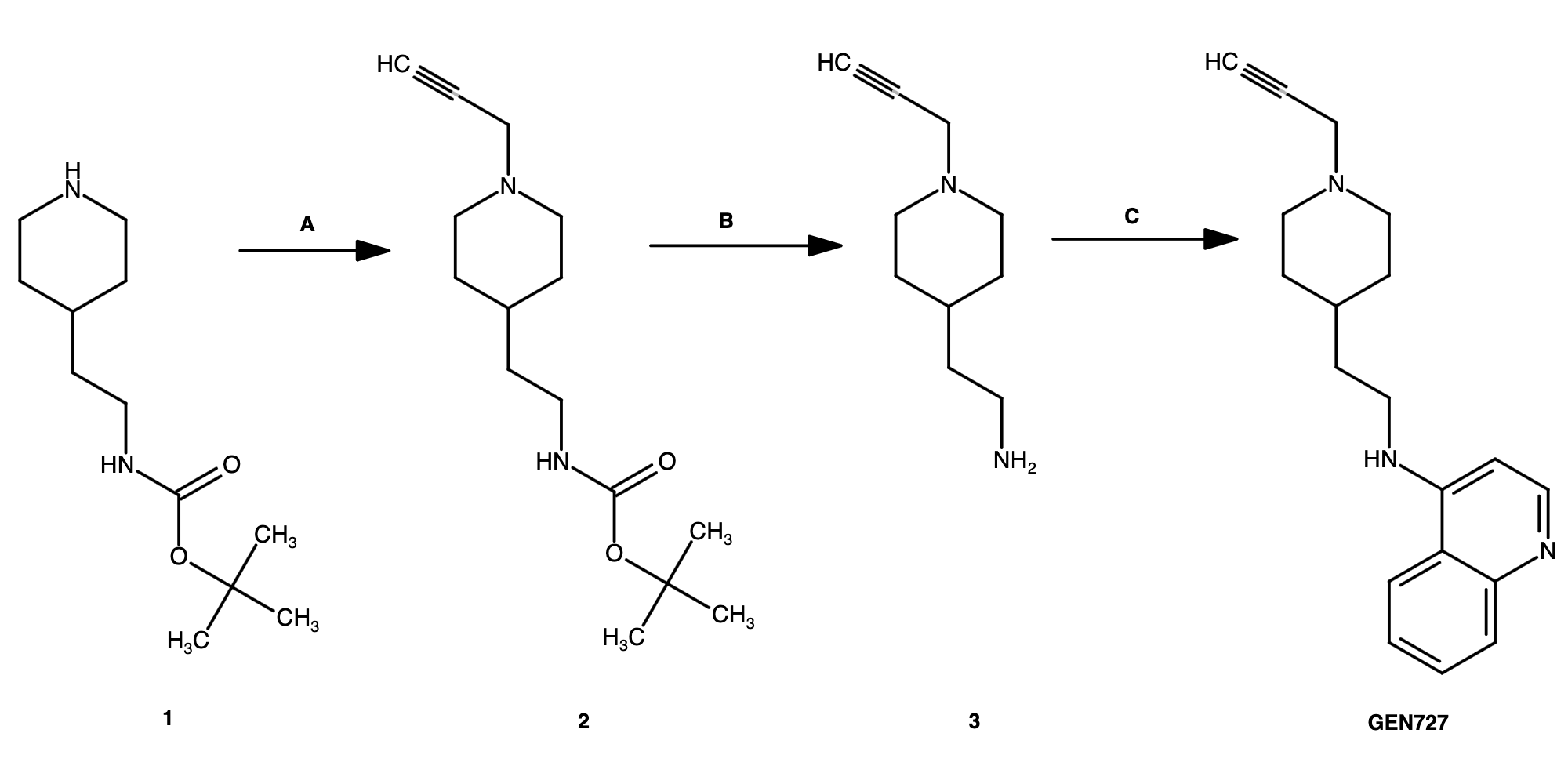}
    \caption{GEN727 synthesis route}
    \label{fig:gen727_synthesis}
\end{figure}
\textbf{Step A: }
A mixture of compound \textbf{1} (0.5 g, 2.2 mmol), propargyl bromide (0.4 g, 3.3 mmol) and potassium carbonate (0.6 g, 4.4 mmol) was suspended in acetonitrile (20 mL) and the reaction mixture was heated to \SI{60}{\degreeCelsius} for 18 h. The solids were removed via filtration and the solvent was removed \textit{in vacuo}. The residue was diluted with an aqueous $\mathrm{NaHSO_4}$ solution (50 mL) and washed with dichloromethane (2 × 20 mL); the aqueous layer was basified with NaOH to pH=14, and extracted with dichloromethane (3 × 30 mL). The organic extracts were combined, dried over $\mathrm{Na_2SO_4}$ and concentrated \textit{in vacuo} to obtain crude \textbf{2} (0.4 g) which was used in the next step without purification.

\textbf{Step B: }
Crude compound \textbf{2} (0.4 g) was dissolved in methanol (10 mL) and a hydrogen chloride solution in dioxane (20 mL) was added. The reaction mixture was stirred for 18 h at \SI{20}{\degreeCelsius}. The volatiles were removed \textit{in vacuo} to obtain crude \textbf{3} (0.32 g) as a hydrochloride salt. 

\textbf{Step C: }
Crude compound \textbf{3} (0.32 g) was dissolved in DMSO (5 mL), 4-chloroquinoline (0.330 g, 2 mmol) and DIPEA (0.65 g, 5 mmol) were added to the solution. The reaction mixture was stirred at \SI{100}{\degreeCelsius} for 48 h and purified via preparative HPLC to obtain \textbf{GEN727} (2 fractions: 0.0257 g and 0.0278 g, overall yield 9\%) as brown solid.

\begin{figure}[h]
    \centering
    \includegraphics[width=0.8\textwidth,trim={0 200px 0 200px},clip]{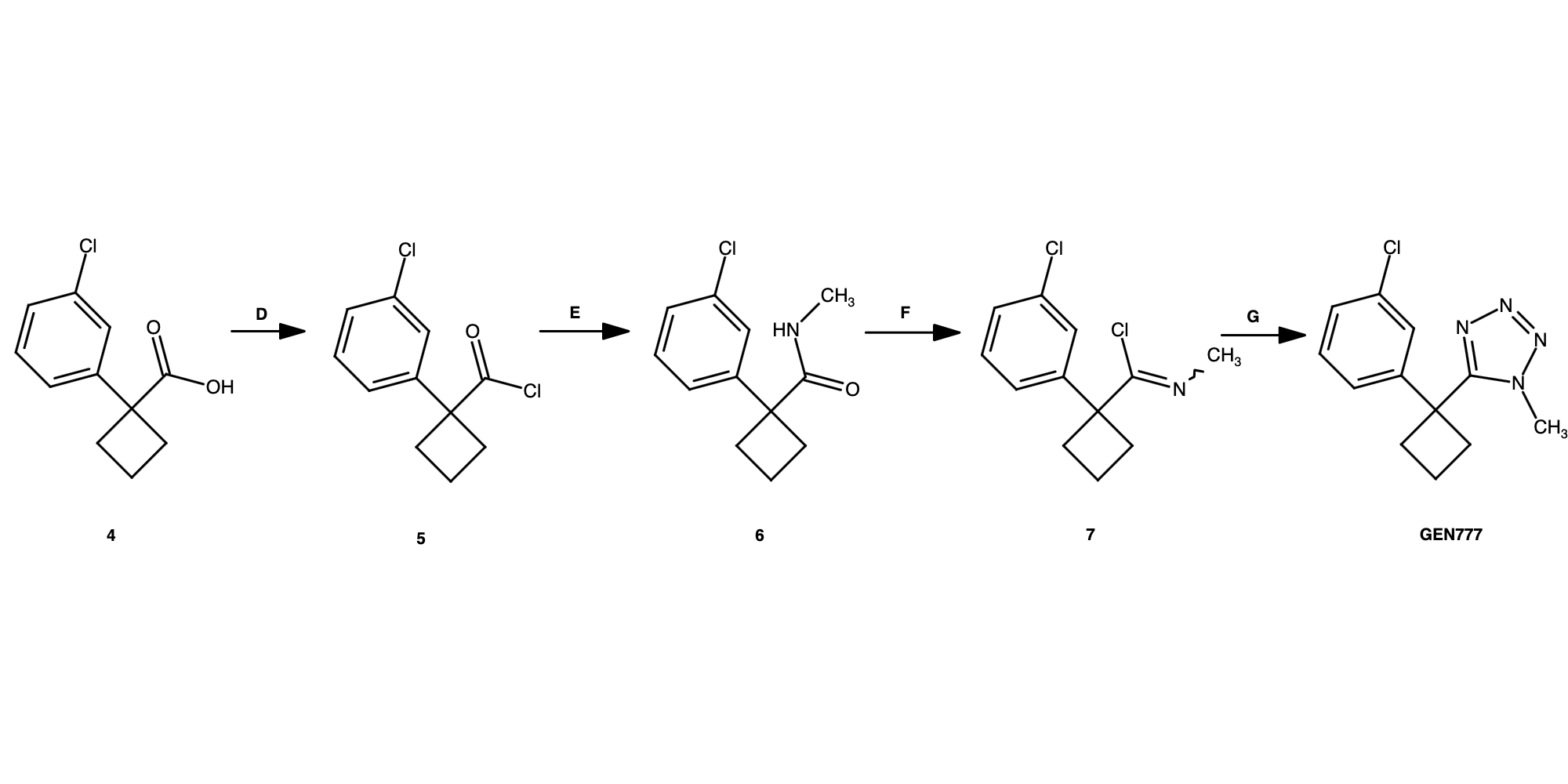}
    \caption{GEN777 synthesis route}
    \label{fig:gen777_synthesis}
\end{figure}
\textbf{Step D: }
Thionyl chloride (3 g, 25.2 mmol) was added to a solution of compound \textbf{4} (1.7 g, 6.6 mmol) in dichloromethane (10 mL) and the mixture was refluxed for 1 h and evaporated under reduced pressure to give compound \textbf{5}.

\textbf{Step E: }
To a saturated solution of aqueous methylamine (5 g), cooled to \SI{0}{\degreeCelsius}, was added compound \textbf{5} (1.8 g, 7.9 mmol). After the completion of the reaction was confirmed, the resulting mixture was extracted with MTBE. The combined organic layers were washed with brine dried over anhydrous $\mathrm{Na_2SO_4}$ and evaporated under reduced pressure to obtain 1 g of compound \textbf{6}, which was used in the next step without further purification.

\textbf{Step F: }
To a solution of compound \textbf{6} (1 g, 4.5 mmol) in dichloromethane (700 mL) was added $\mathrm{PCl_5}$ (1.4 g, 6.72 mmol). The reaction mixture was stirred for 2 h at r.t.\ to obtain the solution contained compound \textbf{7} which was not isolated but directly used in the next step.

\textbf{Step G: }
To the solution of compound \textbf{7} in dichloromethane (from \textbf{Step F}) was added TMSN3 (2.5 g, 21.7 mmol). The reaction mixture was stirred overnight at r.t.\ and evaporated under reduced pressure. The residue was purified by HPLC to give 0.130 g of \textbf{GEN777}.

\begin{figure}[h]
    \centering
    \includegraphics[width=0.7\textwidth,trim={0 100px 0 100px},clip]{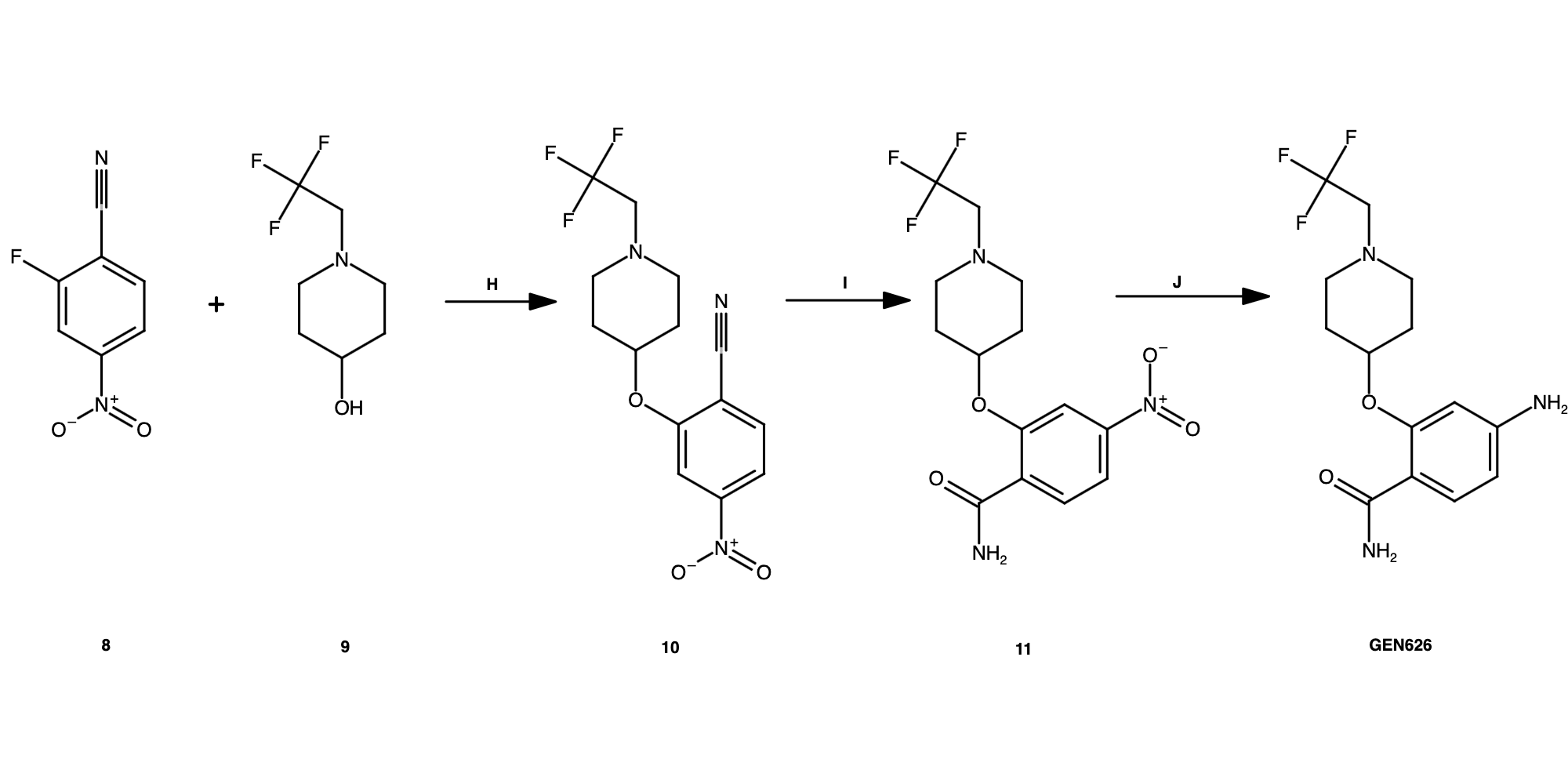}
    \caption{GEN626 synthesis route}
    \label{fig:gen626_synthesis}
\end{figure}
\textbf{Step H: }
To a solution of compound \textbf{9} (0.55 g, 3 mmol) in dry DMF (15 mL), sodium hydride (as 60\% suspension in mineral oil, 0.132 g, 3.3 mmol) was added in one portion. The mixture was stirred at \SI{40}{\degreeCelsius} for 30 min and compound \textbf{8} (0.5 g, 3 mmol) was added. The reaction mixture was stirred at \SI{20}{\degreeCelsius} for 18 h, diluted with water (100 mL), and extracted with ethyl acetate (3 × 30 mL). The combined organic layers were washed with water (4 × 50 mL), dried over $\mathrm{Na_2SO_4}$ and concentrated \textit{in vacuo} to obtain the crude material which was purified via column chromatography ($\mathrm{CHCl_3}$:MeOH 10:1 as eluent) to afford \textbf{10} (0.18 g, 0.55 mmol, 18\% yield) as yellow oil. 

\textbf{Step I: }
Compound \textbf{10} (0.18 g, 0.55 mmol) was suspended in conc. $\mathrm{H_2SO_4}$ (5 mL) and the reaction mixture was heated to \SI{60}{\degreeCelsius} for 2 h, cooled with ice and diluted with an aqueous $\mathrm{Na_2CO_3}$ solution to basic pH. The resulting mixture was extracted with ethyl acetate (3 × 30 mL); the organic layer was dried over $\mathrm{Na_2SO_4}$ and concentrated \textit{in vacuo} to obtain \textbf{11} (0.16 g, 0.46 mmol, 84\% yield) as yellow solid.

\textbf{Step J: }
To a solution of compound \textbf{11} (0.16 g, 0.46 mmol) in methanol (10 mL), Pd/C (10\%w, 0.100 g) was added. The reaction mixture was evacuated and backfilled with hydrogen and then stirred for 18 h. The catalyst was removed via filtration and the solvent was removed \textit{in vacuo} to obtain the crude material which was purified via preparative HPLC to obtain \textbf{GEN626} (0.0614 mg, 42\% yield) as white solid.

\begin{figure}[h]
    \centering
    \includegraphics[width=0.6\textwidth,trim={0 100px 0 100px},clip]{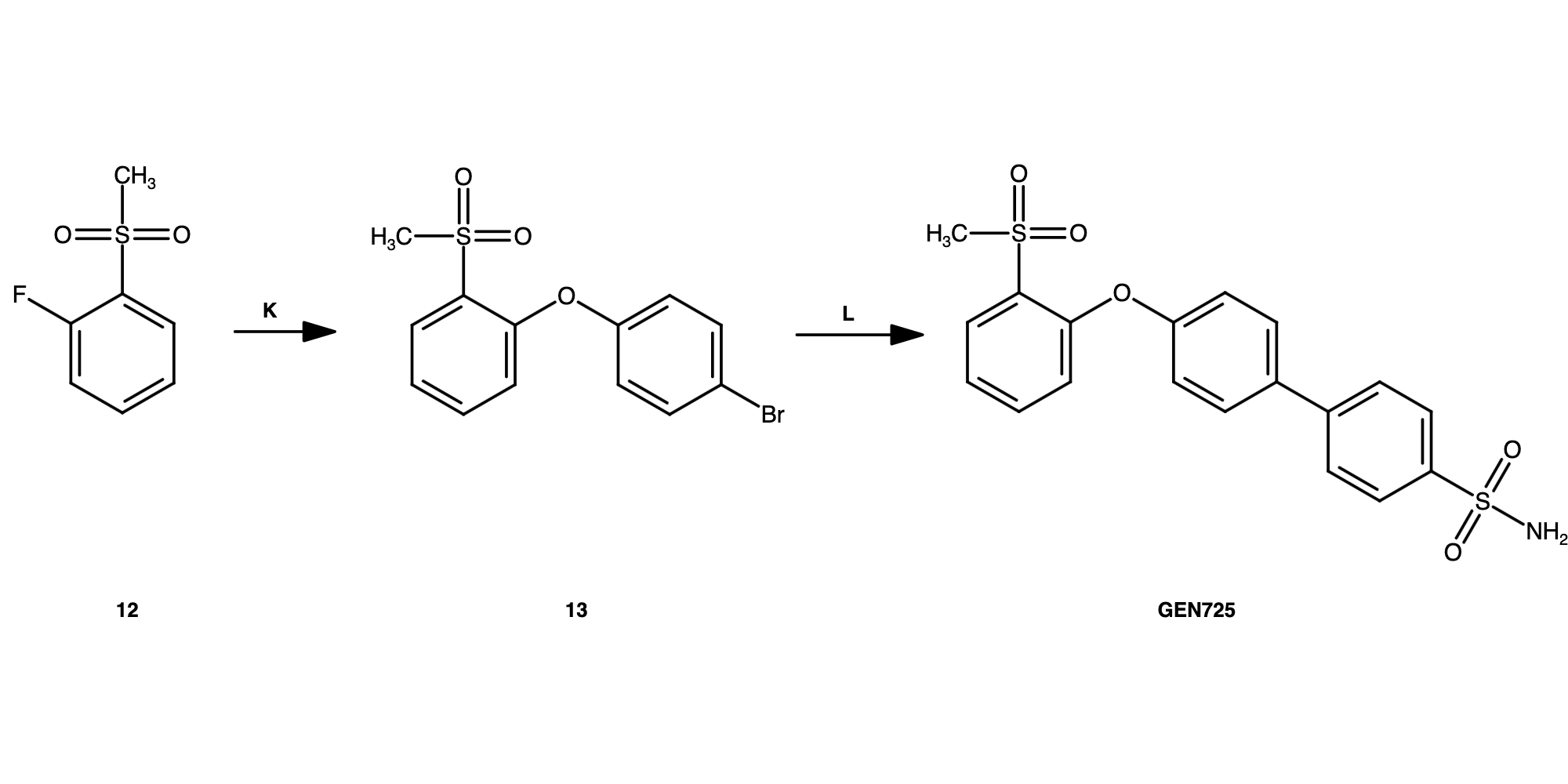}
    \caption{GEN725 synthesis route}
    \label{fig:gen725_synthesis}
\end{figure}
\textbf{Step K: }
To a suspension of NaH (0.250 g, 6.31 mmol, 60\% dispersion in mineral oil) in DMF (5 mL) was added dropwise a solution of 4-bromophenol (1.09 g, 6.31 mmol) in DMF (5 mL). The mixture was stirred for 1 h and compound \textbf{12} (1 g, 5.74 mmol) was added. The reaction mixture was stirred at \SI{100}{\degreeCelsius} overnight, cooled to r.t.\ and poured into ice (100 mL). The precipitate was filtered and washed with water (3 × 10 mL) and with hexanes. The solid was dried \textit{in vacuo} to give \textbf{13} (1.72 g, 92\%).

\textbf{Step L: }
To a mixture of compound \textbf{13} (1 g, 3.06 mmol), 4-(4,4,5,5-tetramethyl-1,3,2-dioxaborolan-2-yl)-benzenesulfonamide (1.04 g, 3.67 mmol) and sodium carbonate (0.81 g, 7.65 mmol) in  a mixture of dioxane and water (9:1, 10 mL) was added XPhos Pd G3 (0.260 g, 0.36 mmol) under an inert atmosphere. The reaction mixture was stirred for 16 h at \SI{95}{\degreeCelsius} (oil bath), cooled to r.t., diluted with water (10 mL) and extracted with EtOAc (2 × 10 mL). The combined organic layers were dried over $\mathrm{Na_2SO_4}$ and evaporated under reduced pressure. The residue was purified by preparative HPLC to give \textbf{GEN725} (0.304 g, 25\%).

\begin{figure}[H]
    \centering
    \includegraphics[width=0.6\textwidth]{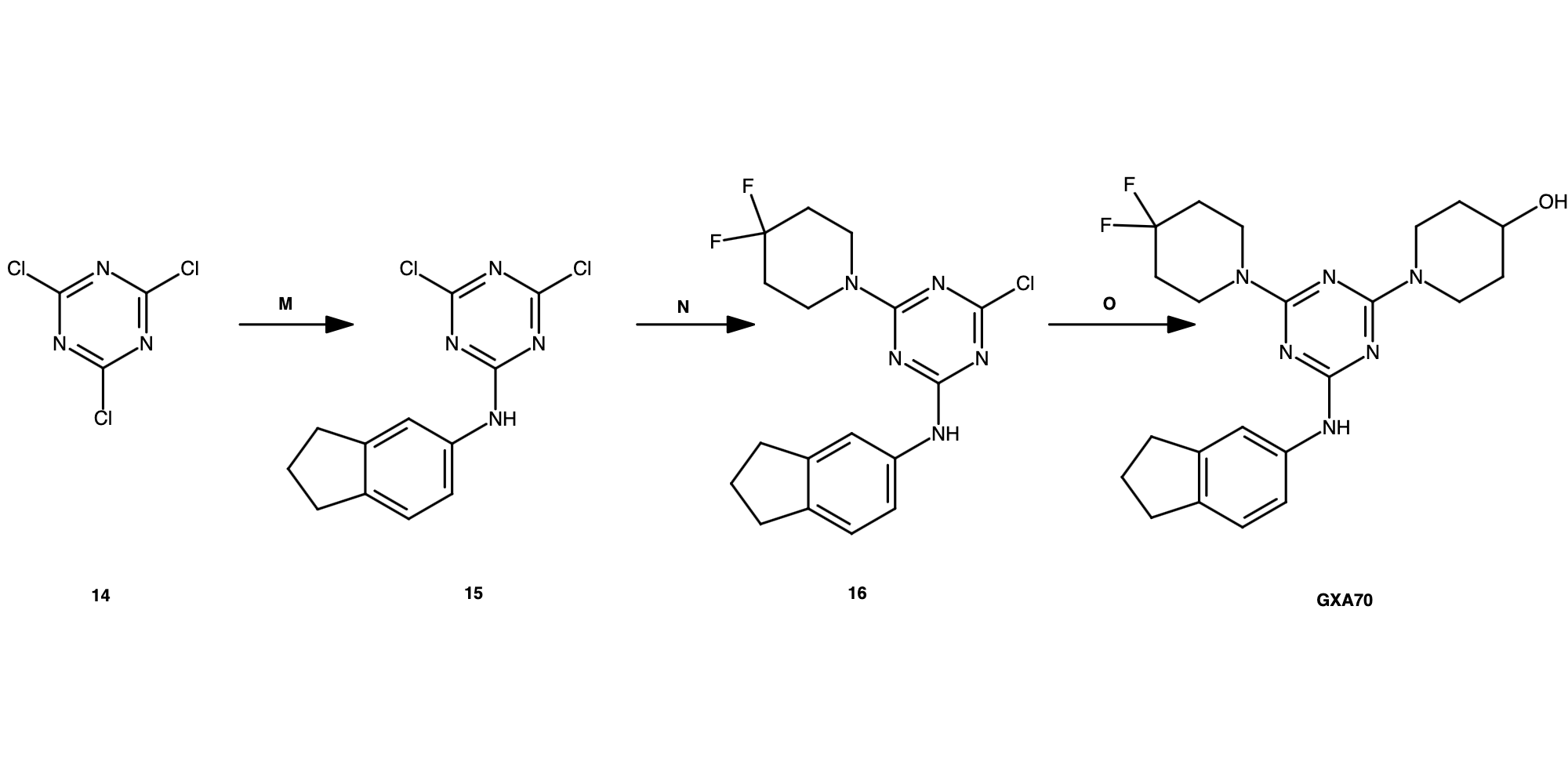}
    \caption{GXA70 synthesis route}
    \label{fig:gxa70_synthesis}
\end{figure}

\textbf{Step M: }
To the solution of compound \textbf{14} (2.0 g, 10.8 mmol, 1 eq) in 30 mL of dichloromethane cooled to \SI{0}{\degreeCelsius}, 1.2 equivalent of DIPEA was added dropwise under continuous stirring. Thereafter 1 eq of 2,3-dihydro-1H-inden-5-amine dissolved in 10 mL of dichloromethane was added. The resulting mixture was stirred at ambient temperature overnight. Thereafter resulting solution was washed with water, 3 × 20 mL. Then organic layer was dried over anhydrous sodium sulfite and evaporated \textit{in vacuo}. Resulting compound \textbf{15} with 90\% purity was used in the next step without additional purification. Yield 92\%, 2.8 g.

\textbf{Step N: }
To the solution of compound \textbf{15} (2.8 g, 9.7 mmol, 1 eq) in 40 mL of dichloromethane 2.2 equivalents of DIPEA was added dropwise at \SI{0}{\degreeCelsius} under continuous stirring. The resulting solution was stirred for additional 30 min and then 4,4-difluoropiperidine hydrochloride was added portionwise (1.1 eq). The resulting mixture was left to stir at ambient temperature overnight. Next day the reaction solution was washed with water, 3 × 20 mL. Resulting organic layer was dried with anhydrous sodium disulfite and evaporated under reduced pressure. The resulting product \textbf{16} with 90\%+ purity was used in the next step without any additional purification. Yield 91\%, 3.3 g.

\textbf{Step O: }
To the solution of compound \textbf{16} (3.3 g, 9.1 mmol, 1 eq) in 40 mL of DMF cooled to \SI{0}{\degreeCelsius}. 1.2 eq of DIPEA was added dropwise under stirring. Then mixture was stirred for additional 30 min and 1.05 eq of the corresponding amine in 10 mL of DMF was added. Resulting reaction mixture was stirred at \SI{80}{\degreeCelsius} overnight. Thereafter all volatiles were evaporated \textit{in vacuo} and residue was washed with water twice. Resulting precipitate was dissolved in 50 mL of dichloromethane, dried with anhydrous sodium sulfate and filtered through the Celite pad. Resulting filtrate was evaporated under reduced pressure to give \textbf{GXA70} with 95\% purity. Yield 70\%, 2.7 g.

\begin{figure}[h]
    \centering
    \includegraphics[width=0.7\textwidth]{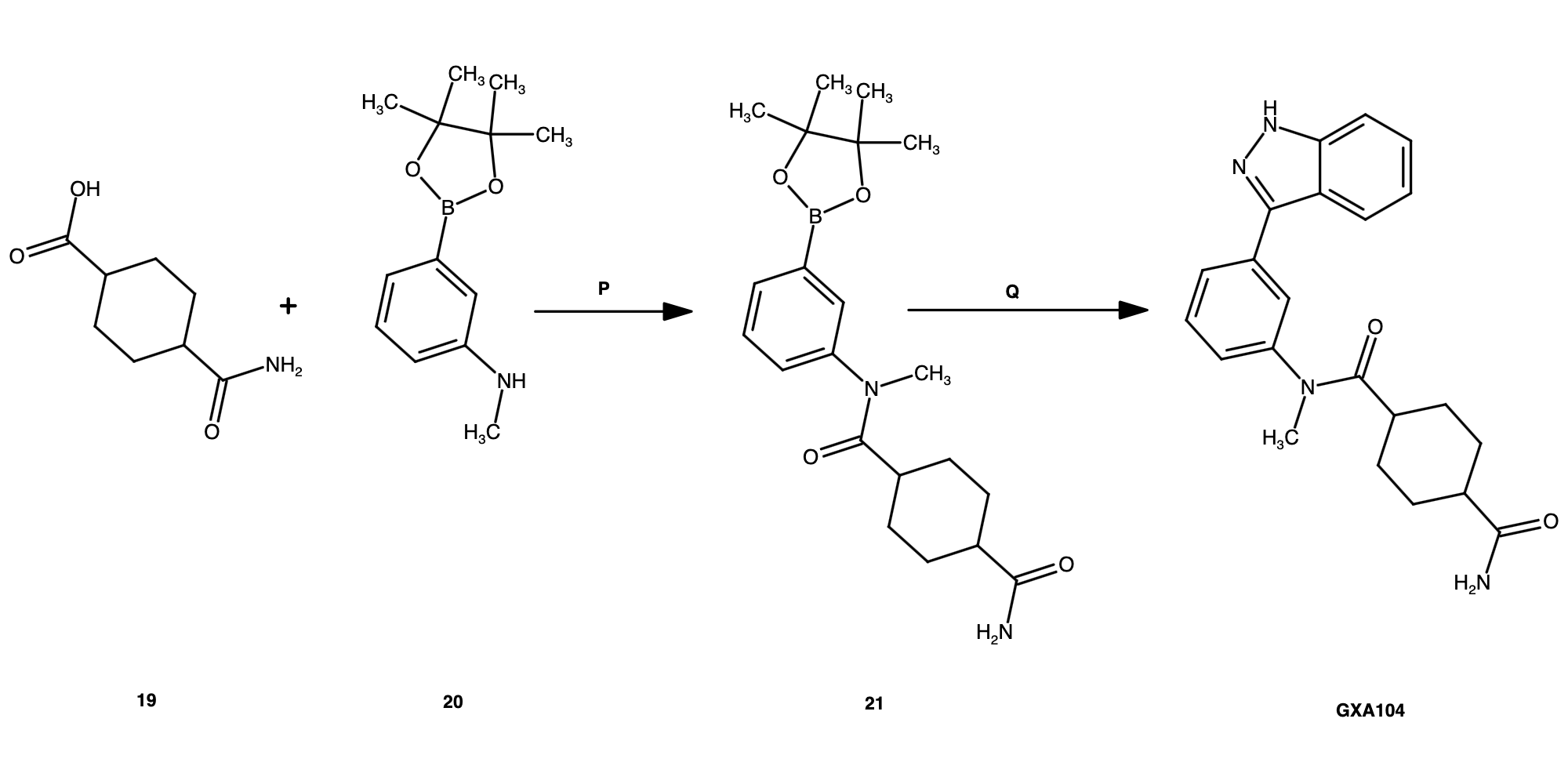}
    \caption{GXA104 synthesis route}
    \label{fig:gxa104_synthesis}
\end{figure}
\textbf{Step P: }
To a solution of compound \textbf{19} (0.975 g, 5.70 mmol), compound \textbf{20} (1.21 g, 5.18 mmol) and HOBt (0.775 g, 5.70 mmol) in dry DMA (10 mL), cooled to \SI{0}{\degreeCelsius}, was added dropwise EDC (0.964 g, 6.31 mmol) and the reaction mixture was stirred overnight at r.t., diluted with water and extracted with ethyl acetate. The combined organic layers were washed with water, dried over anhydrous $\mathrm{Na_2SO_4}$ and evaporated under reduced pressure. The residue was crystallized from the minimum amount of ethyl acetate to obtain 1.26 g of compound \textbf{21} (63\% yield).

\textbf{Step Q: }
A solution of compound \textbf{21} (0.410 g, 1.06 mmol), 3-iodo-1H-indazole (0.259 g, 1.06 mmol), $\mathrm{Pd(PPh_3)_4}$ (0.061 g, 0.05 mmol) and $\mathrm{Na_2CO_3}$ (0.225 g, 2.13 mmol) in a mixture of dioxane/water (4:1) (5 mL) was stirred overnight at \SI{90}{\degreeCelsius} under an argon atmosphere. The cooled mixture was diluted with water and extracted with dichloromethane. The combined organic layers were washed with water, dried over anhydrous $\mathrm{Na_2SO_4}$ and evaporated under reduced pressure. The residue was purified by column chromatography to obtained by HPLC to afford 0.180 g of compound \textbf{GXA104} (45\% yield).

\begin{figure}[h]
    \centering
    \includegraphics[width=0.8\textwidth,trim={0 100px 0 100px},clip]{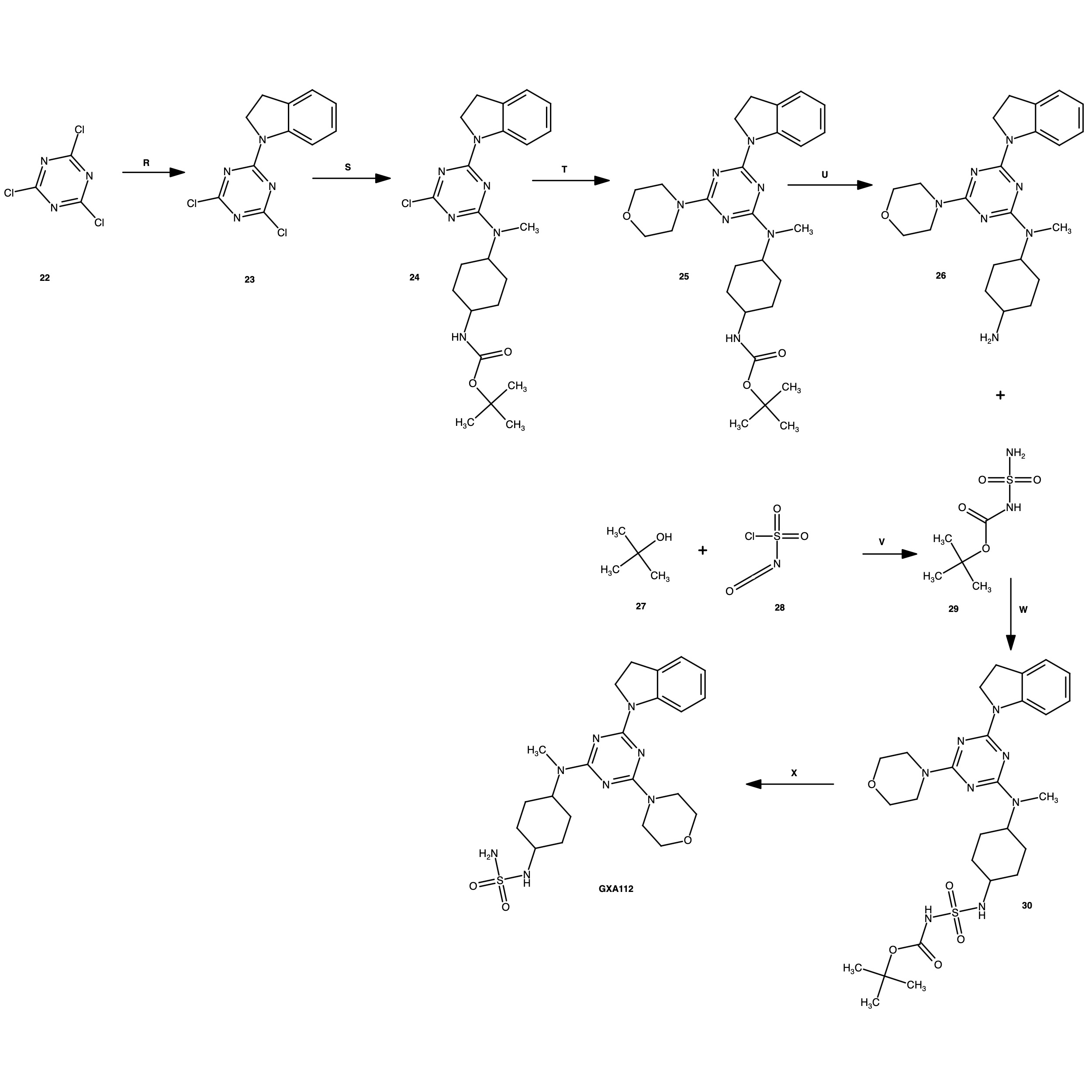}
    \caption{GXA112 synthesis route}
    \label{fig:gxa112_synthesis}
\end{figure}
\textbf{Step R: }
To a stirred solution of compound \textbf{22} (2 g, 11 mmol) in dichloromethane (40 mL) at \SI{0}{\degreeCelsius} were added DIPEA (2.3 mL, 13.2 mmol) and 2,3-dihydro-1H-indole (1.22 mL) and the resulting mixture was stirred at r.t.\ for 16 h. After that the reaction mixture was diluted with water; the organic phase was washed with water and brine, dried over $\mathrm{Na_2SO_4}$ and evaporated to obtain crude product \textbf{23} (1.1 g), which was used in the next step without further purification.

\textbf{Step S: }
To a stirred solution of compound \textbf{23} (1.1 g, 4 mmol) in dichloromethane (40 mL) at \SI{0}{\degreeCelsius} were added DIPEA (0.86 mL, 4.94 mmol) and tert-butyl N-[4-(methylamino)cyclohexyl]carbamate (0.94 g) and the resulting mixture was stirred at r.t.\ for 16 h. After that the reaction mixture was diluted with water; the organic phase was washed with water and brine, dried over $\mathrm{Na_2SO_4}$ and evaporated under reduced pressure to obtain crude product \textbf{24} (1.5 g), which was used in the next step without further purification.

\textbf{Step T: }
To a stirred solution of compound \textbf{24} (1.5 g, 3 mmol) in dichloromethane (30 mL) at r.t.\ were added DIPEA (0.68 mL, 3.90 mmol) and morpholine (0.28 mL, 3.25 mmol) and the resulting mixture was stirred at r.t.\ for 16 h. After that an additional amount of DIPEA (0.68 mL, 3.90 mmol) and morpholine (0.28 mL, 3.25 mmol) was added and the resulting mixture was stirred at r.t.\ for another 16 h. Then the reaction mixture was diluted with water; the organic phase was washed with water and brine, dried over $\mathrm{Na_2SO_4}$ and evaporated under reduced pressure to obtain crude product \textbf{25} (1.7 g), which was used in the next step without further purification.

\textbf{Step U: }
To a stirred solution of compound \textbf{25} (1.7 g, 3 mmol) in dichloromethane (25 mL) was added 4 M HCl solution in dioxane and the resulting mixture was stirred at r.t.\ for 8 h. After that the reaction mixture was evaporated under reduced pressure to obtain crude product \textbf{26} (1.2 g), which was used in the next step without further purification.

\textbf{Step V: }
To a stirred solution of compound \textbf{27} (0.7 mL, 7.4 mmol) in diethyl ether (10 mL) was added compound \textbf{28} (0.15 mL, 0.243 g, 1.7 mmol) at \SI{-78}{\degreeCelsius} and the resulting mixture was stirred at r.t.\ for 1 h. The reaction mixture was evaporated without heating to obtain crude product \textbf{29}, which was immediately used in the next step.

\textbf{Step W: }
To a stirred suspension of compound \textbf{26} (0.8 g, 1.7 mmol) in dichloromethane (10 mL) at \SI{0}{\degreeCelsius} was added $\mathrm{Et_3N}$ (0.76 mL, 5.45 mmol) followed by a solution of compound \textbf{29} in dichloromethane (3 mL) and the resulting mixture was stirred at r.t.\ for 16 h. After that the reaction mixture was diluted with water; the organic phase was washed with water and brine, dried over $\mathrm{Na_2SO_4}$ and evaporated under reduced pressure to obtain crude product \textbf{30} (0.8 g), which was used in the next step without further purification.

\textbf{Step X: }
To a stirred solution of compound \textbf{30} (0.8 g, 1.4 mmol) in dichloromethane (5 mL) was added 4 M HCl solution in dioxane (1 mL) and the resulting mixture was stirred at r.t.\ for 8 h. Then the reaction mixture was evaporated under reduced pressure, the obtained residue was diluted with water, basified with a $\mathrm{NaHCO_3}$ solution and extracted with dichloromethane. The combined organic phase was washed with water, dried over $\mathrm{Na_2SO_4}$ and evaporated under reduced pressure to obtain crude product. The crude product was purified by HPLC to obtain 0.01 g of \textbf{GXA112}.

\begin{figure}[H]
    \centering
    \includegraphics[width=0.7\textwidth]{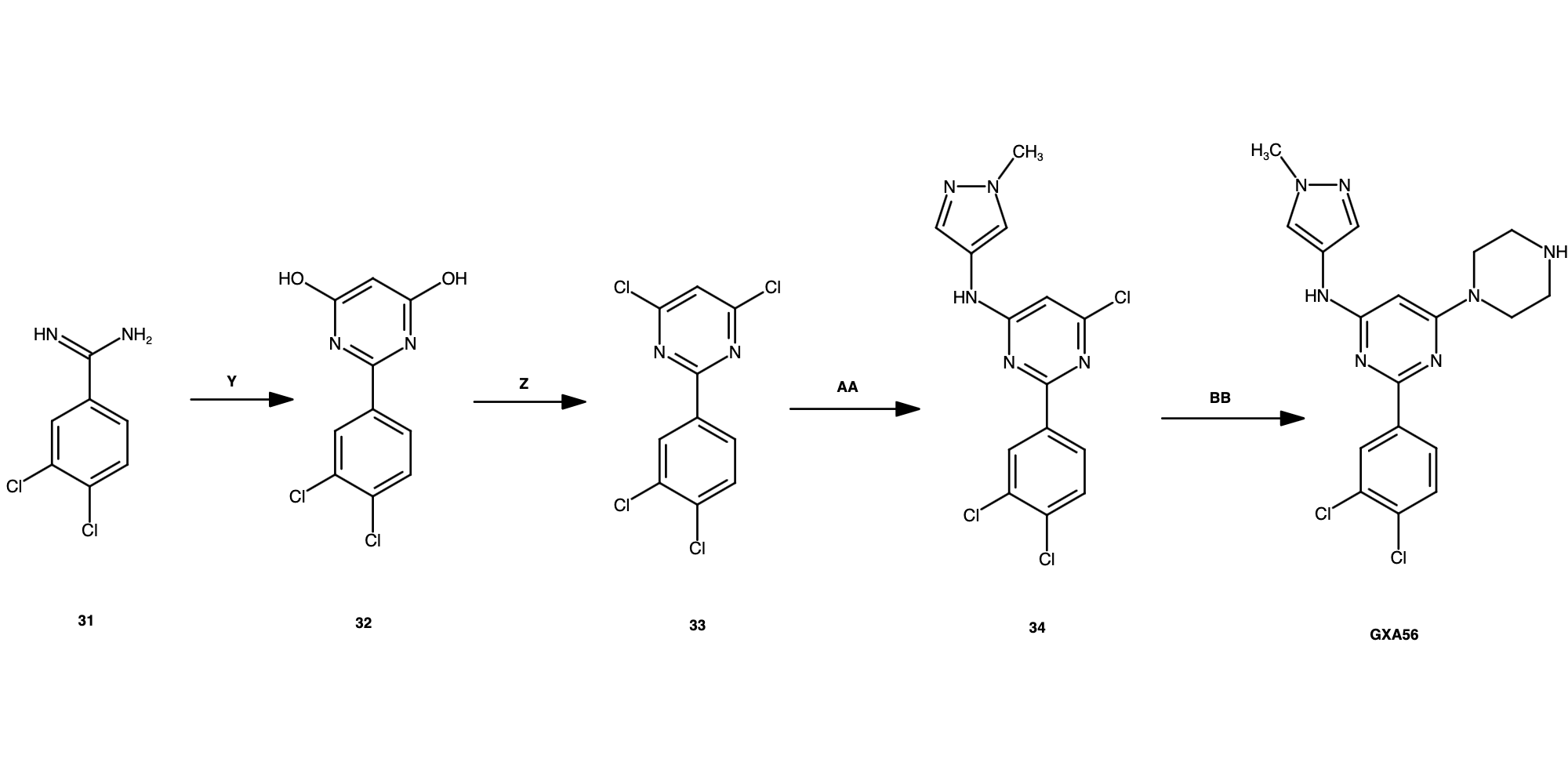}
    \caption{GXA56 synthesis route}
    \label{fig:gxa56_synthesis}
\end{figure}

\textbf{Step Y: }
Metallic sodium (0.47g, 2.2 eq) was dissolved portionwise in 50 mL of dry methanol. Then \textbf{31} (2g, 1eq) and diethylmalonate (1.43g, 1eq) were added thereto. Resulting mixture was stirred at \SI{60}{\degreeCelsius} overnight. Formed precipitate was filtered off, dissolved in water and acidified with sodium hydrosulphate to pH 2, then stirred for 20 min and filtered to obtain compound \textbf{32} as yellow solid. Yield 66\%, 1.8 g.

\textbf{Step Z: }
To compound \textbf{32} (1.8g, 1 eq) in 15 mL of $\mathrm{POCl_3}$ was added 0.15 mL of DIPEA and resulting mixture was stirred at reflux for 3 hours. The resulting mixture was evaporated, quenched with ice and saturated solution of anhydrous potassium carbonate up to pH 12. Then the solution was left to stir at ambient temperature for 20 min. The resulting precipitate was filtered off and washed with water several times to obtain compound \textbf{33}. Yield 26\%, 0.53 g.

\textbf{Step AA: }
1-Methyl-1H-pyrazol-3-amine (0.175 g, 1 eq), sodium iodide (0.27 g, 1 eq) and DIPEA (0.46 g, 2 eq) were added subsequently to a solution of compound \textbf{33} (0.5 g, 1 eq) in 10 mL of dry DMF. The resulting mixture was stirred at \SI{80}{\degreeCelsius} overnight. After mixture was cooled to r.t.\ and then diluted with water, formed precipitate was filtered and washed with water to give compound \textbf{34}. Yield 58\%, 0.35g.

\textbf{Step BB: }
Compound \textbf{34} (0.35 g, 1 eq) together with piperazine (0.17 g, 2 eq) and anhydrous potassium carbonate (0.27 g, 2 eq) was mixed in 15 mL of dry DMF and heated up to \SI{120}{\degreeCelsius} overnight. Thereafter a mixture was cooled, and insoluble material was filtered out. Then organic layer was evaporated and purified by HPLC to give \textbf{GXA56} as a white solid. Yield 22.5\%, 0.08g.

\end{document}